\documentclass[12pt]{article}

\usepackage{amssymb}
\usepackage{amsmath}

\newtheorem{thm}{Theorem}[section]
\newtheorem{cor}[thm]{Corollary}
\newtheorem{lem}[thm]{Lemma}
\newtheorem{prop}[thm]{Proposition}
\newtheorem{rem}[thm]{Remark}
\newtheorem{defi}[thm]{Definition}
\newtheorem{conj}[thm]{Conjecture}

\newcommand{\pen}{\circ \! |\! \circ}

\begin{document}

\title{Adsorption of Lattice Polymers with Quenched Topologies}
\author{Neal Madras\footnote{Supported in part by a Discovery Grant from the Natural Sciences
and Engineering Research Council of Canada.}
\\ Department of Mathematics and Statistics \\ York University \\ 4700 Keele Street \\
Toronto, Ontario  M3J 1P3,  Canada \\   {\tt madras@yorku.ca}  
\\   ORCID iD:    0000-0003-2981-3577  }
\maketitle

\begin{abstract}
We introduce a framework for adsorption
of a single polymer in which the topology of the 
polymer is quenched before adsorption, in contrast to more standard adsorption models having 
annealed topology.      
Our ``topology'' refers either to the precise branching structure of a branched polymer (in
any dimension), or else to the knot type of a ring polymer in three dimensions.  
The quenched topology is chosen uniformly at random from 
all lattice polymers 
of a given size in one of four classes (lattice animals, trees, combs, or rings), and we then consider 
adsorption of the subclass of configurations that have the quenched topology.
When the polymer-surface attraction increases without bound, the quenched topological structure 
keeps a macroscopic fraction of monomers off the surface, 
in contrast with annealed models that asymptotically 
have 100\% of monomers in the surface.
We prove properties of the limiting free energy and the critical point in each model, although 
important open questions remain.
We pay special attention to the class of comb polymers, which admit some rigorous answers
to questions that otherwise remain open.
Since the class of all combs was not previously examined rigorously in full generality,
we also prove the existence of its growth constant and its limiting free energy for annealed adsorption.
\end{abstract}



\section{Introduction}
  \label{sec.intro}

This paper considers the effect of polymer topology  on the adsorption transition of polymers.  
Phenomena of polymer adsorption at surfaces play an important role in chemical physics and 
physical chemistry \cite{Fleer,Rad}.
Theoretical models of adsorption provide fertile ground for the mathematical study of
phase transitions \cite{DeLook,Gi,denH,KlSk,RyWh,SoWh}.

Our treatment will include several different variants of models in a lattice of dimension two or more. 
This introductory section provides a general overview, while  
Section \ref{sec-adsmain} will provide more details.

Firstly, the class of polymers
that we consider could be \textit{branched polymers} or \textit{linear polymers}
(Figure \ref{fig.shapes}).
Branched polymer configurations
are modelled by lattice trees or lattice animals, while linear polymers are modelled by self-avoiding
walks \cite{JvR2,MS,Va}.  Sites and edges of a lattice polymer respectively correspond to monomers 
and chemical bonds of a polymer molecule.
A sub-variant of the linear polymer is the \textit{ring polymer}, modelled by self-avoiding
polygons.  A special kind of branched polymer is the \textit{comb polymer}, 
which consists of several side chains attached to a long backbone chain at sites of degree 3. 
We shall discuss lattice combs in some detail in this paper.

\setlength{\unitlength}{.6mm}
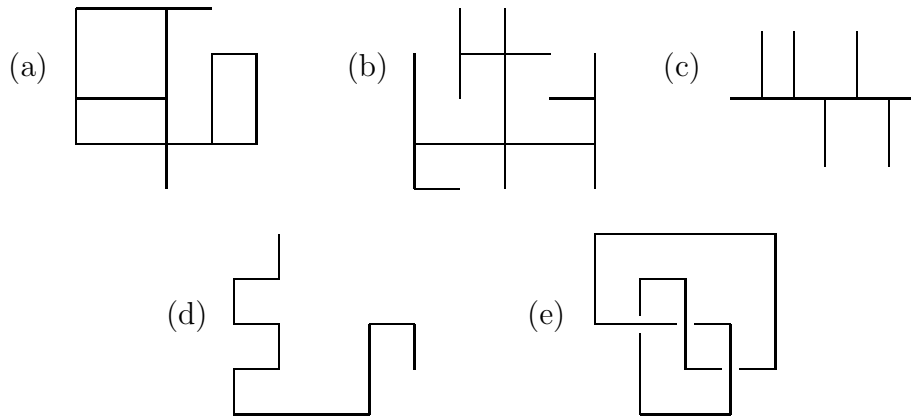
\begin{figure}
\begin{center}
\begin{picture}(200,90)
%   
% animal
\put(0,75){(a)}
\put(15,50){\begin{picture}(40,40)
\put(0,10){\line(1,0){40}}
\put(0,20){\line(1,0){20}}
\put(0,40){\line(1,0){30}}
\put(30,30){\line(1,0){10}}
\put(0,10){\line(0,1){30}}
\put(20,0){\line(0,1){40}}
\put(30,10){\line(0,1){20}}
\put(40,10){\line(0,1){20}}
\end{picture}
}
%
% tree
\put(75,75){(b)}
\put(90,50){\begin{picture}(40,40)
\put(0,0){\line(1,0){10}}
\put(0,10){\line(1,0){40}}
\put(10,30){\line(1,0){20}}
\put(30,20){\line(1,0){10}}
\put(0,0){\line(0,1){30}}
\put(20,0){\line(0,1){40}}
\put(10,20){\line(0,1){20}}
\put(40,0){\line(0,1){30}}
\end{picture}
}
%
%   comb
\put(145,75){(c)}
\put(160,50){\begin{picture}(40,40)
\put(0,20){\line(1,0){40}}
\put(7,20){\line(0,1){15}}
\put(14,20){\line(0,1){15}}
\put(21,20){\line(0,-1){15}}
\put(28,20){\line(0,1){15}}
\put(35,20){\line(0,-1){15}}
\end{picture}
}
%
%   SAW
\put(35,20){(d)}
\put(50,0){\begin{picture}(40,40)
\put(0,0){\line(1,0){30}}
\put(0,10){\line(1,0){10}}
\put(0,20){\line(1,0){10}}
\put(0,30){\line(1,0){10}}
\put(30,20){\line(1,0){10}}
\put(0,0){\line(0,1){10}}
\put(0,20){\line(0,1){10}}
\put(10,10){\line(0,1){10}}
\put(10,30){\line(0,1){10}}
\put(30,0){\line(0,1){20}}
\put(40,10){\line(0,1){10}}
\end{picture}
}
%
% SAP
\put(115,20){(e)}
\put(130,0){\begin{picture}(40,40)
\put(0,40){\line(1,0){40}}
\put(0,20){\line(1,0){18}}
\put(10,0){\line(1,0){20}}
\put(10,30){\line(1,0){10}}
\put(30,20){\line(-1,0){8}}
\put(20,10){\line(1,0){8}}
\put(40,10){\line(-1,0){8}}
\put(0,20){\line(0,1){20}}
\put(40,10){\line(0,1){30}}
\put(10,0){\line(0,1){18}}
\put(10,30){\line(0,-1){8}}
\put(20,10){\line(0,1){20}}
\put(30,0){\line(0,1){20}}
\end{picture}
}
\end{picture}
\end{center}
\caption{\label{fig.shapes} Examples of different classes of polymers:
(a) lattice animal, (b) lattice tree, (c) lattice comb, (d) self-avoiding walk, and (e) self-avoiding polygon
(shown forming a three-dimensional trefoil knot).
}
\end{figure}

Secondly, the surface to which the polymer adsorbs could either be
\textit{impenetrable}, such as 
the wall of the solution's container or
a solid plate inserted into the solution,  
or \textit{penetrable}, 
representing an interfacial layer between two fluids, as in an emulsion (Figure \ref{fig.adsorpex}).  
Our surface in three dimensional space will be a 
plane.  In $d$ dimensions, our surface will be a $(d-1)$-dimensional hyperplane. 
Roughly speaking, we say that 
adsorption (respectively, desorption) occurs when polymers are likely (respectively, unlikely) 
to have a macroscopic fraction of  sites touching the surface.  We shall give a formal definition below.

\setlength{\unitlength}{.6mm}
\begin{figure}
\begin{center}
\begin{picture}(140,70)
%   
% animal
\put(0,40){(a)}
\put(15,0){\begin{picture}(40,70)
\put(20,10){\line(1,0){20}}
\put(20,20){\line(1,0){20}}
\put(0,30){\line(1,0){30}}
\put(0,40){\line(1,0){20}}
\put(0,60){\line(1,0){30}}
\put(40,10){\line(0,1){10}}
\put(0,30){\line(0,1){30}}
\put(20,10){\line(0,1){50}}
\multiput(-1.5,0)(0,4){18}{\line(0,1){1.5}}
\multiput(1.5,0)(0,4){18}{\line(0,1){1.5}}
\multiput(0,30)(0,10){4}{\circle*{2}}
\end{picture}
}
\put(85,40){(b)}
\put(100,0){\begin{picture}(40,70)
\put(0,10){\line(1,0){20}}
\put(0,20){\line(1,0){20}}
\put(0,30){\line(1,0){30}}
\put(0,40){\line(1,0){20}}
\put(0,60){\line(1,0){20}}
\put(0,10){\line(0,1){10}}
\put(0,30){\line(0,1){30}}
\put(20,20){\line(0,1){50}}
\put(20,10){\line(0,1){10}}
\multiput(8.5,0)(0,4){18}{\line(0,1){1.5}}
\multiput(11.5,0)(0,4){18}{\line(0,1){1.5}}
\multiput(10,10)(0,10){4}{\circle*{2}}
\put(10,60){\circle*{2}}
\end{picture}
}
\end{picture}
\end{center}
\caption{\label{fig.adsorpex} Examples of a lattice animal at a surface in the impenetrable (a) and
penetrable (b) cases.   The surface is indicated by the parallel pair of dashed lines.  Each solid dot
represents a monomer of the animal that lies in the surface, and hence interacts energetically with the
surface.  Observe that each of these two animals has the same topology 
as the animal in Figure \ref{fig.shapes}(a).
}
\end{figure}
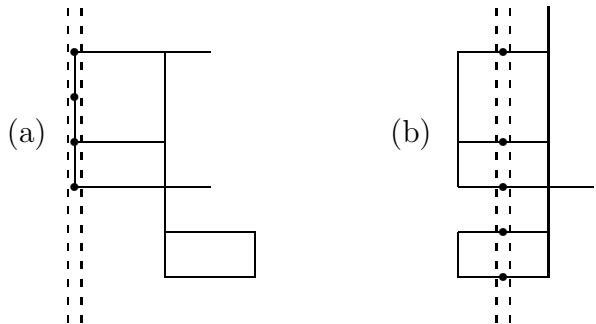

Thirdly, the topology of the polymer could be  either \textit{quenched} or \textit{annealed}.  
For a branched lattice polymer, the ``topology'' 
corresponds to the \textit{abstract graph} of which the given tree, animal, or comb is an embedding in 
the given lattice; we shall describe this more carefully soon.
(For now, we note that this graph determines not only the branching structure but also the length of 
each branch.)
For a linear polymer, this interpretation of topology would be trivial:  every $N$-step 
self-avoiding walk is an 
embedding of the path with $N$ edges, 
and every $N$-step self-avoiding polygon is an embedding of the cycle with $N$ edges.   
Instead, for linear polymers, we shall consider ring polymers specifically in three
dimensions, and the ``topology'' of a ring polymer will be its \textit{knot type} \cite{Ad,OrWh}.  
In the quenched topology model that we introduce in this paper, the topology is first chosen 
uniformly at random from a given class of polymers of a given size, and then we consider 
the possible adsorption of all polymers of that precise topology.  In the annealed model,
the topology of the adsorbing polymer essentially emerges according to 
reweighted probabilities given by the surface interaction.  The details are described more 
fully below.

Numerous researchers have contrasted quenched branching structure to annealed branching,
 mostly in the context of tree polymers.  
Gutin, Grosberg, and Shakhonovich \cite{GGS} used quenched branching to 
approximate the effect of secondary structure in RNA.  
Cui and Chen \cite{CuiCh} found different scaling laws for quenched and annealed branching.  
Everaers et al.\ \cite{EGRR} examined Flory theory for quenched versus annealed branching,
while Rosa and Everaers \cite{RoEv} used simulation to compare radius of gyration, topological
structure, and related observables in quenched versus annealed models.

\smallskip
The rest of this introductory section presents an overview of our adsorption models with 
quenched and annealed topologies.
We shall consider a general class $\mathcal{P}$  of polymers    
which could represent branched polymers (lattice
animals, denoted $\mathcal{A}$, or lattice trees, denoted $\mathcal{T}$), ring polymers 
(self-avoiding polygons, $\mathcal{R}$),
or comb polymers ($\mathcal{C}$).  

To focus ideas, we shall first consider the
case of adsorption of polymers in a three-dimensional solution
to an impenetrable two-dimensional surface (Figure \ref{fig.adsorpex}(a)). 
We work in the  3-dimensional simple cubic lattice $\mathbb{L}^3$, 
with sites $x=(x_1,x_2,x_3)\in \mathbb{Z}^3$ and edges connecting nearest neighbours.  
Our impenetrable surface corresponds to the coordinate plane $\{x:x_1=0\}$.  Our polymers 
will live in the part of $\mathbb{L}^3$ in the half-space $x_1\geq 0$, which we shall denote 
$\mathbb{L}^3_+$.
Let $\mathcal{P}^+_N$ be the set of polymers in $\mathcal{P}$ of size $N$ that are contained in 
$\mathbb{L}^3_+$ and that include the origin (0,0,0) as one of their sites.

Adsorption depends on the interaction between individual monomers of a polymer and the 
molecules of the surface.  We can think of 
each monomer that touches the surface receiving a ``reward'' in terms of energy.   
We use the term ``visit''  to denote a monomer that touches the surface.  
For each polymer $\rho$ in $\mathcal{P}^+_N$, let $\sigma(\rho)$ be the number of its visits, 
i.e.\ the number of sites of $\rho$ that lie in the surface $x_1=0$.  
The Gibbs formalism of statistical mechanics says that for a given set of possible polymer 
configurations, 
\begin{equation}
    \label{eq.assump**}
    \begin{array}{l}
    \hbox{the probability of observing a particular configuration $\rho$ is }
    \\
    \hbox{proportional to $\exp(\beta \sigma(\rho))$, where $\beta$ is a given real parameter.}  
    \end{array}
\end{equation}
In physical terms, the dimensionless parameter $\beta$
corresponds to $-\epsilon_a/kT$, where 
$\epsilon_a$ is the adsorption energy per monomer, $T$ is absolute temperature, and $k$ is 
Boltzmann's constant. 
If $\beta>0$, then the surface is attractive, and configurations with more visits are more likely 
than those with fewer visits.
If $\beta<0$, then the surface is repulsive, and configurations with fewer visits are preferred.  
It is a fairly general result 
that adsorption does not occur when $\beta$ is negative, although a positive value of $\beta$ may
not be sufficient for adsorption to occur.

The assumption (\ref{eq.assump**}) 
applied to the set of configurations  $\mathcal{P}^+_N$ determines a 
probability distribution $\Pr_{\beta}(\,\cdot \,|\, \mathcal{P}^+_N)$ on this set, namely 
\begin{equation}
   \label{eq.probP+}
     \text{Pr}_{\beta}(\rho \,|\,\mathcal{P}_N^+)  \;=\; \frac{e^{\beta \sigma(\rho)}}{
        Z_N(\beta\,|\,\mathcal{P}^+)}
     \hspace{15mm}(\rho \in \mathcal{P}_N^+)  \,,      
\end{equation}
where the normalizing constant $Z_N$ is the partition function 
(see Equation (\ref{eq.defZPgen}) below).
Observe that when $\beta=0$, this is the uniform distribution on $\mathcal{P}^+_N$. 
When $\beta$ is positive (respectively, negative), the probability is larger 
for configurations that have more (respectively, fewer) sites in the surface.    
Significantly, as $\beta$ tends to $+\infty$ under this model,  the expected fraction of a large polymer's 
sites that lie in the surface converges to 1.   
More precisely, writing $E_{\beta}$ for the expectation operator corresponding to
probability distribution of Equation (\ref{eq.probP+}), we have (see Theorem \ref{thm-slopelim})
\begin{equation}
  \label{eq.limsigeq1}
     \lim_{\beta\rightarrow\infty} \liminf_{N\rightarrow\infty}  E_{\beta}\left(  \frac{\sigma(\rho)}{N} \,
     \left| \rule[-4mm]{0mm}{8mm}   \right.    \,
        \mathcal{P}^+_N\right)    \;=\;  1 \,.
\end{equation}

Equation (\ref{eq.limsigeq1}) seems to conflict with known mathematical
properties of branched and ring polymers.  
For example,  lattice animals and lattice trees
in the cubic lattice $\mathbb{L}^3$ can have many sites of degree 6.  Indeed, the pattern theorem
of \cite{Ma99} implies that there is a constant $\alpha>0$ such that all but exponentially few $N$-site 
animals (or trees) have at least $\alpha N$ sites of degree 6.   But any site of degree 6 that lies
in the surface $x_1=0$ must have at least 2 neighbouring sites that are not in the surface.  
So if the proportion of sites in the surface tends to 1 as $\beta$ tends to infinity, then the 
proportion of sites of degree 6 must tend to 0.  That is, the polymers that adsorb to the surface
have different topological properties from those that exist in a good solution (i.e., under 
the uniform distribution on $\mathcal{A}_N^+$ or $\mathcal{T}_N^+$).

There is an analogous dilemma for ring polymers if we consider their knotting behaviour.
In particular, since a non-trivial knot cannot be flattened to lie completely in a plane without
intersecting itself, it follows that, for large $\beta$ and large $N$,
the distribution $\Pr_{\beta}(\cdot\,|\,\mathcal{R}^+_N)$
gives most of the weight to polygons that are unknotted or have low knot complexity.
This contrasts with the theorem that under the uniform distribution, most large ring polymers have 
large knot complexity \cite{SSW}. 
More formally:  For any $\epsilon>0$, let $\mathcal{R}_{N,[\geq \epsilon N]}^+$ be the class of 
self-avoiding
polygons in $\mathcal{R}_N^+$ whose knot type includes the product of at least $\epsilon N$ trefoil
knots.  Then it can be shown that  
\begin{equation}
   \label{eq.Prepstrefoil0}  
   \lim_{\beta\rightarrow\infty} \lim_{N\rightarrow\infty}
   \text{Pr}_{\beta}(\mathcal{R}_{N,[\geq \epsilon N]}^+\,|\,\mathcal{R}_N^+)\;=\;0
   \hspace{5mm}\hbox{for every }\epsilon>0.
\end{equation}
but that
\begin{equation}
   \label{eq.Prepstrefoil1}
      \lim_{N\rightarrow\infty}
   \text{Pr}_{0}(\mathcal{R}_{N,[\geq \alpha N]}^+\,|\,\mathcal{R}_N^+)\;=\;1 
   \hspace{5mm}\hbox{for some }\alpha>0.
\end{equation}
(In fact, Vanderzande \cite{Va95} proved that for each $\beta\geq 0$, Equation (\ref{eq.Prepstrefoil1}) holds for $\Pr_{\beta}$ for some $\alpha$).

One physical interpretation of the phenomena behind 
the above apparent conflicts 
is the following.   
When large branched polymers (respectively, ring polymers) 
form in a good solvent, without any significant interaction with 
a surface,
then we expect that most of them will have a macroscopic proportion of high-degree sites 
(respectively, a complex knot type).  
However, when large polymers form in the presence of a strongly attracting surface, then 
we expect to obtain fewer high-degree sites (or simpler knot types).

In summary, the preceding adsorption model, governed by the family of probability distributions 
of Equation (\ref{eq.probP+})
over the entire ensemble $\mathcal{P}^+_N$, does not pay explicit attention to the 
topological differences between the distributions for $\beta=0$ and for large $\beta$.
We can think of allowing the topology to change as $\beta$ changes (perhaps by some 
isomerization process or rearrangement reaction).
We shall say that  the preceding model has \textit{annealed} topology.  

\smallskip
In contrast to the preceding annealed topology,
we now introduce our model(s) with \textit{quenched} topology.  Again, we consider 
three-dimensional polymers near a two-dimensional surface.
Suppose that large branched (or ring) polymers form in a good solvent, free from any interaction with
the walls of the solution's container, say.   The polymers formed are assumed to be 
drawn \textit{uniformly} at random from the set  $\mathcal{P}_N^+$. 
Next, we take these polymers and place them close to an attracting wall.
In this situation, the topology of a polymer is predetermined, 
and we want to see whether and how it adsorbs onto the surface.  We allow the bond angles to 
change as needed during the adsorption process, but bond connections cannot be changed.
In particular, a ring polymer cannot change its knot type, and a branched polymer cannot
change the underlying abstract graph that describes the molecule's topology (as opposed to 
its geometry, which is flexible).

To formalize the principle for constructing our quenched model for $\mathcal{P}_N^+$
at a given real value of $\beta$,
here is the description of the corresponding quenched probability distribution $P_{\beta}^Q$. 
First, choose a configuration $\tilde{\rho}$ uniformly at random from $\mathcal{P}_N^+$
(i.e., according to the probability distribution $P_0(\cdot |\mathcal{P}_N^+)$.)
Next, let $\tilde{\tau}$ be the topology of $\tilde{\rho}$ (either the underlying abstract graph
or the knot type).  Let $\mathcal{P}^+_N[\tilde{\tau}]$ be the set of all configurations in 
$\mathcal{P}^+_N$ that have this particular topology $\tilde{\tau}$.  Finally, choose
a configuration $\tilde{\psi}$ from the set $\mathcal{P}^+_N[\tilde{\tau}]$ with probability
proportional to $\exp[\beta \sigma(\tilde{\psi})]$, i.e., according to the probability distribution
$P_{\beta}(\cdot\,|\,\mathcal{P}_N^+[\tilde{\tau}])$ as in Equation (\ref{eq.probP+}).
We can express the resulting {\em quenched} probability distribution on $\mathcal{P}_N^+$ as
\begin{equation}
    \label{eq.probPQ+}
      P_{\beta}^Q(\rho \,|\,\mathcal{P}_N^+)   \;=\;  \sum_{\tau} 
       \frac{| \mathcal{P}^+_N[\tau]|}{|\mathcal{P}_N^+|}  \,
       \frac{e^{\beta \sigma(\rho)} }{Z_N(\beta\,|\,\mathcal{P}^+[\tau])}
        \hspace{8mm}(\rho \in \mathcal{P}_N^+),
\end{equation}
where the sum is over all possible topologies $\tau$ (i.e, those $\tau$'s for which 
$\mathcal{P}^+_N(\tau)$ is nonempty).
We write $E_{\beta}^Q$ for the corresponding expected value operator.
See Section \ref{sec-quench} for a complete description of the definitions and notation for the 
quenched model.

In contrast to Equation (\ref{eq.limsigeq1}) for the annealed model, we prove that our 
quenched models for animals, trees, and rings satisfy
\begin{equation}
  \label{eq.limsiglt1}
     \lim_{\beta\rightarrow\infty} \limsup_{N\rightarrow\infty}  E^Q_{\beta}\left(  \frac{\sigma(\rho)}{N} \,
     \left| \rule[-4mm]{0mm}{8mm}   \right.    \,
        \mathcal{P}^+_N\right)    \;<\;  1 
\end{equation}
(see Theorem \ref{thm-slopelim}).
That is, a nonvanishing fraction of sites remain off the surface, even as $\beta$ tends to infinity.
For combs, however, Equation (\ref{eq.limsiglt1}) is true only when the spatial dimension is two; 
for three or more dimensions, the left side equals 1. 

So far, the theory for penetrable surfaces is the same as for impenetrable surfaces.  
However, they differ when it comes to determining the adsorption transition point.
The transitions in the annealed models are reasonably well understood; see Section \ref{sec-anneal}
for details and references.  (Results for general lattice combs are new to the present paper.) 
Briefly, we consider
the \textit{limiting free energy per site} defined by
\begin{equation}
   \label{eq.limFP+}
   \mathcal{F}(\beta \,|\, \mathcal{P}^+)  \;:=\;  \lim_{N\rightarrow\infty} \frac{1}{N} \,\log Z_N(\beta \,|\,
      \mathcal{P}^+)  \,,
\end{equation}
which is a convex, continuous, nondecreasing function of $\beta$.  It is also known that 
$\mathcal{F}(\beta \,|\, \mathcal{P}^+) \,=\, \mathcal{F}(0 \,|\, \mathcal{P}^+)$ for every negative
$\beta$ (Proposition \ref{prop-negbeta}).   
That is, when the polymer-surface interaction  is repulsive, the surface has no
macroscopic effect on the energy of the system, which is the same as for a completely
desorbed polymer.
(Note that $\mathcal{F}(0 \,|\, \mathcal{P}^+) \,=\,
\log \,[\lim_{N\rightarrow\infty}|\mathcal{P}_N^+|^{1/N}]$.)   It turns out that desorption
also occurs in our annealed models with impenetrable boundary for 
some strictly positive values of $\beta$ (Theorem \ref{thm.betacanngt0}).   
Accordingly, we define the {\em critical value} $\beta_c(\mathcal{P}^+)$ to be
\begin{equation}
   \label{eq.betacP+def}
   \beta_c(\mathcal{P}^+)  \;:=\;  \sup \left\{  \beta \,: \,  \mathcal{F}(\beta\,|\, \mathcal{P}^+) \,=\, 
       \mathcal{F}(0 \,|\, \mathcal{P}^+) \right\} \,.
\end{equation}
We know that $\beta_c(\mathcal{P}^+)$ is finite  (Proposition \ref{prop.betacboundann}), 
and therefore it is a point where the limiting 
free energy is non-analytic, which is a classical hallmark of a phase transition.  
For a penetrable surface, the same definitions and results
hold except that it is generally believed that the critical point equals 0.  (See \cite{Ma} for further
discussion of this issue and for strong theoretical grounds in support of this belief for animals, trees,
and self-avoiding walks.)

One significant consequence of $\beta_c$ being strictly positive for an impenetrable boundary
is the following.
Recall that $\beta$ corresponds to $-\epsilon_a /kT$.   
Thus, it only takes a change in the system's temperature $T$ to turn the situation from one of
adsorption into one of desorption. Thus, even if polymers are forming in solution close to a 
wall, then the attraction of the wall will have a minor influence when the temperature is 
high enough.

The phase transition of the quenched models is more subtle than for the annealed model.
As we shall discuss more fully in Section \ref{sec-quench},
the quenched analogue of the free energy per site is the average of the free energy of the 
topologically restricted sets of configurations: 
\begin{eqnarray}
   \nonumber
     F_N^Q(\beta\,|\,\mathcal{P}^{+})  & = &  E_0\left(\frac{1}{N} \,\log Z_N(\beta\,|\, 
        \mathcal{P}^{+}[\tilde{\tau}] )  \right)
     \\
     \label{eq.FQP+deffull}
    & = &  \sum_{\tau}  \frac{|\mathcal{P}^{+}_N[\tau]|}{|\mathcal{P}^{+}_N|} 
      \,\frac{1}{N} \,\log Z_N(\beta\,|\,   \mathcal{P}^{+}[\tau]) \,,
\end{eqnarray}
where $E_0$ is the expected value corresponding to the uniform distribution on all of 
$\mathcal{P}_N^+$, and the sum is over all topologies $\tau$.
Unfortunately, we are not (yet) able to prove that 
$\lim_{N\rightarrow\infty}F_N^Q(\beta\,|\,\mathcal{P}^{+})$ exists for our models, even in the
special case $\beta=0$.   A proof would seem to require some deeper knowledge of 
topological behaviour of our class $\mathcal{P}$.  Fortunately, the closely related limit 
\begin{equation}
    \label{eq.FminusF}
    \mathcal{F}^Q_{\Delta}(\beta\,|\,\mathcal{P}^+)  \;:=\;
        \lim_{N\rightarrow\infty}\left(   F_N^Q(\beta\,|\,\mathcal{P}^{+}) \,-\,  F_N^Q(0\,|\,\mathcal{P}^{+})
         \right)        
\end{equation}        
is a bit easier to deal with.  
We cannot prove that the limit $\mathcal{F}^Q_{\Delta}(\beta\,|\,\mathcal{P}^+)$
exists for every $\beta$, but we can prove that  it exists and equals 0 for every negative $\beta$
(Proposition \ref{prop.negbetaQ}).
This of course is the quenched analogue of the property that 
$\mathcal{F}(\beta\,|\,\mathcal{P}^+)\,=\,  \mathcal{F}(0\,|\,\mathcal{P}^+)$ for all negative $\beta$
in the annealed model.       
Accordingly, we define the critical value in the quenched model for 
$\mathcal{P}^+$ to be 
\begin{equation}
   \label{eq.betacQP+def}
   \beta_c^Q(\mathcal{P}^+)  \;:=\;  \sup \left\{  \beta \,: \,  \mathcal{F}^Q_{\Delta}(\beta\,|\,\mathcal{P}^+) 
   \, =\,0    \right\} 
\end{equation}
(where of course the existence of the limit (\ref{eq.FminusF}) is necessary for the assertion
$\mathcal{F}^Q_{\Delta}(\beta\,|\,\mathcal{P}^+) \, =\,0$ to hold).
We shall also present analogous definitions for penetrable surfaces, where the technical difficulties are 
pretty similar.

Here is what we can and cannot prove about the quenched critical values of our models.
\\
$\bullet$  We know that $\beta_c^Q \geq \beta_c\geq 0$ for each of our models, both impenetrable
and penetrable (Corollary \ref{cor.negbetaquench}).  
In particular, for an impenetrable surface, the quenched critical value is strictly positive.
\\
$\bullet$ For self-avoiding polygons in $\mathbb{L}^3$, we can prove that 
$\beta_c^Q$ is finite (for both penetrable and impenetrable surfaces).  
See Proposition \ref{prop.ringQfin}. 
\\
$\bullet$ For lattice animals and trees in two dimensions, we conjecture that 
$\beta_c^Q$ is infinite (for both penetrable and impenetrable surfaces).  
This is equivalent to the assertion that for any finite $\beta$,
the expected fraction of sites in the surface converges to 0 as $N$ tends to infinty.
This is due to the fact that the longest path in a uniformly chosen tree or animal is 
believed to have expected length  $o(N)$.  See Conjecture \ref{conj.2dnoads} and the discussion
following it.  
\\
$\bullet$  For lattice combs, in contrast, we can prove that $\beta_c^Q$ is finite in every dimension 
$d\geq 2$ for both penetrable and impenetrable surfaces (Proposition \ref{prop.negbetaQcomb}).  
\\
$\bullet$ We conjecture that $\beta_c^Q$ is finite for trees and animals in three or more dimensions,
but this seems challenging to prove.

\medskip
We also investigate the commonest topology, and find an important difference between ring polymers
and branched polymers.  We consider the commonest knot type for each size $N$, namely 
the knot type that has the most $N$-step self-avoiding polygons of that type. 
Then the number of self-avoiding polygons having the commonest knot type for size $N$ grows 
exponentially at the same rate as does the number of all self-avoiding polygons of size $N$; 
that is, the commonest knot type
is not exponentially rare (see Proposition \ref{prop.MNeqmu}).  
This disproves a conjecture of \cite{RR} that the 
commonest knot type should consist of some number of left-handed and right-handed trefoils.
In contrast, the commonest topology among $N$-site trees is exponentially rare (Proposition
\ref{prop.MNtree}).   These results have implications for the possible values of the
quenched free energy.

\smallskip
The rest of the paper is organized as follows.
Our models are set up in detail in Section \ref{sec-adsmain}.  In particular, 
Section \ref{sec-latpoly} defines basic notation and
specifies the various classes of lattice polymers that we shall consider.
Section \ref{sec-adsgen} presents the general formalism for adsorption phenomena. 
Section \ref{sec-anneal} reviews the results for annealed adsorption models,
which, except for comb models, are already in the literature.
Section \ref{sec-quench} introduces our quenched models in detail, first for knotted ring polymers
and then for branched polymers.  This subsection also describes our results for quenched models
and contrasts them with the behaviour of annealed models.
Section \ref{sec.proofs1} presents proofs of general results on quenched and annealed adsorption.
Some of these use familiar methods adapted to the more general framework which now includes
lattice combs.  Section \ref{sec.comblimits} contains (rather lengthy) 
proofs of fundamental results for lattice combs:  the existence of the growth constant, 
and the existence of the (annealed) limiting free energy of adsorption.  
This section can be omitted on first reading without loss of continuity.
Section \ref{sec.combprops} proves several further results about lattice combs and their
adsorption properties.
Section \ref{sec-ringtree} proves some properties of rings and trees:
that the critical value for quenched ring polymers in three dimensions
is finite (Section \ref{sec-ringQfin}); and 
that the commonest knot type is
not exponentially rare among all ring polymers, but that the commonest branching topology is
exponentially rare among all trees  (Section \ref{sec-sizes}).
Section  \ref{sec.discussion} contains a concluding discussion.

\section{Lattice Models of Polymer Adsorption}
     \label{sec-adsmain}

\subsection{Classes of lattice polymers}
     \label{sec-latpoly}

First, we establish some basic notation.    
We write $\mathbb{Z}^d$ for
the set of points $(x_1,\ldots,x_d)$ in  $\mathbb{R}^d$ whose coordinates $x_i$ are all integers.
The $d$-dimensional hypercubic lattice $\mathbb{L}^d$ 
is the infinite graph embedded in $\mathbb{R}^d$
whose sites are the points of $\mathbb{Z}^d$ and whose edges join each pair of sites that
are distance 1 apart.
We denote the standard basis of $\mathbb{R}^d$ by $e^{(1)},\ldots,e^{(d)}$; that is, $e^{(i)}$ is the 
unit vector in the $+x_i$ direction.

We shall discuss several classes of lattice polymers.  We model the configuration of a lattice polymer
by a finite subgraph of the lattice (possibly with one or two sites labelled).
Here are the five fundamental classes of polymer models in $\mathbb{L}^d$ ($d\geq 2$)
that we shall discuss (Figure \ref{fig.shapes}).
\begin{itemize}
\item \underline{Lattice animals.}  A lattice animal is a finite connected subgraph of $\mathbb{L}^d$.
We write $\mathcal{A}_N$ for the set of all lattice animals with $N$ sites.
\item \underline{Lattice trees.}  A lattice tree is a lattice animal with no cycles.
We write $\mathcal{T}_N$ for the set of all lattice trees with $N$ sites (and hence $N-1$ edges).
\item \underline{Self-avoiding walks.}  A self-avoiding walk (SAW) is a path $\omega$ in $\mathbb{L}^d$
(i.e., a walk with all sites distinct), in which
the sites are labelled $\omega(0),\omega(1),\ldots,\omega(N)$, with an edge joining $\omega(i-1)$
to $\omega(i)$ for each $i=1,\ldots,N$.  
Equivalently, for $N\geq 1$, an $N$-edge SAW $\omega$ is a lattice tree 
having $N+1$ sites of which exactly two have
degree 1, with one site of degree 1 labelled $\omega(0)$ and the other labelled 
$\omega(N)$.
The $j$-th coordinate of $\omega(i)$ is written $\omega_j(i)$.
We write   $\mathcal{W}_N$ for the set of all SAWs with $N$ edges 
(and hence $N+1$ sites).  In contrast to the other four classes, the quenched topology of 
SAWs is trivial, and will not be treated.  However, SAWs will be needed throughout this paper.
\item \underline{Ring polymers, or self-avoiding polygons.}  These are lattice animals that consist of
a single cycle and nothing else.  
We write $\mathcal{R}_N$ for the set of all ring polymers with $N$ sites (or equivalently with $N$
edges).
We shall focus on knots in ring polymers, and for that reason we only consider the three-dimensional
lattice $\mathbb{L}^3$.  Note that $\mathcal{R}_N$ is empty when $N$ is odd or when $N$ equals 2.
\item\underline{Lattice combs.}  A lattice comb is a lattice tree $\rho$ in $\mathbb{L}^d$ with two distinguished vertices 
$\rho_A$ and $\rho_B$ such that the following properties hold:
\\
(a)  $\rho_A$ and $\rho_B$ have degree one in $\rho$;
\\
(b) no site of $\rho$ has degree more than 3; and
\\
(c) every site of degree 3 is on the unique path in $\rho$ joining $\rho_A$ to $\rho_B$.
See Figure \ref{figcomb2} for an example.
The path  in $\rho$ joining $\rho_A$ to $\rho_B$ is called the \textit{backbone} of $\rho$.
A \textit{side chain} of the comb $\rho$ is a path in $\rho$ that contains no edge of the backbone and 
joins a site of degree 3 to a site of degree 1 (respectively called the attached end and the free end
of the side chain)\footnote{The side chains are often called {\em teeth} in the polymer literature.}.  
Note that every edge of $\rho$ is either in the backbone or in a side chain.
We write $\mathcal{C}_N$ for the set of all lattice combs with $N$ edges (and hence $N+1$ sites; 
note that, in contrast, we count trees by sites).
\end{itemize}

The coordinate hyperplane $\{x:x_1=0\}$ is the surface with which our polymers will interact.
Let $\mathbb{L}_+^d$ be the part of $\mathbb{L}^d$ that lies in the half-space $\{x:x_1\geq 0\}$.
If the surface is impenetrable, then we shall only consider polymers that live in $\mathbb{L}^d_+$.

We say that $(x_1,\ldots ,x_d)$ is \textit{lexicographically smaller} 
than $(y_1,\ldots,y_d)$ if there exists a $k\in \{1,\ldots,d\}$ such that  $x_i=y_i$ for 
all $i<k$ and $x_k<y_k$.  The lexicographic order is a total order of 
the points of $\mathbb{L}^d$ (in fact, of $\mathbb{R}^d$).

\smallskip
Let the polymer model $\mathcal{P}$ be one of $\mathcal{A}$, $\mathcal{T}$, 
$\mathcal{W}$, $\mathcal{R}$, or $\mathcal{C}$.  
Let $\mathcal{P}_N^{\pen}$ be the set of all $\rho\in \mathcal{P}_N$ such 
that the origin is a site of $\rho$, and let  $\mathcal{P}_N^{+}$ be the set of all 
$\rho\in \mathcal{P}_N^{\pen}$  such that $\rho$ is contained in $\mathbb{L}^d_+$.
(So this defines $\mathcal{A}_N^{\pen}$, $\mathcal{A}^+_N$, $\mathcal{T}_N^{\pen}$, and 
so on.)
The symbol $\pen$ was chosen to suggest a polymer that could be on both sides of a vertical plane.
 We shall use $\mathcal{P}^{\pen}$ for models with a penetrable surface, and 
$\mathcal{P}^+$ for models with an impenetrable surface.

If $A\subset \mathbb{R}^d$ (or if $A$ is a subgraph of $\mathbb{L}^d$)
and $x\in \mathbb{Z}^d$, then the translation of $A$ by the vector $x$ is denoted $A+x$.

For  $\mathcal{P}$ being one of $\mathcal{A}$, $\mathcal{T}$, 
$\mathcal{W}$, $\mathcal{R}$, or $\mathcal{C}$,  
let $\overline{\mathcal{P}}_N$ be the set of polymers in $\mathcal{P}_N$ whose lexicographically
smallest site is the origin.  Each polymer in $\mathcal{P}_N$ is the translation of a unique member of
$\overline{\mathcal{P}}_N$.  Then we have 
\[
    \overline{\mathcal{P}}_N\;\subseteq \;\mathcal{P}_N^+\;\subseteq \; \mathcal{P}_N^{\pen} 
      \;\subseteq \; \mathcal{P}_N    \;;
   \hspace{10mm}\hbox{also,}
\]
\begin{eqnarray*}
   |{\cal{P}}^{\pen}_N|\,=\, N\,|\overline{\cal{P}}_N|     \hspace{4mm}
   & \hbox{for} &  \mathcal{P}\,=\,\mathcal{A}, \;
       \mathcal{T}, \;\mathcal{R}, \quad \hbox{and}
   \\
   |{\cal{P}}^{\pen}_N|\,=\, (N+1)\,|\overline{\cal{P}}_N| & \hbox{for} &  \mathcal{P}\,=\,\mathcal{W}, \;
       \mathcal{C} \,.      
\end{eqnarray*}
Note that each element of $\overline{\mathcal{P}}_N$ corresponds to
an equivalence class of all polymers in $\mathcal{P}_N$ modulo translation
(i.e., to the set of all of the translations of that element by vectors in $\mathbb{Z}^d$).

\smallskip

We now consider the cardinalities of each class of lattice polymers modulo translation.
For each $N\geq 0$, let
\[
    a_N\;=\; |\overline{\cal{A}}_N|\,,  \quad 
     t_N\;=\; |\overline{\cal{T}}_N|\,,  \quad 
   w_N\;=\; |\overline{\cal{W}}_N|\,,  \quad  
  r_N\;=\; |\overline{\cal{R}}_N|\,,  \quad 
    c_N\;=\; |\overline{\cal{C}}_N|\,.
\footnote{Note that other references, such as \cite{MS}, use the notation $c_N$ 
to count self-avoiding walks.}
\]
Since the smallest graph has just one site, we have $a_0=t_0=r_0=0$ and 
$w_0=c_0=1$.  Moreover, since a cycle in $\mathbb{L}^d$ must have even length, we 
have $r_N=0$ whenever $N$ is odd; also, $r_2=0$.  Henceforth, when discussing ring polymers
(as in the expression $\lim_{N\rightarrow\infty}r_N^{1/N}$ below),
we shall implicitly restrict attention to even values of $N$.

Each of the above quantities has been proved to grow exponentially, in the sense that the 
following limits exist and are finite for each dimension $d\geq 2$:
\begin{eqnarray}
    \label{eq.animallim}
        \lambda_{d,A}  & = & \lim_{N\rightarrow\infty}a_N^{1/N}  
        \hspace{5mm}  \hbox{(Klarner, 1967 \cite{Kla})},
        \\
            \label{eq.treelim}
        \lambda_{d,T}  & = & \lim_{N\rightarrow\infty}t_N^{1/N}  
          \hspace{5mm}  \hbox{(Klein, 1981 \cite{Kle})},
                  \\
    \label{eq.sawlim}
      \lambda_{d,W} & = &  \lim_{N\rightarrow\infty}w_N^{1/N} 
       \;\; = \;\; \lim_{N\rightarrow\infty}r_N^{1/N}  \;\; = \;\;   \mu_{d}  
        \\
        & &
        \nonumber 
          \hspace{2mm} 
           \hbox{(Hammersley and Morton, 1954 \cite{HM}; Hammersley, 1961 \cite{Ham})}, 
        \\
    \label{eq.comblim}
        \lambda_{d,C}  & = & \lim_{N\rightarrow\infty}c_N^{1/N}  
          \hspace{5mm}  \hbox{(this paper, Proposition \ref{prop.combgrowth})}.        
\end{eqnarray}
The notation $\mu_d$ is fairly standard for the SAW limit, and we shall use that as well as
$\lambda_{d,W}$, as is convenient.  We know that $\lambda_{d,A}>\lambda_{d,T}>\lambda_{d,C}$
by the pattern theorem of \cite{Ma99}, and $\lambda_{d,C}>\lambda_{d,W}$ by Proposition 
\ref{prop.combneqsaw}.

\begin{rem}
\label{rem.dimension}
{\em In general, we do not insist that the dimension $d$ is 3.  While
3 is the most natural dimension from a chemist's point of view,
the case $d=2$ is also important, representing polymers in a very thin film of liquid, 
with adsorption to a 1-dimensional ``surface'' (a line).  
And dimensions above 3 are frequently studied by theoretical physicists as well as
mathematicians.  Papers such as \cite{HTW}, \cite{RY}, and \cite{Ma} 
consider adsorption of lattice polymers in arbitrary dimension. 
However, for ring polymers we shall restrict our attention to three dimensions because that is where 
knotting of closed curves is meaningful.}
\end{rem}

\subsection{Adsorption models:  generalities}
     \label{sec-adsgen}

For a subgraph $\rho$ of $\mathbb{L}^d$, let $\mathcal{H}(\rho)$ be the set of sites $x$ of $\rho$
such that $x_1=0$.  Each point of $\mathcal{H}(\rho)$ is called a visit (of the polymer to the surface). 
Define the quantity
$\sigma(\rho)$ to be $|\mathcal{H}(\rho)|$, the cardinality of $\mathcal{H}(\rho)$.  (The notation
$\sigma(\cdot)$ was introduced in Section \ref{sec.intro}.)

We are still guided by the formalism (\ref{eq.assump**}) 
that the probability of observing a particular configuration
$\rho$ is proportional to $\exp(\beta \sigma(\rho))$ where $\beta$ is a real parameter.
For models with an \textit{impenetrable} 
boundary, we shall consider only sets of configurations that lie in $\mathbb{L}^d_+$,  which 
are the sets of the form $\mathcal{P}^+_N$.
For models with a \textit{penetrable} boundary, we shall consider the larger sets of
configurations  $\mathcal{P}_N^{\pen}$.   

Consider a lattice polymer model $\mathcal{P}^X$,  
which is described by sets of the form $\mathcal{P}_N^{X}$ ($N\geq 1$),
where the symbol $X$ is either $+$ or $\pen$.   
We define the partition function 
$Z_N(\beta\,|\,\mathcal{P}^X)$ for each $N$ and each real $\beta$ by the (finite) sum
\begin{equation}
     \label{eq.defZPgen}
      Z_N(\beta\,|\,\mathcal{P}^X)  
  \; := \;   \sum_{\rho\in {\mathcal{P}_N^X} } \exp(\beta \sigma(\rho))  \,.
\end{equation}
We also define the corresponding probability distribution on $\mathcal{P}_N^X$ 
\begin{equation}
   \label{eq.probgen}
     \text{Pr}_{\beta}(\rho \,|\,\mathcal{P}_N^X)  \;=\; \frac{e^{\beta \sigma(\rho)}}{
        Z_N(\beta\,|\,\mathcal{P}^X)}
     \hspace{9mm}(\rho \in \mathcal{P}_N^X)  \,.      
\end{equation}
Observe that when $\beta=0$, this is the uniform distribution on $\mathcal{P}_N^X$ (in particular,
$Z_N(0\,|\,\mathcal{P}^X)=|\mathcal{P}_N^X|$).  We also define the associated free energy function
\begin{equation}
    \label{eq.freedef}
     F_N(\beta \,|\,\mathcal{P}^X)   \; :=\;  \frac{1}{N} \,\log Z_N(\beta \,|\, \mathcal{P}^X) \,
\end{equation}
which is a convex function of $\beta$ (e.g.\ Lemma 4.1.2 of \cite{MS}).  Equivalently, we
can say that $Z_N$ is log-convex in $\beta$.

We shall write the expectation operator of $\Pr_{\beta}(\cdot \,|\,\mathcal{P}_N^X)$ as
$E_{\beta}(\cdot \,|\,\mathcal{P}_N^X)$.  An important special case is the expected number of 
visits (essentially, the expected negative energy), which is given by
\begin{eqnarray}
    \nonumber
    E_{\beta}\left(\sigma(\rho)\,|\,\mathcal{P}_N^X\right)  & = & \sum_{\rho\in \mathcal{P}_N^X}
       \frac{\sigma(\rho) \,e^{\beta \sigma(\rho)}}{   Z_N(\beta\,|\,\mathcal{P}^X)} 
    \\
    \label{eq.expvisitsgen}
   & = & \frac{  d  \, \log Z_N(\beta\,|\,\mathcal{P}^X)}{d \, \beta} \,.
\end{eqnarray}
One property of a convex function is that its first derivative is non-decreasing.  This shows that
$E_{\beta}\left(\sigma(\rho)\,|\,\mathcal{R}_N^+\right)$ is a non-decreasing function of $\beta$
(which makes intuitive sense:  the larger $\beta$ is, the more visits one expects).

\subsection{Annealed adsorption models}
    \label{sec-anneal}

The \textit{limiting free energy} of the annealed model $\mathcal{P}^X$ is defined by
\begin{equation}
   \label{eq.Fgen}
    {\cal F}(\beta\,|\,\mathcal{P}^X)  
      \;:=\;  \lim_{N\rightarrow\infty} F_N(\beta \,|\,\mathcal{P}^X  ) 
            \;:=\;  \lim_{N\rightarrow\infty}\frac{1}{N}\,\log Z_N(\beta \,|\,\mathcal{P}^X  ) 
\end{equation}
if the limit exists.   

\begin{thm}
    \label{thm.Fannexists}
Let $\mathcal{P}$ be one of $\mathcal{A}$, $\mathcal{T}$, $\mathcal{W}$, $\mathcal{R}$, 
or $\mathcal{C}$, and let $X$ be one of $+$ or $\pen$.    
Then the  limit ${\cal F}(\beta\,|\,\mathcal{P}^X)$ exists for every real $\beta$.
\end{thm}
\textbf{Proof:}
This was proved for SAWs in \cite{HTW}.  
It was proved for trees in \cite{RY} (or see Theorem 11.15 of \cite{JvR2}),
and the proof for animals is essentially the same.
The proof for self-avoiding polygons at an impenetrable surface is due to 
\cite{Sot}, and the method can be extended to the penetrable case; see also  
Theorem 9.21 of \cite{JvR2} for a different exposition.
The result for combs is new to the present paper, and is proved in Propositions \ref{prop.Fcombneglim}
and \ref{prop.Fcombann}.
\hfill  $\Box$

\medskip

The limit  ${\cal F}(\beta\,|\,\mathcal{P}^X)$ is a finite non-decreasing function of $\beta$
that is convex 
and hence continuous.
Recalling Equation (\ref{eq.expvisitsgen}) and 
assuming that we can interchange limit and derivative here, we can heuristically interpret
the derivative $\mathcal{F}^{\,\prime}(\beta\,|\,\mathcal{P}^X)$ to be
\[    \lim_{N\rightarrow\infty} \frac{1}{N} \,\frac{  d  \, \log Z_N(\beta\,|\,\mathcal{P}^X)}{d \, \beta}
   \;=\;    \lim_{N\rightarrow\infty} E_{\beta}\left(\left. \frac{\sigma(\rho)}{N}\,\right|\,\mathcal{P}_N^X\right). 
\]   
On the one hand,
if $\mathcal{F}^{\,\prime}(\beta\,|\,\mathcal{P}^X)=0$, then we say that the polymer is 
\textit{desorbed} at this value of $\beta$, 
meaning that on average a negligible fraction of the polymer's sites are on the surface.
On the other hand,
if $\mathcal{F}^{\,\prime}(\beta\,|\,\mathcal{P}^X)>0$, then we say that the polymer is \textit{adsorbed} 
at this value of $\beta$, 
meaning that we expect a macroscopic fraction of the polymer's sites to be on the surface.
The following proposition says that in all of our main annealed models, desorption occurs for all 
negative values of $\beta$.

\begin{prop}
    \label{prop-negbeta}
Let $\mathcal{P}$ be any one of $\mathcal{A}$,
$\mathcal{T}$, $\mathcal{W}$, $\mathcal{R}$,  or $\mathcal{C}$, and
let $X$ be either  $+$ or $\pen$. 
Then $\mathcal{F}(\beta\,|\,\mathcal{P}^X) \,=\,\mathcal{F}(0 \,|\,\mathcal{P}^X)\,=\,
\lim_{N\rightarrow\infty} \frac{1}{N} \log|\mathcal{P}^X_N|$ for all $\beta \leq 0$.
\end{prop}
\textbf{Proof:}
This is proved for SAWs in \cite{HTW}, which together with \cite{Sot} implies the result for rings.
The proof for  trees in \cite{RY}
also works for animals. The result for  combs is proved in Proposition \ref{prop.Fcombneglim}
of  the present paper.
\hfill  $\Box$

\smallskip
Proposition \ref{prop-negbeta} motivates the following definition of a {\em critical point}  
$\beta_c(\mathcal{P}^X)$ for each model $\mathcal{P}^X$:
\begin{equation}
   \label{eq.defbetac}
   \beta_c(\mathcal{P}^X)   \; :=\;  \sup \left\{  \beta\,:\;  \mathcal{F}(\beta\,|\,\mathcal{P}^X) \,=\,
       \mathcal{F}(0 \,|\,\mathcal{P}^X) \right\} \,.
\end{equation}
Thus, we can say that adsorption occurs in a model when $\beta>\beta_c$, and desorption 
occurs when $\beta<\beta_c$.  

The following result describes basic properties of annealed critical points, due to \cite{HTW} for SAWs.
In Section \ref{sec-quench}, we shall see that nonnegativity holds for the quenched analogue, 
but finiteness is much subtler.

\begin{prop}
    \label{prop.betacboundann}
In each of our annealed models, we have $0\leq \beta_c(\mathcal{P}^X) <\infty$.
\end{prop}
\textbf{Proof:}
Proposition \ref{prop-negbeta} says that $\beta_c(\mathcal{P}^X)\geq 0$ for each model.
Finiteness is a consequence of Proposition \ref{prop.betacfiniteann}.
\hfill  $\Box$

\medskip
It is believed that  $\beta_c(\mathcal{P}^{\pen})$ equals 0 for each of our penetrable models; 
see \cite{Ma} for discussion and a theoretical explanation of why this should be true.
In contrast, for an impenetrable boundary, it is known rigorously that the critical point is 
{\em not} zero.

\begin{thm}
   \label{thm.betacanngt0}
We have $\beta_c(\mathcal{P}^+)\,>\,0$ when $\mathcal{P}$ is any of $\mathcal{A}$, $\mathcal{T}$,
$\mathcal{W}$, $\mathcal{R}$, or $\mathcal{C}$.
\end{thm}
\textbf{Proof:}
This result was first proved for self-avoiding walks in 1982 by \cite{HTW}. 
The result for self-avoiding polygons is a direct corollary:  indeed, for every $\beta$ between 0 and 
$\beta_c(\mathcal{W}^+)$, we have 
\[   Z_N(0\,|\,\mathcal{R}^+)  \;\leq \; Z_N(\beta \,|\,\mathcal{R}^+) 
      \;\leq \; Z_{N-1}(\beta \,|\,\mathcal{W}^+)  
\]
(since for any $N$-step self-avoiding polygon that contains the origin, we can remove one
edge to produce an $(N-1)$-step self-avoiding walk that starts at the origin), and from this we obtain
\begin{eqnarray*}
  \mathcal{F}(0 \,|\, \mathcal{R}^+)   \;\leq \; \mathcal{F}(\beta \,|\,\mathcal{R}^+) 
      & \leq &  \mathcal{F}(\beta \,|\,\mathcal{W}^+) 
      \\
       & = &   \mathcal{F}(0 \,|\, \mathcal{W}^+)  
      \;  =\;  \log \mu_d \;=\;   \mathcal{F}(0 \,|\, \mathcal{R}^+)    
\end{eqnarray*}
(where the last equality is from  Equation (\ref{eq.sawlim})).  
Hence $\beta_c(\mathcal{R}^+)\geq \beta_c(\mathcal{W}^+)$.
The result for trees is due to \cite{RY}.
A different argument for walks, trees, and animals was given more recently by \cite{Ma}, and
 is applied to combs in the present paper
(Proposition \ref{prop.Fcombcrit}).
\hfill $\Box$

\medskip
For very large positive $\beta$, the probability distribution $\Pr_{\beta}(\cdot\,|\,\mathcal{P}^X_N)$
of Equation (\ref{eq.probgen}) gives
high probability to configurations with a very high proportion of their sites on the surface, and 
this proportion converges to 1 as $N$ tends to infinity
(see Theorem \ref{thm-slopelim} below).
This raises some conundrums for branched polymers and knotted ring polymers, as discussed in 
Section \ref{sec.intro} after Equation (\ref{eq.limsigeq1}), 
which was our motivation for formulating the quenched model.

\subsection{Quenched adsorption models}
    \label{sec-quench}
    
First we describe the quenched model for adsorption of knotted ring polymers at an 
impenetrable surface.   
Following the thinking outlined in Section \ref{sec.intro}, we consider the following situation.
Suppose that large ring polymers form in a good solvent, free from the influence of
any surface.   The polymers formed are assumed to be 
drawn uniformly at random from the set of all $N$-site self-avoiding polygons  in $\mathbb{L}^3$
up to translation, which corresponds to sampling from $\overline{\mathcal{R}}_N$.  
Next, we take these ring polymers and place them close to an attracting surface
(this could be done by lowering the temperature of the system, changing conditions to make a
neutral surface attracting, or by dipping a new surface into the solution).  
In this situation, the knot type of the polymer is predetermined, 
and we want to see whether and how this knot adsorbs onto the surface.

The ``knot type'' of a self-avoiding polygon can be described as the product (or composition, or sum)
of a collection of prime knots.    
For a readable introduction to basic ideas of knot theory, see \cite{Ad} or \cite{OrWh}.
 Examples of knot types with their
standard notation are the unknot $0_1$,
the trefoil\footnote{Chirality is significant.  E.g.\ a left-handed trefoil is different from its mirror image,
a right-handed trefoil.  But we will not need to deal with chirality explicitly in this paper.}
$3_1$, the product of four trefoils $3_1\sharp 3_1\sharp 3_1\sharp 3_1$ (also written $(3_1)^4$), 
and  the  four-crossing knot $4_1$, called the ``figure-eight knot.''

Let $K$ be a knot type, and let $\mathcal{R}_N^+[K]$ be the set of all polygons in $\mathcal{R}_N^+$
that have knot type $K$.  These represent the possible configurations that can occur 
when an $N$-site ring polymer of knot type $K$ comes into contact with the surface.  Accordingly, let
\begin{equation}
   \label{eq.Zknot}
  Z_N(\beta\,|\,K)  \;\equiv \; Z_N\left(\beta\,\left|\,\mathcal{R}^+[K] \right. \right)  
   \; :=\;  \sum_{\rho\in {\mathcal{R}^+_N[K]} } \exp(\beta \sigma(\rho))  \,.
\end{equation}
Analogously to Equation (\ref{eq.probgen}), for each real $\beta$, the principle (\ref{eq.assump**}) 
gives us  a probability distribution $\Pr_{\beta}(\cdot \,|\,\mathcal{R}_N^+[K])$
 on $\mathcal{R}_N^+[K]$ by the formula
\begin{equation}
   \label{eq.probknot}
     \text{Pr}_{\beta}(\rho \,|\,\mathcal{R}_N^+[K])  \;=\; \frac{e^{\beta \sigma(\rho)}}{
        Z_N(\beta\,|\,\mathcal{R}^+[K])}
     \hspace{10mm}(\rho \in \mathcal{R}_N^+[K])  \,.      
\end{equation}

We can now consider properties of the partition function and probability distribution for particular
knot types $K$.  For example, one could choose $K$ to be the unknot and proceed with an 
analysis similar to what was done for the annealed model. 
But as we mentioned earlier, 
the unknot is rare among large self-avoiding polygons.  So it would be more physically interesting 
to let $K$ be a  knot type that is more typical for large $N$.  However, the complexity of the 
typical knot increases as $N$ increases, so a knot type $K$ that is typical for $N=10^4$ may not
be typical for $N=10^7$.  
Our way of dealing with this issue is  to choose the knot type {\em randomly}
according to its frequency in the set of all $N$-step self-avoiding polygons.
Here is the formal description.

For a self-avoiding polygon $\rho$, let $K(\rho)$ be the knot type of $\rho$.
Let $\tilde{\rho}$ be a random polygon chosen from the uniform distribution on all self-avoiding
polygons of size $N$ (modulo translation).  
Then the knot type of $\tilde{\rho}$ is $K(\tilde{\rho})$, and the 
associated partition function for the polygons of this knot type is 
$Z_N(\beta\,|\,K(\tilde{\rho}))$  (that is, 
$Z_N\left(\beta\,\left|\,\mathcal{R}^+[K(\tilde{\rho})] \right. \right)$ ).
Since $K(\tilde{\rho})$ is a \textit{randomly chosen knot type}, we can describe 
$Z_N(\beta\,|\,K(\tilde{\rho}))$ as a \textit{randomly chosen partition function}.
Similarly, $\Pr_{\beta}(\cdot \,|\,\mathcal{R}_N^+[K(\tilde{\rho})])$ is a \textit{randomly chosen 
probability distribution}.

We shall use a slight variation on this procedure that will be more convenient to work with and to 
generalize.
Instead of sampling uniformly over all $N$-step 
self-avoiding polygons, we  sample $\tilde{\rho}$ uniformly over all polygons in $\mathcal{R}^+_N$.
(This is essentially the same as sampling from the set of all polygons with probability 
proportional to the number of sites with minimum $x_1$ coordinate on the polygon.)  
Then the probability that $K(\tilde{\rho})$ equals a particular knot type $K_0$ is 
$|\mathcal{R}_N^+[K_0]|/|\mathcal{R}_N^+|$.

The random selection of a polymer configuration $\tilde{\rho}$ restricts the knot type of the 
adsorbed configuration to be $K(\tilde{\rho})$.  This is what we mean when we
refer to the knot type as being {\em quenched}.

For fixed $N$, we understand $F_N(\beta\,|\,K(\tilde{\rho})) \,\equiv\,
\frac{1}{N}\log Z_N(\beta\,|\,K(\tilde{\rho}))$ to be a
\textit{randomly chosen free energy function}.  The average with respect to $\tilde{\rho}$ 
of this function over $\mathcal{R}_N^+$
is the \textit{average quenched free energy} $F_N^Q(\beta\,|\,\mathcal{R}^+)$: 
\begin{eqnarray}
   \nonumber
     F_N^Q(\beta\,|\,\mathcal{R}^+)  & = &  E_0\left(\frac{1}{N} \,\log Z_N(\beta\,|\, K(\tilde{\rho}))  \right)   
      \quad = \;  E_0\left( F_N(\beta\,|\, K(\tilde{\rho}))  \right) 
            \\
     \nonumber
     & = &   \sum_{\rho\in \mathcal{R}_N^+} \frac{1}{|\mathcal{R}^+_N|} \,F_N\left(\beta\,|\, K(\rho)\right)
     \\
     \label{eq.FQknotdef}
    & = &  \sum_K  \frac{|\mathcal{R}^+_N[K]|}{|\mathcal{R}^+_N|}   \,F_N(\beta\,|\, K) \,,
\end{eqnarray}
where $E_0$ is expectation over $\tilde{\rho}$ uniformly drawn from $\mathcal{R}^{+}_N$
and the final sum is over all knot types $K$.

The definitions for quenched adsorption of ring polymers at a penetrable surface are exactly
analogous, but with the superscript $\pen$ instead of $+$.
For a knot type  $K$, let $\mathcal{R}_N^{\pen}[K]$ be the set of all polygons in $\mathcal{R}_N^{\pen}$
that have knot type $K$.  These represent the possible configurations when an $N$-bond ring polymer 
of knot type $K$ comes into contact with the penetrable surface.  Accordingly, we define
the partition function
\begin{equation}
   \label{eq.Zknotfull}
    Z_N\left(\beta\,\left|\,\mathcal{R}^{\pen}[K] \right. \right)  
   \; =\;  \sum_{\rho\in {\mathcal{R}_N^{\pen}[K]} } \exp(\beta \sigma(\rho))  \,,
\end{equation}
the probability distribution (for $\beta\in \mathbb{R}$)
\begin{equation}
   \label{eq.probknotfull}
     \text{Pr}_{\beta}(\rho \,|\,\mathcal{R}^{\pen}_N[K])  \;=\; \frac{e^{\beta \sigma(\rho)}}{
        Z_N(\beta\,|\,\mathcal{R}^{\pen}[K])}
     \hspace{5mm}(\rho \in \mathcal{R}^{\pen}_N[K])  \,,    
\end{equation}
and
the average quenched free energy 
\begin{eqnarray}
   \nonumber
     F_N^Q(\beta\,|\,\mathcal{R}^{\pen})  & = &  E_0\left(\frac{1}{N} \,\log Z_N(\beta\,|\, 
        \mathcal{R}^{\pen}[K(\tilde{\rho})] )  \right)
     \\
     \nonumber
     & = &   \sum_{\rho\in \mathcal{R}^{\pen}_N} \frac{1}{|\mathcal{R}^{\pen}_N|} 
     \,\frac{1}{N} \,\log Z_N\left(\beta\,|\, 
      \mathcal{R}^{\pen}[K(\rho)]\right)
     \\
     \label{eq.FQknotdeffull}
    & = &  \sum_K  \frac{|\mathcal{R}^{\pen}_N[K]|}{|\mathcal{R}^{\pen}_N|} 
      \,\frac{1}{N} \,\log Z_N(\beta\,|\,   \mathcal{R}^{\pen}[K]) \,.
\end{eqnarray}

\medskip
We now turn to branched polymers.  For $N\in \mathbb{N}$,   
let $\mathcal{G}_N$ be the set of all unlabelled connected
graphs with exactly $N$ sites.  Here we consider abstract graphs, rather than subgraphs 
of $\mathbb{L}^d$.  We only need to consider simple graphs, i.e.\ graphs with no multiple edges and 
no one-edge loops.  
Also, let $\mathcal{G}^T_N$ be the set of all trees in  $\mathcal{G}_N$, i.e.\ graphs with
no cycles.  See Figure \ref{fig.graphs}.

Each lattice animal $\rho$ in $\mathcal{A}_N$ is isomorphic as a graph\footnote{We emphasize that 
graph isomorphism preserves the number of edges and sites.  Edges cannot be subdivided.
E.g.\ in Figure \ref{fig.graphs}, any animal isomorphic to $w$, $x$, or $y$ must have exactly 
four edges and five sites.} 
to a unique graph in $\mathcal{G}_N$, which we denote $G(\rho)$.  
If $g=G(\rho)$, we say that $\rho$ is an embedding of $g$ (in $\mathbb{L}^d$).
(In our discussion of lattice animals in $\mathbb{L}^3$, we ignore knotting behaviour.  See Section
\ref{sec-discother} for alternate approaches.) 
If $\rho\in \mathcal{T}_N$, then $G(\rho)\in \mathcal{G}_N^T$.
For a given $g\in \mathcal{G}_N$,  we would like to consider the set of all embeddings of $g$ 
in $\mathbb{L}^d$, that is the set of all lattice animals $\alpha$ such that $G(\alpha)=g$.
Accordingly, we define the set of embeddings of $g$ at a penetrable surface, 
$\mathcal{A}^{\pen}_N[g]$,         
to be 
\begin{equation}
    \mathcal{A}^{\pen}_N[g] \;=\;  \{  \alpha \in \mathcal{A}^{\pen}_N\,:  \,  G(\alpha)\,=\,g\} 
     \;=\; G^{-1}(g)\,;
\end{equation} 
that is, $\mathcal{A}^{\pen}_N[g]$ is the inverse image of $g$ under $G$, where it is
implicitly understood that the domain of $G$ is being restricted to $\mathcal{A}_N^{\pen}$.
If $g$ is a tree with $N$ vertices, then $\mathcal{A}^{\pen}_N[g]\,\subseteq \,\mathcal{T}^{\pen}_N$.

For example, in the case $N=5$ and the simple cubic lattice $\mathbb{L}^3$, the embeddings
of the graph $w$ from Figure \ref{fig.graphs} are the (undirected) 4-step self-avoiding walks that contain
the origin.  Thus we can compute that $|\mathcal{A}^{\pen}_5[w]|\,=\, 1815$.  
(Table C.2 in \cite{MS} says 
that there are 726 4-step SAWs in $\mathbb{L}^3$ that start at the origin; each such SAW is
a translation of 5 walks (including itself) that contain the origin; 
each embedding of $w$ gives two SAWs 
since either endpoint can be the starting point of the walk; and finally, $726\times 5 \div 2=1815$.)
For the other two trees $x$ and $y$ in Figure \ref{fig.graphs}, we can compute 
$|\mathcal{A}^{\pen}_5[x]|\,=\, 5\binom{6}{4} \,=\, 75$ and $|\mathcal{A}^{\pen}_5[y]| \,=\, 1500$.  
Since $\mathcal{G}^T_5=\{w,x,y\}$, it follows that 
$|\mathcal{T}^{\pen}_5|\,=\, 1815+75+1500 \,=\, 3390$ in $\mathbb{L}^3$.

\setlength{\unitlength}{.6mm}
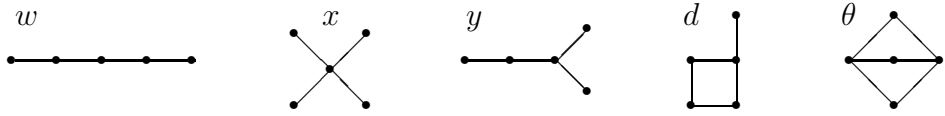
\begin{figure}
 \newsavebox{\dddott}
\savebox{\dddott}(0,0){\circle*{1.5}}
\begin{center}
\begin{picture}(205,22)
\put(0,18){$w$}
\put(0,10){\line(1,0){40}}
\multiput(0,10.2)(10,0){5}{\usebox{\dddott}}
\put(100,18){$y$}
\put(100,10){\line(1,0){20}}
\put(120,10){\line(1,1){7}}
\put(120,10){\line(1,-1){7}}
\multiput(100.5,10.2)(10,0){3}{\usebox{\dddott}}
\multiput(127.5,3.2)(0,14){2}{\usebox{\dddott}}
\put(68,18){$x$}
\put(62,0){\line(1,1){16}}
\put(62,16){\line(1,-1){16}}
\multiput(62.6,0.2)(8,8){3}{\usebox{\dddott}}
\multiput(62.6,16.2)(16,-16){2}{\usebox{\dddott}}
\put(148,18){$d$}
\put(150,10){\line(1,0){10}}
\put(150,0){\line(1,0){10}}
\put(150,0){\line(0,1){10}}
\put(160,0){\line(0,1){20}}
\put(160.6,20.2){\usebox{\dddott}}
\multiput(150.6,10.2)(10,0){2}{\usebox{\dddott}}
\multiput(150.6,0.2)(10,0){2}{\usebox{\dddott}}
\put(183,18){$\theta$}
\put(185,10){\line(1,0){20}}
\put(185,10){\line(1,1){10}}
\put(185,10){\line(1,-1){10}}
\put(195,20){\line(1,-1){10}}
\put(195,0){\line(1,1){10}}
\multiput(185.6,10.2)(10,0){3}{\usebox{\dddott}}
\multiput(195.6,0.2)(0,20){2}{\usebox{\dddott}}
\end{picture}
\end{center}
\caption{\label{fig.graphs} Some of the graphs in $\mathcal{G}_5$.  Observe that 
$\mathcal{G}_5^T$, the set of trees with 5 sites, contains the graphs $w$, $x$, and $y$, 
but no other graphs.}
\end{figure}

Similarly, we define
\begin{eqnarray*}
   \mathcal{T}^{\pen}_N[g] & = &  \{  \alpha \in \mathcal{T}^{\pen}_N\,:  \,  G(\alpha)\,=\,g\} \,,
   \\
     \mathcal{A}_N^+[g] & = &  \{  \alpha \in \mathcal{A}^+_N\,:  \,  G(\alpha)\,=\,g\} \,,
     \\
        \mathcal{T}_N^+[g] & = &  \{  \alpha \in \mathcal{T}^+_N\,:  \,  G(\alpha)\,=\,g\} \,.
\end{eqnarray*}
(The case of combs will be treated separately below.)
Observe that if $g\in \mathcal{G}_N$ and $g$ is a tree, then $\mathcal{T}_N[g]=\mathcal{A}_N[g]$ 
and  $\mathcal{T}_N^+[g]=\mathcal{A}_N^+[g]$.
However, if  $g\in \mathcal{G}_N$ but $g$ is not a tree, then 
$\mathcal{T}^{\pen}_N[g] =\emptyset = \mathcal{T}_N^+[g]$.

We now consider the case of combs.  There are two minor technical differences between trees 
and combs:  each comb has two special sites that are labelled, and we count combs by edges rather
than by sites.  (These two differences also hold between trees and self-avoiding walks.)

Let $\mathcal{G}_N^C$ be the set of (abstract) graphs with $N$ edges and two labeled vertices 
that have the topology of a comb.  
We can list the graphs in $\mathcal{G}_N^C$ explicitly as follows.  
For a given nonnegative integer $b$, let $\vec{n}=(n_0,n_1,\ldots,n_b)\in \mathbb{N}^{b+1}$ and
$\vec{s}=(s_1,s_2,\ldots,s_b)\in \mathbb{N}^b$ be vectors of strictly positive integers.
Let $N_k=\sum_{i=0}^{k-1} n_i$ for each $k$.
Then let $\langle b;\vec{n};\vec{s}\rangle \equiv \langle b; n_0,\ldots,n_b;s_1,\ldots,s_b\rangle$
denote the graph defined as follows (see Figure \ref{figcomb1}):
\begin{itemize}
\item[($i$)]  there is a path $P$ with $N_{b+1}$ steps whose endpoints are labelled, say, 
$v_A$ and $v_B$;
%\\
\item[($ii$)] for each $k=1,\ldots,b$, there is a path with $s_k$ steps, of which one endpoint is the 
point of $P$ that is $N_{k}$ steps from $v_A$;
%\\
\item[($iii$)]  the paths in ($ii$) have no other vertices in common with one another or with $P$.
\end{itemize}
In our terminology, the path $P$ is the backbone of the graph, and the $b$ paths of part ($ii$) are 
the side chains.
For $b=0$, the graph $\langle 0; n_0;\rangle$ is just a single path of $n_0$ steps.
Then $\mathcal{G}_N^C$ is the set of all graphs $\langle b;\vec{n};\vec{s}\rangle$ with $b\geq 0$
such that $n_0+\sum_{i=1}^b(n_i+s_i)\,=\,N$.

\setlength{\unitlength}{1mm}
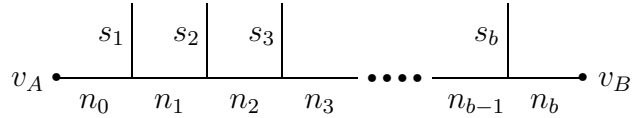
\begin{figure}[h]
\begin{center}
\begin{picture}(90,24)(-10,4)
\put(0,10){\line(1,0){40}}
\put(10,10){\line(0,1){10}}
\put(20,10){\line(0,1){10}}
\put(30,10){\line(0,1){10}}
\put(60,10){\line(0,1){10}}
\put(50,10){\line(1,0){20}}
\multiput(42,10)(2,0){4}{\circle*{1.0}}
\put(3,6){$n_0$}
\put(13,6){$n_1$}
\put(23,6){$n_2$}
\put(33,6){$n_3$}
\put(52,6){$n_{b-1}$}
\put(63,6){$n_b$}
\put(5.5,15){$s_1$}
\put(15.5,15){$s_2$}
\put(25.5,15){$s_3$}
\put(55.5,15){$s_b$}
\put(0,10.2){\circle*{1.2}}
\put(70,10.2){\circle*{1.2}}
\put(-6,9){$v_A$}
\put(72,9){$v_B$}
\end{picture}
\end{center}
\caption{\label{figcomb1}   Schematic sketch of a typical comb topology 
$\langle b; \vec{n};\vec{s}\rangle$.  The numbers of steps in the segments are given by
the components of $\vec{n}=(n_0,n_1,\ldots,n_b)$ and $\vec{s}=(s_1,s_2,\ldots,s_b)$.
The endpoints of the backbone are labelled $v_A$ and $v_B$.}
\end{figure}
\setlength{\unitlength}{1mm}
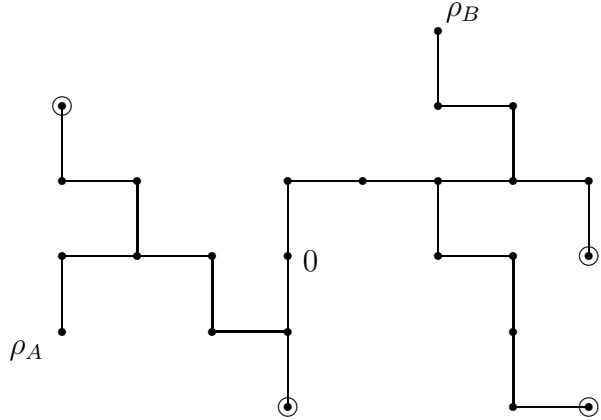
\begin{figure}[h]
\begin{center}
\begin{picture}(80,55)(0,0)
\put(10,10){\line(0,1){10}}
\put(20,20){\line(0,1){10}}
\put(10,20){\line(1,0){20}}
\put(10,30){\line(0,1){10}}
\put(30,10){\line(0,1){10}}
\put(10,30){\line(1,0){10}}
\put(30,10){\line(1,0){10}}
\put(40,0){\line(0,1){30}}
\put(40,30){\line(1,0){40}}
\put(60,20){\line(0,1){10}}
\put(60,20){\line(1,0){10}}
\put(70,0){\line(0,1){20}}
\put(70,0){\line(1,0){10}}
\put(70,30){\line(0,1){10}}
\put(60,40){\line(0,1){10}}
\put(60,40){\line(1,0){10}}
\put(80,20){\line(0,1){10}}
\multiput(10,10)(0,10){4}{\circle*{1.2}}
\multiput(20,20)(0,10){2}{\circle*{1.2}}
\multiput(30,10)(0,10){2}{\circle*{1.2}}
\multiput(40,0)(0,10){4}{\circle*{1.2}}
\multiput(50,30)(10,0){4}{\circle*{1.2}}
\multiput(60,40)(10,0){2}{\circle*{1.2}}
\multiput(70,0)(0,10){3}{\circle*{1.2}}
\put(60,50){\circle*{1.2}}
\put(60,20){\circle*{1.2}}
\put(80,0){\circle*{1.2}}
\put(80,20){\circle*{1.2}}
\put(10,40){\circle{2.4}}
\put(40,0){\circle{2.4}}
\put(80,0){\circle{2.4}}
\put(80,20){\circle{2.4}}
\put(3,7){$\rho_A$}
\put(61,52){$\rho_B$}
\put(42,18){$0$}
\end{picture}
\end{center}
\caption{\label{figcomb2}   A comb $\rho$ in $\mathcal{C}^{\pen}_{24}[4;2,3,4,1,3;3,1,5,2]$. 
The ends of the 
backbone are labelled $\rho_A$ and $\rho_B$.  The free end of each of the four side chains is 
encircled.}
\end{figure}

Since there are $\binom{N-1}{2b}$ ways to choose a sequence of $2b+1$ positive integers whose 
sum is $N$, the number of topologies for $N$-edge combs is
\[     \sum_{b=0}^{\lfloor (N-1)/2\rfloor} \binom{N-1}{2b}    \;=\;  \frac{1}{2}\,2^{N-1}   \;=\; 2^{N-2}.
\]
(The first equality is well known:  since $\sum_{t=0}^M\binom{M}{t}(-1)^t=(1-1)^M=0$, we have 
$\sum_{t \hbox{ }even}\binom{M}{t} \,=\, \sum_{t \hbox{ }odd}\binom{M}{t}$.)

\medskip
For each graph $\langle b;\vec{n};\vec{s}\rangle$ in $\mathcal{G}_N^C$, the set of 
combs with this topology in $\mathcal{C}^X_N$ (where $X\in \{+,\pen\}$) is given by
$\mathcal{C}^X_N[ \langle b;\vec{n};\vec{s}\rangle ]$,
which we shall write more simply as $\mathcal{C}^X_N[ b;\vec{n};\vec{s} ]$.  
See Figure \ref{figcomb2}.

\smallskip
We now return to more general branched models $\mathcal{P}^X$, where $\mathcal{P}$ can be 
$\mathcal{A}$, $\mathcal{T}$, or $\mathcal{C}$.
For $g$ in $\mathcal{G}_N$ (or $\mathcal{G}_N^C$) and $\beta$ in $\mathbb{R}$, 
we define the partition function on $\mathcal{P}^X_N[g]$ by
\begin{equation}
   \label{eq.Zgraphgen}
    Z_N\left(\beta\,\left|\,\mathcal{P}^X[g] \right. \right)  
   \; :=\;  \sum_{\rho\in {\mathcal{P}^X_N[g]} } \exp(\beta \sigma(\rho))  \,.
\end{equation}
We also define 
the probability distribution $\Pr_{\beta}(\cdot \,|\,\mathcal{P}^X_N[g])$
on $\mathcal{P}^X_N[g]$ by 
\begin{equation}
   \label{eq.probclassgen}
     \text{Pr}_{\beta}(\rho \,|\,\mathcal{P}_N^X[g])  \;=\; \frac{e^{\beta \sigma(\rho)}}{
        Z_N(\beta\,|\,\mathcal{P}^X[g])}
     \hspace{5mm}(\rho \in \mathcal{P}^X_N[g])  \,.      
\end{equation}
Now consider a random lattice animal $\tilde{\alpha}$ chosen uniformly from 
$\mathcal{P}^X_N$.   We then interpret $G(\tilde{\alpha})$ as a randomly chosen graph (topology).  
For fixed $N$, we understand $F_N( \beta\,|\,\mathcal{P}^X[G(\tilde{\alpha})])
\,= \, \frac{1}{N}\log Z_N(\beta\,|\,\mathcal{P}^X[G(\tilde{\alpha})])$ to be a
\textit{randomly chosen free energy function}.  The average with respect to $\tilde{\alpha}$ 
of this function over $\mathcal{P}^X_N$
is the \textit{average quenched free energy} $F_N^Q(\beta\,|\,\mathcal{P}^X)$:
\begin{eqnarray}
   \nonumber
     F_N^Q(\beta\,|\,\mathcal{P}^X)  & = &  E_0\left( F_N(\beta\,|\, 
        \mathcal{P}^X[G(\tilde{\alpha})] )  \right)
     \\
     \nonumber
     & = &   \sum_{\alpha\in \mathcal{P}^X_N} 
     \frac{1}{|\mathcal{P}^X_N|} \,\frac{1}{N} \,\log Z_N\left(\beta\,|\, 
      \mathcal{P}^X[G(\alpha)]\right)
     \\
     \label{eq.FQbranchdef}
    & = &  \sum_{g}  \frac{|\mathcal{P}^X_N[g]|}{|\mathcal{P}^X_N|}   \,\frac{1}{N} \,\log Z_N(\beta\,|\, 
       \mathcal{P}^X[g]) \,,   
\end{eqnarray}
where $E_0$ is expectation over all $\tilde{\alpha}$ uniformly drawn from $\mathcal{P}^X_N$,
and the second sum is over all graphs (or trees) $g$ with $N$ sites, 
or all combs $g$ in $\mathcal{G}_N^C$.

The notation for quenched models of animals, trees, and combs coincides with that for knots in
self-avoiding polygons.  We do not consider quenched models for self-avoiding walks, 
which have limited topological interest for us.  Henceforth, when we use a phrase such as 
``all of our quenched models,'' we are referring to the eight models $\mathcal{P}^X$ 
where $\mathcal{P}$ is one of $\mathcal{A}$, $\mathcal{T}$, $\mathcal{C}$, or $\mathcal{R}$, 
and $X$ is one of $+$ or $\pen$.

We shall use the variable $\tau$ for a generic topology, either a knot type for rings or an
abstract graph for branched models.  With this understanding,
we introduce the following expectation for each quenched model $\mathcal{P}^X$:
\begin{equation}
  \label{eq.PbQdef}
   E_{\beta}^Q\left(f(\rho)\,|\,\mathcal{P}^X_N\right)  \;=\;  
   \sum_{\tau}  \frac{|\mathcal{P}^X_N[\tau]|}{|\mathcal{P}^X_N|} 
        \, E_{\beta}\left(f(\rho)\,|\,\mathcal{P}^X_N[\tau]\right)  
\end{equation}
where $f$ is a function and
$E_{\beta}\left( \,\cdot\,|\,\mathcal{P}^X_N[\tau]\right)$ is the expectation operator 
corresponding to the probability distribution of Equation (\ref{eq.probclassgen})
or (\ref{eq.probknot}) or (\ref{eq.probknotfull}), i.e.\
\begin{equation}
  \label{eq.Pbclassdef}
    E_{\beta}\left(f(\rho)\,|\,\mathcal{P}^X_N[\tau]\right)  \;=\;  \sum_{\rho\in\mathcal{P}^X_N[\tau]}
      f(\rho) \,\frac{e^{\beta \sigma(\rho)}}{
        Z_N(\beta\,|\,\mathcal{P}^X[\tau])}  \,.
\end{equation}
Notice that with this definition, 
\begin{equation}
   \label{eq.expvisitsquench}
    \frac{1}{N} E_{\beta}^Q\left(\sigma(\rho)\,|\,\mathcal{P}^X_N\right) \;=\;
         \frac{d}{d\beta} F_N^Q(\beta\,|\,\mathcal{P}^X), 
\end{equation}
analogously to  Equation (\ref{eq.expvisitsgen}).

One important difference between the annealed and the quenched models is in terms of the 
expected number of surface contacts for large $\beta$, as given in the following result.
(See also the discussions around Equations (\ref{eq.limsigeq1})--(\ref{eq.Prepstrefoil1}).)

\begin{thm}
    \label{thm-slopelim}
Let $X$ be one of $+$ or $\pen$.
Let $\mathcal{P}$ be one of $\mathcal{A}$ or $\mathcal{T}$ in dimension $d\geq 2$, or 
$\mathcal{R}$ in 3 dimensions.  Then    
\begin{eqnarray}
    \label{eq.Elimanneal}
    \lim_{\beta\rightarrow\infty}  \liminf_{N\rightarrow\infty}
    \frac{E_{\beta}\left(\sigma(\rho)\,|\,\mathcal{P}^X_N\right)}{N}  & = & 1
          \hspace{3mm}\text{for each annealed model, and}
          \\
      \label{eq.Elimquench}
 \lim_{\beta\rightarrow\infty} 
      \limsup_{N\rightarrow\infty}
      \frac{ E_{\beta}^Q\left(\sigma(\rho)\,|\,\mathcal{P}^X_N\right) }{N} & < & 1
          \hspace{3mm}\text{for each quenched model.}
\end{eqnarray}     
When $\mathcal{P}=\mathcal{C}$, Equation (\ref{eq.Elimanneal}) always holds, 
Equation (\ref{eq.Elimquench}) 
holds for $d=2$, and
\begin{equation}
      \label{eq.Elimquenchcomb}
 \lim_{\beta\rightarrow\infty} 
      \liminf_{N\rightarrow\infty}
      \frac{ E_{\beta}^Q\left(\sigma(\rho)\,|\,\mathcal{C}^X_N\right) }{N} \; = \; 1
          \hspace{3mm}\text{for $d\geq 3$.}
\end{equation}
\end{thm}

\begin{rem}
    \label{rem.ringexp2}
{\em   Rings in $d=2$ behave differently.  See Remark \ref{rem.ringslb1}. }
\end{rem}

The proof of Theorem \ref{thm-slopelim} for quenched combs is Proposition \ref{prop.negbetaQcomb}, 
and the rest of the proof is Proposition \ref{prop-slopelimATR}.

\medskip

Unfortunately, it appears difficult to prove that $F_N^Q(\beta\,|\,\mathcal{P}^X)$ has a limit 
as $N\rightarrow\infty$, even in the special case $\beta=0$.  A proof for quenched models 
appears to require more than the  concatenation arguments 
that were used to prove the existence of the limiting free energy for annealed models
(for instance, see the discussion following Proposition \ref{prop.MNtree}).
We can say more for the relative free energies 
$F_N^Q(\beta\,|\,\mathcal{P}^X) - F_N^Q(0\,|\,\mathcal{P}^X)$, although the results here remain
incomplete.  The following two results are fundamental.

\begin{prop}
   \label{prop.jensen}
For each of our models $\mathcal{P}^X$ and every real $\beta$, 
\[    F_N(\beta\,|\,\mathcal{P}^X)\,-   F_N(0\,|\,\mathcal{P}^X)   \;\geq \;  
     F^Q_N(\beta\,|\,\mathcal{P}^X)\,-   F^Q_N(0\,|\,\mathcal{P}^X) \,.
\]
\end{prop}
The proof of the above proposition is an application of Jensen's Inequality.  It appears 
in Section  \ref{sec.proofs1}.

\smallskip

For each of our models $\mathcal{P}^X$ and each real $\beta$, we define the limiting relative 
quenched free energy 
\begin{equation}
    \label{eq.FminusFgen}
    \mathcal{F}^Q_{\Delta}(\beta\,|\,\mathcal{P}^X)  \;:=\;
        \lim_{N\rightarrow\infty}\left(   F_N^Q(\beta\,|\,\mathcal{P}^{X}) \,-\,  F_N^Q(0\,|\,\mathcal{P}^{X})
         \right)        
\end{equation}        
if this limit exists.
The next result says that we can prove that the limit $\mathcal{F}^Q_{\Delta}(\beta\,|\,\mathcal{P}^X)$ 
exists for negative $\beta$.
It essentially tells us that desorption occurs whenever the boundary is repulsive.

\begin{prop}
   \label{prop.negbetaQ}
For each of our models $\mathcal{P}^X$, the limit $\mathcal{F}^Q_{\Delta}(\beta\,|\,\mathcal{P}^X)$
exists and equals 0  for every $\beta\leq 0$.
\end{prop}
The proof appears in Section \ref{sec.proofs1}.

\medskip
The above result motivates the following definition of the critical point for the adsorption transition
in the quenched models.

\begin{defi}
    \label{def.betacQ}
For each model $\mathcal{P}^X$, let 
\begin{equation}
    \label{eq.betacQdef}
    \beta^Q_c(\mathcal{P}^X)  \;=\;  
     \sup\left\{ \beta \,:\,   
     \mathcal{F}^Q_{\Delta}(\beta\,|\,\mathcal{P}^X) 
      \;=\; 0 \right\} \,.
\end{equation}
Here, existence of the limit of Equation (\ref{eq.FminusFgen}) is implicit in the assertion that
$\mathcal{F}^Q_{\Delta}(\beta\,|\,\mathcal{P}^X)=0$ for a given $\beta$.
\end{defi}

\begin{cor}
   \label{cor.negbetaquench}
For each model $\mathcal{P}^X$, 
\[    \beta_c^Q(\mathcal{P}^X)\;\geq \; \beta_c(\mathcal{P}^X) \;\geq \; 0 \,.
\]
In particular, $\beta^Q_c(\mathcal{P}^+)>0$ for each model at an impenetrable surface.
\end{cor}
\textbf{Proof:}  This follows from Propositions \ref{prop.betacboundann},       
\ref{prop.jensen} and \ref{prop.negbetaQ}, 
together with the fact that $F^Q_N(\beta\,|\,\mathcal{P}^X)\,-   F^Q_N(0\,|\,\mathcal{P}^X) \;\geq\; 0$
for all $\beta\geq 0$.  The final assertion of the corollary
follows from Theorem \ref{thm.betacanngt0}.
\hfill  $\Box$

\bigskip
Although it is not hard to see  that $\beta_c$ is finite for each of our annealed models 
(Proposition \ref{prop.betacboundann}),
it is less obvious whether $\beta_c^Q$ is finite in the quenched models.
To appreciate the difficulties, consider the following.
\begin{conj}
  \label{conj.2dnoads}
Let $d=2$.  Then $\beta_c^Q(\mathcal{P}^X)=\infty$ for the models $\mathcal{T}^{\pen}$, 
$\mathcal{T}^+$, $\mathcal{A}^{\pen}$, and $\mathcal{A}^+$.
\end{conj}
That is, we conjecture that \textit{adsorption cannot occur in these quenched two-dimensional
models of branched polymers,}  namely lattice trees or animals
in the square lattice being adsorbed
at the (penetrable or impenetrable) line $x_1=0$.  We shall explain the reasoning now
for the model of lattice trees and a penetrable line.

For any lattice tree $\rho\in \mathcal{T}_N$, let $LP(\rho)$ be the length of 
the longest path in $\rho$ (this is also called the diameter of the graph $\rho$, but 
we avoid this term because the 
diameter of a subset of $\mathbb{R}^d$ has other possible interpretations).  Observe that 
$\sigma(\rho)\leq LP(\rho)+1$, since the set $\mathcal{H}(\rho)$ of visits is a subset of the line
$x_1=0$, and the maximum distance between two points of $\mathcal{H}(\rho)$ is an upper bound 
for $\sigma(\rho)-1$ ($=|\mathcal{H}(\rho)|-1$) and is a lower bound for $LP(\rho)$.
Then for $\beta>0$, 
\[      Z_N\left(\beta\,|\, \mathcal{T}^{\pen}[G(\rho)] \right)  \;\leq \;
       |\mathcal{T}^{\pen}[G(\rho)]|   \,\exp\left(\beta  (LP(\rho)+1)\right) \,.
\]
Taking the expectation of $\rho$ with respect the uniform distribution $E_0$ on 
$\mathcal{T}^{\pen}_N$ gives, for all $\beta>0$,
\begin{eqnarray}
\nonumber
    F_N^Q(\beta \,|\,\mathcal{T}^{\pen})   & = & 
         E_0\left(   \frac{1}{N} \log Z_N\left(\beta\,|\, \mathcal{T}^{\pen}[G(\rho)] \right)   \right)
     \\
     \nonumber
    & \leq &E_0\left( \frac{1}{N} \log |\mathcal{T}^{\pen}[G(\rho)]| \right)  \,+\, 
    \beta\, \frac{E_0(LP(\rho))+1}{N}
    \\
    & = &  F_N^Q(0\, |\,\mathcal{T}^{\pen}) \,+\, \beta\, \frac{E_0(LP(\rho))+1}{N} \,.
    \label{eq.2dnoads}
\end{eqnarray}
According to heuristic arguments and Monte Carlo results \cite{GdG,RM}, it is believed that 
$E_0(LP(\rho))$ is asymptotically proportional to $N^p$ for some power $p<1$, which 
is estimated to be close to 0.75 in $\mathbb{L}^2$.  
Since $F_N^Q(\beta \,|\,\mathcal{T}^{\pen})\geq F_N^Q(0 \,|\,\mathcal{T}^{\pen})$ for $\beta>0$, it
follows that $\lim_{N\rightarrow\infty}F_N^Q(\beta \,|\,\mathcal{T}^{\pen})-
F_N^Q(0 \,|\,\mathcal{T}^{\pen}) \,=\,0$
for all $\beta>0$.  This is the conclusion of Conjecture \ref{conj.2dnoads}.

\smallskip
In contrast to Conjecture \ref{conj.2dnoads}, we 
can prove that adsorption does occur for quenched comb polymers in all dimensions.
Indeed, Proposition \ref{prop.negbetaQcomb}  tells us that for $X\in \{+,\pen\}$, 
\begin{equation}
     \label{eq.betacQcombbound}
        \beta_c^Q(\mathcal{C}^X) \; \begin{cases}
         \leq \;  \log \lambda_{d,C}   & \hbox{for $d\geq 3$, \; \;and}  \\
         <  \;  \infty &  \hbox{for $d=2$.}
         \end{cases}
\end{equation}
As well, Proposition \ref{prop.ringQfin} proves that adsorption occurs for quenched ring polymers
in three dimensions, i.e.\ that $\beta_c^Q(\mathcal{R}^X)<\infty$.

\section{Proofs of general properties}
   \label{sec.proofs1}

We begin with a simple lemma whose importance was originally demonstrated in \cite{HTW} for SAWs. 
First, we introduce the dimension-dependent notation 
$\mathcal{P}^{X,d}$ to specify that we are discussing the $d$-dimensional version of the
model $\mathcal{P}^X$.
This is necessary in the following lemma where we discuss the model simultaneously in dimensions
$d$ and $d-1$.  We shall write $\lambda_{d,P}$ for the corresponding 
growth constant defined in Equations (\ref{eq.animallim}-\ref{eq.comblim}).

\begin{lem}
   \label{lem.Fbounds}
Consider the annealed model $\mathcal{P}^{X,d}$ where $\mathcal{P}$ is one of $\mathcal{A}$, 
$\mathcal{T}$, $\mathcal{W}$, or $\mathcal{C}$ in dimension $d\geq 2$, or 
$\mathcal{P}=\mathcal{R}$ in dimension $d\geq 3$.  Then
\begin{equation}
   \label{eq.Z2bounds}
   \left|\mathcal{P}_N^{\pen,d-1}\right|\,e^{\beta N}  \;\leq \;  Z_N(\beta\,|\,\mathcal{P}^{X,d})   
   \; \leq \;     \left|\mathcal{P}_N^{X,d}\right|\,e^{\beta(N+1)}   
   \hspace{6mm}\hbox{for all }\beta \geq 0\,,
\end{equation}
and hence
\begin{equation}
    \label{eq.Fboundsann}
    \log \lambda_{d-1,P}  \,+\,\beta   \;\leq \;  \mathcal{F}(\beta \,|\,\mathcal{P}^{X,d})
        \;\leq \;     \log \lambda_{d,P}  \,+\,\beta 
        \hspace{6mm}\hbox{for all }\beta \geq 0\,.
\end{equation}
\end{lem}
\textbf{Proof:} The upper bound on $Z_N$
is obvious because $\sigma(\rho)\leq N+1$ for every $\rho\in \mathcal{P}^{X,d}_N$, and
the lower bound results from restricting the sum defining $Z_N$ to 
members of $\mathcal{P}^{X,d}_N$ that are entirely contained in the hyperplane $x_1=0$.
The bounds on $\mathcal{F}$ follow immediately.
\hfill  $\Box$

\begin{rem}[due to Soteros \cite{Sot}]
     \label{rem.ringslb1}  
{\em The lower bound of Equation (\ref{eq.Z2bounds}) is zero for rings in 2 dimensions.  
This is because
the line $x_1=0$ in $\mathbb{L}^2$ cannot contain a ring, 
so  $\mathcal{R}_N^{X,1}$ is empty for every $N$.
Indeed, 
no $N$-step ring in $\mathbb{L}^2$ can have more than $N/2$ sites in the line $x_1=0$;
and for each even $N$, there exists a rectangular polygon $\rho\in \mathcal{R}_N^{X,2}$
with $\sigma(\rho)=N/2$.  This leads to the modified bounds
$1 \cdot e^{\beta N/2} \,\leq \, Z_N(\beta\,|\,\mathcal{R}^{X,2})\,\leq \, 
|\mathcal{R}_N^{X,2}| \, e^{\beta N/2}$ for $\beta\geq 0$ for even $N$.   Therefore 
$\beta/2 \,\leq \, \mathcal{F}(\beta\,|\,\mathcal{R}^{X,2})\,\leq  \, \log\lambda_{2,W}+\beta/2$,
which is Equation (4.10) of \cite{Sot}.}
\end{rem}

Our first application of Lemma \ref{lem.Fbounds} is to prove the following result,
which is required for Proposition \ref{prop.betacboundann}.  
It was first proved in  \cite{HTW} for SAWs.
Recall the definition of $\beta_c$ from Equation (\ref{eq.defbetac}).

\begin{prop}
    \label{prop.betacfiniteann}
Consider the annealed model $\mathcal{P}^{X,d}$ where $\mathcal{P}$ is one of $\mathcal{A}$, 
$\mathcal{T}$, $\mathcal{W}$, or $\mathcal{C}$ in dimension $d\geq 2$, or 
$\mathcal{P}=\mathcal{R}$ in dimension $d\geq 3$. 
The critical value satisfies 
$\beta_c(\mathcal{P}^{X,d}) \,\leq \,  \log \lambda_{d,P} -\log \lambda_{d-1,P}$. 
\end{prop}
\textbf{Proof:}
The left hand inequality of Equation (\ref{eq.Fboundsann}) implies that
$\mathcal{F}(\beta\,|\,\mathcal{P}^{X,d}) \,>\,  \mathcal{F}(0\,|\,\mathcal{P}^{X,d})\,=\,\log \lambda_{d,P}$
whenever  $\beta>  \log \lambda_{d,P} -\log \lambda_{d-1,P}$. 
The proposition follows.
\hfill $\Box$

\begin{rem}
    \label{rem.betacring2}
{\em Remark \ref{rem.ringslb1} implies that
$\beta_c(\mathcal{R}^{X,2}) \,\leq \, 2\log \lambda_{2,W}$.}
\end{rem}

We shall also use Lemma \ref{lem.Fbounds}  together with Equation (\ref{eq.expvisitsgen}), which 
says that 
\begin{equation}
   \label{eq.expvisitsgen2}
   E_{\beta}\left( \left.\frac{\sigma(\rho)}{N} \,\right|\,\mathcal{P}_N^X\right)   \;=\;
    \frac{  d  \, \log F_N(\beta\,|\,\mathcal{P}^X)}{d \, \beta} \,,
\end{equation}
as well as its quenched analogue, Equation   (\ref{eq.expvisitsquench}). 
These ingredients will be linked by the following lemma.

\begin{lem}
    \label{lem.convderiv}
Let $f$ be a convex differentiable function on $\mathbb{R}$, and assume that there
exist constants $\underline{a}$, $\overline{a}$, $\underline{m}$ and $\overline{m}$ such that
\begin{equation}
    \label{eq.convexbound}
      \underline{a}\,+\,\underline{m}\,\beta  \;\leq \; f(\beta)  \;\leq \;    \overline{a}\,+\,\overline{m}\,\beta    
   \hspace{5mm}\hbox{for all }\beta\geq 0.
\end{equation}
Then the derivative of $f$ satisfies
\begin{equation}
    \label{eq.convexbound}
      \underline{m}\,+\,\frac{\underline{a}-\overline{a}}{\beta}  \;\leq \; f^\prime(\beta)  \;\leq \;   \overline{m}   
   \hspace{5mm}\hbox{for all }\beta > 0.
\end{equation}
\end{lem}
\textbf{Proof:}    
Let $\beta>0$.  By basic properties of convexity, we have
\[       f^{\prime}(\beta)  \;\geq \;  \frac{f(\beta)-f(0)}{\beta}  \;\geq \;  
 \frac{ \underline{a}\,+\,\underline{m}\,\beta \,-\,\overline{a}}{\beta} \,.
\]
This proves the lower bound of Equation (\ref{eq.convexbound}).  Also, for every $x>0$ we have
\[       f^{\prime}(\beta)  \;\leq \;  \frac{f(\beta+x)-f(\beta)}{x}  \;\leq \;  
 \frac{ \overline{a}\,+\,\overline{m}\,(\beta+x) \,-\,\underline{a}\,-\,\underline{m}\,\beta}{x} \,.
\]
Taking the limit $x\rightarrow\infty$ shows that $f^{\prime}(\beta)\,\leq \,\overline{m}$.
\hfill  $\Box$

\bigskip
The following result proves part of Theorem \ref{thm-slopelim}.   
The rest of Theorem \ref{thm-slopelim},
concerning quenched combs, is treated in Proposition \ref{prop.negbetaQcomb}.

\begin{prop}
    \label{prop-slopelimATR}
Let $\mathcal{P}$ be one of $\mathcal{A}$ 
or $\mathcal{T}$ with $d\geq 2$, or 
$\mathcal{P}=\mathcal{R}$ with $d= 3$.  Let $X$ be one of $+$ or $\pen$.      
Then    
\begin{eqnarray}
    \label{eq.ElimannealATR}
    \lim_{\beta\rightarrow\infty}  \liminf_{N\rightarrow\infty}
    \frac{E_{\beta}\left(\sigma(\rho)\,|\,\mathcal{P}^X_N\right)}{N}  & = & 1
          \hspace{3mm}\text{for each annealed model, and}
          \\
      \label{eq.ElimquenchATR}
 \lim_{\beta\rightarrow\infty} 
      \limsup_{N\rightarrow\infty}
      \frac{ E_{\beta}^Q\left(\sigma(\rho)\,|\,\mathcal{P}^X_N\right) }{N} & < & 1
          \hspace{3mm}\text{for each quenched model.}
\end{eqnarray}   
Also, if $\mathcal{P}$ is $\mathcal{C}$ with $d\geq 2$, then Equation (\ref{eq.ElimannealATR}) holds.    
\end{prop}

\medskip
\noindent
\textbf{Proof:} 
First we consider any of the specified annealed models $\mathcal{P}^{X,d}$.
From Equation (\ref{eq.Z2bounds}), we see that for every nonnegative $\beta$ we have
\begin{equation}
   \nonumber
    \log \left|\mathcal{P}_N^{\pen,d-1}\right|^{1/N}\,+\,\beta  \;\leq \;  F_N(\beta\,|\,\mathcal{P}^{X,d})   
   \; \leq \;     \log \left|\mathcal{P}_N^{X,d}\right|^{1/N}\,+\,\beta\, \frac{N+1}{N}   \,.
\end{equation}
By Lemma \ref{eq.convexbound} and  Equation (\ref{eq.expvisitsgen2}), we find that
\[
     1 \,+\, \frac{  \log \left|\mathcal{P}_N^{\pen,d-1}\right|^{1/N} \,-\,
     \log \left|\mathcal{P}_N^{X,d}\right|^{1/N} }{\beta}  
     \; \leq \;     \frac{E_{\beta}\left(\sigma(\rho)\,|\,\mathcal{P}^X_N\right)}{N} 
      \;\leq \;  \frac{N+1}{N}  
\]
for every $\beta>0$.  Since the above lower bound converges to 
$1+(\log\lambda_{d-1,P}-\log\lambda_{d,P})/\beta$ as $N\rightarrow\infty$, Equation 
(\ref{eq.ElimannealATR}) follows.

Next, we shall prove Equation (\ref{eq.ElimquenchATR}) for quenched trees and animals.  
We shall use the pattern theorem of \cite{Ma99}, as follows.  
Let $H$ be the subgraph of $\mathbb{L}^d$ consisting of the origin,
its $2d$ neighbouring sites, and the $2d$ edges that connect the origin to each neighbour.  
In the terminology of \cite{Ma99}, the pair $(H,\emptyset)$ is a proper pattern for trees or animals
(essentially meaning that $H$ can occur as a subgraph of arbitrarily large trees or animals).  
Note however that if $\rho\in \mathcal{P}^X_N$ and $\sigma(\rho)=N$, then $\rho$ lies in the 
hyperplane $\{x_1=0\}$ and therefore \textit{cannot contain any translate of $H$.}

For $\delta>0$, let $\mathcal{P}^X_N({\leq} \delta)$ be the set of $\rho\in \mathcal{P}^X_N$ 
such that $\rho$ contains 
at most $\delta N$ translates of $H$ (we do not require the translates to be disjoint when we count).
By the pattern theorem Theorem 2.1 of \cite{Ma99}, there exists a $\delta>0$ such that 
\begin{equation}
  \label{eq.pattthmX}
    \limsup_{N\rightarrow\infty}|\mathcal{P}^X_N({
    \leq} \delta)|^{1/N}  \;<\;
    \lim_{N\rightarrow\infty}|\mathcal{P}^X_N|^{1/N}  .
\end{equation} 
Observe that  $\sigma(\rho)\leq N-\delta N$ for every $\rho$ in the complement of
$\mathcal{P}^X_N({\leq} \delta)$.     
For each graph $g$ in $\mathcal{G}_N$, the set $\mathcal{P}^X_N[g]$ (if nonempty) 
is either completely contained
in $\mathcal{P}^X_N({\leq} \delta)$ or else is contained in 
$\mathcal{P}_N({\leq} \delta)^c$, depending on the number of vertices of degree $2d$ in  $g$.    
Recalling the notation of Equations (\ref{eq.PbQdef})--(\ref{eq.Pbclassdef}), we have
\begin{eqnarray}
  \nonumber
    E_{\beta}^Q\left(\sigma(\rho)\,|\,\mathcal{P}^X_N\right) & = & 
     \sum_{g\,:\, \mathcal{P}^X_N[g]\,\subseteq \,\mathcal{P}^X_N({\leq} \delta)}
       \frac{|\mathcal{P}^X_N[g]|}{|\mathcal{P}^X_N|} 
        \, E_{\beta}\left(\sigma(\rho)\,|\,\mathcal{P}^X_N[g]\right)     
        \\
        \nonumber
        & & \hspace{6mm} + \; 
        \sum_{g\,:\, \mathcal{P}^X_N[g]\,\subseteq \,\mathcal{P}^X_N({\leq} \delta)^c}
       \frac{|\mathcal{P}^X_N[g]|}{|\mathcal{P}^X_N|} 
        \, E_{\beta}\left(\sigma(\rho)\,|\,\mathcal{P}^X_N[g]\right)     
        \\
         \label{eq.ENglessX}        
  & \leq & \frac{ |\mathcal{P}^X_N({\leq} \delta)|}{|\mathcal{P}^X_N|} N  \,+\,
  \left(1- \frac{ |\mathcal{P}^X_N({\leq} \delta)|}{|\mathcal{P}^X_N|} \right) \! (1{-}\delta) N.
\end{eqnarray}
As $N\rightarrow\infty$, Equation (\ref{eq.pattthmX}) says that the ratio 
$|\mathcal{P}^X_N({\leq} \delta)|/|\mathcal{P}^X_N|$ decays to 0 exponentially rapidly.  Therefore 
we conclude from Equation (\ref{eq.ENglessX}) that
\begin{equation}
   \label{eq.limsupallb}
    \limsup_{N\rightarrow\infty}
      \frac{ E_{\beta}^Q\left(\sigma(\rho)\,|\,\mathcal{P}^X_N\right) }{N} \;\leq \; 1-\delta
    \hspace{5mm}  \text{for every real $\beta$}.   
\end{equation}
The left hand side of Equation (\ref{eq.limsupallb}) is increasing in $\beta$ by  
Equation (\ref{eq.expvisitsquench}) and log-convexity of $F_N^Q$.  
Therefore the limit as $\beta\rightarrow\infty$ exists, and
Equation (\ref{eq.ElimquenchATR}) for trees and animals follows.

The argument for quenched knots in self-avoiding polygons is very similar, 
except we define $\mathcal{R}^X_N(\leq \delta)$ to be the set of 
rings in $\mathcal{R}^X_N$ whose knot type has crossing number at most $\delta N$. 
The crossing number of a knot is the fewest number of crossings in any planar drawing of the knot.
Therefore if $\rho\in \mathcal{R}^X_N\setminus \mathcal{R}^X_N(\leq \delta)$, we see
that $\sigma(\rho)< N-\delta N$.  
Corollary 3.2 and Theorem 3.3 of \cite{SSW} imply that the anaologue of Equation 
(\ref{eq.pattthmX}) holds here.  Equation  (\ref{eq.ElimquenchATR}) for rings follows as above.
\hfill  $\Box$

\medskip

Next we shall prove Proposition \ref{prop.jensen}, which says that for every model $\mathcal{P}^X$
and every real $\beta$,
\[    F_N(\beta\,|\,\mathcal{P}^X)\,-   F_N(0\,|\,\mathcal{P}^X)   \;\geq \;  
     F^Q_N(\beta\,|\,\mathcal{P}^X)\,-   F^Q_N(0\,|\,\mathcal{P}^X) \,.
\]
\textbf{Proof of Proposition \ref{prop.jensen}}:
We write the generic topology (graph or knot) for the model $\mathcal{P}^X$ as $\tau$.
Since
\begin{eqnarray*}
    Z_N(\beta\,|\,\mathcal{P}^X)  & = &   \sum_{\rho\in \mathcal{P}_N^X}  e^{\beta\sigma(\rho)}
   \\
    & = & \sum_{\tau}  \sum_{\rho\in \mathcal{P}_N^X[\tau]} e^{\beta\sigma(\rho)}  
   \\
     & = & \sum_{\tau}  Z_N(\beta \,|\, \mathcal{P}^X[\tau])\,,
\end{eqnarray*}
we have
\begin{eqnarray}
  \nonumber
    \frac{Z_N(\beta\,|\,\mathcal{P}^X)}{Z_N(0\,|\,\mathcal{P}^X)}  & = &   \frac{1}{|\mathcal{P}_N^X|}
      \sum_{\tau}  Z_N(\beta \,|\, \mathcal{P}^X[\tau])
      \\
      \nonumber
      & = & 
      \sum_{\tau}       \frac{|\mathcal{P}^X_N[\tau]|}{|\mathcal{P}_N^X|}
       \left( \frac{Z_N(\beta \,|\, \mathcal{P}^X[\tau])}{Z_N(0\,|\,\mathcal{P}^X[\tau])} \right) 
      \\
      \label{eq.EKZgen}
      & = & E_{\tau}\left( \frac{Z_N(\beta \,|\, \mathcal{P}^X[\tau])}{Z_N(0\,|\,\mathcal{P}^X[\tau])} \right) 
\end{eqnarray}
where $E_{\tau}$ denotes expectation with respect to the probability distribution that assigns 
the probability $|\mathcal{P}^X_N[\tau]|/|\mathcal{P}_N^X|$ to the topology $\tau$.
Then for every real $\beta$ we have
\begin{eqnarray}
   \nonumber
   \lefteqn{F_N(\beta\,|\, \mathcal{P}^X)\,-\,F_N(0 \,|\, \mathcal{P}^X) )  }
   \\
   \nonumber
   & = & \frac{1}{N}\log Z_N(\beta\,|\, \mathcal{P}^X) \,-\,\frac{1}{N}\log Z_N(0\,|\, \mathcal{P}^X)
       \\
       \nonumber
       & = & \frac{1}{N}  \log \left(   \frac{Z_N(\beta\,|\, \mathcal{P}^X)}{Z_N(0\,|\, \mathcal{P}^X)} \right)  
       \\
       \nonumber
       & = & \frac{1}{N}  \log \left(  E_{\tau}\left( \frac{Z_N(\beta \,|\,  \mathcal{P}^X[\tau]  )}{
           Z_N(0\,|\, \mathcal{P}^X[\tau]  )} \right)   \right)
          \hspace{11mm}\text{(by Equation (\ref{eq.EKZgen}))}
       \\
       \nonumber
       & \geq &  \frac{1}{N}  E_{\tau} \left(  \log\left( \frac{Z_N(\beta \,|\, \mathcal{P}^X[\tau])}{
           Z_N(0\,|\,\mathcal{P}^X[\tau])} \right)   \right)       
       \\
       \nonumber
       & & \hspace{8mm}\text{(by Jensen's Inequality, since log is concave)}
       \\
       \nonumber
       & = & \frac{1}{N} E_{\tau}\left(\log Z_N(\beta\,|\,\mathcal{P}^X[\tau])   \,-\,
            \log Z_N(0\,|\,\mathcal{P}^X[\tau]) \right) 
              \\
       \label{eq.jensenknot}
       & = & F_N^Q(\beta \,|\,\mathcal{P}^X)\,-\,F_N^Q(0\,|\,\mathcal{P}^X) 
            \hspace{8mm}\hbox{(by Equation (\ref{eq.FQbranchdef}))}\,.
\end{eqnarray}       
This completes the proof.
\hfill  $\Box$       

\medskip
Next we shall prove Proposition \ref{prop.negbetaQ}, which says that
the limit $\mathcal{F}^Q_{\Delta}(\beta\,|\,\mathcal{P}^X)$ of Equation (\ref{eq.FminusFgen})
exists and equals 0  for every $\beta\leq 0$ for each of our models $\mathcal{P}^X$.

\medskip
\noindent
\textbf{Proof of Proposition \ref{prop.negbetaQ}:}
We first give the proof for a penetrable surface.  We  present the proof for animals, although 
 it clearly works for rings, trees, and combs, with obvious modifications.
  
For $\rho\in \mathcal{A}_N$ and $j\in \{1,\ldots,d\}$, define
\begin{eqnarray*}
     \max{}_j(\rho)  & = & \max\{ x_j\,: (x_1,\ldots,x_d) \text{ is a site of }\rho \}
     \\
        \min{}_j(\rho)  & = & \min\{ x_j\,: (x_1,\ldots,x_d) \text{ is a site of }\rho \}
     \\
     \text{span}_j(\rho) & = & 1\,+\, \max{}_j(\rho) \,-\,\min{}_j(\rho).
\end{eqnarray*}
Thus the $N$-site animal $\rho$ is contained in the box 
\[    \{x\in \mathbb{R}^d: \min{}_j(\rho)\leq x_j\leq \max{}_j(\rho), \,j=1,\dots,d\}.
\]    
Since this box contains
$\prod_{j=1}^d\text{span}_j(\rho)$ lattice sites, there must exist a $J\in \{1,\ldots,d\}$ such that 
span$_J(\rho) \,\geq \,N^{1/d}$.
Therefore there exists an integer $u\in [\min{}_J(\rho),\max{}_J(\rho)]$ such that the number of 
sites of $\rho$ that lie in the hyperplane $x_J=u$ is at most $N/\text{span}_J(\rho)$, which 
in turn is at most $N^{(d-1)/d}$.

Fix $\beta\leq 0$.
For each graph $g\in \mathcal{G}_N$, let $\mathcal{A}^{\pen}_N[g; \langle j\rangle]$ be the 
set of animals in $\mathcal{A}^{\pen}_N[g]$ that have at most $N^{(d-1)/d}$ sites in the hyperplane 
$\{x\,:x_j=0\}$.  By the conclusion of the preceding paragraph, every animal in
$\mathcal{A}^{\pen}_N[g]$ is a translate of 
some member of $\mathcal{A}^{\pen}_N[g; \langle j\rangle]$ for some $j$.  Therefore
\begin{equation}
   \label{eq.Ascriptj}
    |\mathcal{A}^{\pen}_N[g]| \;\; \leq \; \;
    N \,\left|  \bigcup_{j=1}^d \mathcal{A}^{\pen}_N[g; \langle j\rangle ] \right|
      \; \; \leq \;\; dN \,\left|  \mathcal{A}^{\pen}_N[g; \langle 1\rangle ] \right| \,    
\end{equation}    
the second inequality following from symmetry.
Hence, since $\beta\leq 0$, we have
\begin{equation}
     \label{eq.ZgAlowerbound}
   Z_N(\beta\,|\,\mathcal{A}^{\pen}[g])  \;\geq \; 
    \left|  \mathcal{A}^{\pen}_N[g; \langle 1\rangle ] \right| \, e^{\beta N^{(d-1)/d}}
    \; \geq \;  \frac{1}{dN}\, \left|  \mathcal{A}^{\pen}_N[g ] \right| \, e^{\beta N^{(d-1)/d}}.
\end{equation}
Therefore
\begin{eqnarray}
  \nonumber
  F_N^Q(\beta|\mathcal{A}^{\pen}) & =  & \sum_{g\in\mathcal{G}_N}  
     \frac{|\mathcal{A}^{\pen}_N[g]|}{|\mathcal{A}^{\pen}_N|}   \,\frac{1}{N} \,\log Z_N(\beta\,|\, 
       \mathcal{A}^{\pen}[g])
       \\
       \nonumber
       & \geq &  \frac{1}{N}  \sum_{g\in\mathcal{G}_N}  
        \frac{|\mathcal{A}^{\pen}_N[g]|}{|\mathcal{A}^{\pen}_N|}   
          \left(  \log |\mathcal{A}^{\pen}_N[g]| \,-\,\log (dN)  \,+\beta N^{(d-1)/d}   \right)  
          \\
          \label{eq.FQAbetaneg}
          & = & F_N^Q(0\,|\,\mathcal{A}^{\pen})   \,-\,\frac{\log (dN)}{N}  \,+\,\beta N^{-1/d} \,.
\end{eqnarray}
Finally, since $F_N^Q(0|\mathcal{A}^{\pen})\geq F_N^Q(\beta |\mathcal{A}^{\pen})$, 
we conclude from Equation (\ref{eq.FQAbetaneg}) that $\lim_{N\rightarrow\infty}
F_N^Q(\beta|\mathcal{A}^{\pen})- F_N^Q(0 |\mathcal{A}^{\pen}) \,=\,0$.  This proves the result
for $\mathcal{A}^{\pen}$.

\smallskip
Now consider the case of an impenetrable surface.  
In this part of the proof, we shall write 
$\mathcal{P}$ to represent one of $\mathcal{A}$, $\mathcal{T}$, $\mathcal{R}$, or $\mathcal{C}$.
By Theorem \ref{thm.betacanngt0},
the critical point $\beta_c(\mathcal{P}^+)$ of the corresponding impenetrable 
\textit{annealed} model is strictly positive.  
In particular there exists a $\beta_+>0$ such that 
\begin{equation}
    \label{eq.FNbeta+}
   \lim_{N\rightarrow\infty}F_N(\beta_+\,|\,\mathcal{P}^+)-F_N(0\,|\,\mathcal{P}^+) \,=\,0 \,.
\end{equation}
Since $F_N^Q(\cdot|\mathcal{P}^+)$ is an increasing convex function, we see that for every $\beta<0$ 
\begin{eqnarray}
     0 & \leq & \frac{F_N^Q(0\,|\,\mathcal{P}^+)-F_N^Q(\beta\,|\,\mathcal{P}^+)}{0-\beta}
         \label{eq.FQsecant}
         \\
         & \leq & \frac{F_N^Q(\beta_+\,|\,\mathcal{P}^+)-F_N^Q(0\,|\,\mathcal{P}^+)}{\beta_+-0}
         \nonumber
         \\
         & \leq & \frac{F_N(\beta_+\,|\,\mathcal{P}^+)-F_N(0\,|\,\mathcal{P}^+)}{\beta_+}
           \hspace{15mm}\hbox{(by Prop.\ \ref{prop.jensen})}\,.
        \label{eq.FNtoFNQ}
\end{eqnarray}
The desired result now follows from Equations (\ref{eq.FNbeta+})--(\ref{eq.FNtoFNQ}).
\hfill  $\Box$

\begin{rem}
    \label{rem.endsec3}
{\em     The above proof for the impenetrable case cannot be used for the 
penetrable case, where it is believed that adsorption occurs for every strictly positive $\beta$
(see for example \cite{Ma}).}
\end{rem}

\section{Combs:  Existence of limits}
    \label{sec.comblimits}
    
This section is devoted to the rather lengthy proofs of the existence of two related limits:  the 
growth constant $\lambda_{d,C}=\lim_{N\rightarrow\infty}c_N^{1/N}$ and the 
annealed free energy  
$\mathcal{F}(\beta \,|\,\mathcal{C}^X)  \,=\, \lim_{N\rightarrow\infty}F_N(\beta \,| \,\mathcal{C}^X)$,
where $X$ is either $+$ or $\pen$.
Of course, the existence of $\lambda_{d,C}$ is equivalent to the existence of the annealed 
free energy at $\beta=0$, and so the proof of the former (in Section \ref{sec.comblambda})
result could be omitted and obtained as a corollary of the latter.  
Nevertheless, we have included both proofs, because the proof of the 
limiting free energy is complicated by the loss of full translation invariance due to the special
role of the surface, and it is best presented as an adaptation of the proof of the growth constant.
Moreover, the lengthy proof of the free energy's existence is mainly needed when $\beta$ is 
nonnegative (see Section \ref{sec.combbetapos}); 
the existence of $\mathcal{F}(\beta \,|\,\mathcal{C}^X)$ for negative $\beta$ is
a fairly direct consequence of the existence of $\lambda_{d,C}$ (see Section \ref{sec.combbetaneg}).

\smallskip
The methods of this section are not used in the rest of the paper, so the reader can easily read later 
sections before this one.

\subsection{Existence of the growth constant for combs}
    \label{sec.comblambda}

In this subsection, we prove the existence of the limit of Equation (\ref{eq.comblim})
which defines  the growth constant $\lambda_{d,C}$ for combs.

Previous published results \cite{LipW} imply that if we only consider $N$-site combs
with $o(N/\log N)$ branch points, then the growth constant would equal that of the SAW, 
namely $\mu_d$.  
We shall prove in Section \ref{sec.combprops} that $\mu_d<\lambda_{d,C}$ (Proposition 
\ref{prop.combneqsaw}),
and that in fact for some $\delta>0$, the fraction of $N$-site combs with fewer than 
$\delta N$ branch points is exponentially small  (Lemma \ref{lem.manychains}).

\medskip
\begin{prop}
  \label{prop.combgrowth}
For each dimension $d\geq 2$, the limit $\lambda_{d,C} \,=\,\lim_{N\rightarrow\infty}  c_N^{1/N}$ exists.
\end{prop}

\medskip
The following definitions will be used in the proof of Proposition \ref{prop.combgrowth}.

\begin{defi}
    \label{def.comb2}
(a) For a bounded subset $A$ of $\mathbb{Z}^d$, let Top$(A)$ be the set
\[    \hbox{Top}(A) \;:= \;  \{ v\in A  \,|\, v_d\geq x_d \hbox{ for every $x$ in $A$} \}\,.
\]
If $G$ is a finite subgraph of $\mathbb{L}^d$ with vertex set $V(G)$, then Top$(G)$ is defined
to be Top$(V(G))$.  
\\
(b)  Let $\mathcal{C}^*_N$ be the set of combs $\rho$ in $\overline{\mathcal{C}}_N$ such that 
Top$(\rho)$ contains at least one site  in $\rho$ that has degree 1 or at least one site on the backbone 
of $\rho$ that has degree $2$. 
\end{defi}
The idea behind the condition in (b) is that there is a site in Top$(\rho)$ where we can attach a 
SAW to $\rho$ to make a bigger comb.

Consistently with the definitions of Section \ref{sec-latpoly}, we shall view $w_N$ as the 
number of $N$-step SAWs that start at the origin.

\begin{defi}
    \label{def.saws}
Fix $d\geq 2$.   Let $N\geq 0$.   
\\
(a)  A \textit{half-space SAW} is a SAW $(\omega(0),\ldots,\omega(N))$
such that $\omega_d(i)>\omega_d(0)$ for every $i=1,\ldots,N$.  Let $h_N$ be the number 
of $N$-step half-space SAWs that start at the origin. 
\\
(b)  A \textit{bridge} is a SAW $(\omega(0),\ldots,\omega(N))$
such that $\omega_d(N)\geq \omega_d(i)>\omega_d(0)$ for every $i=1,\ldots,N$.  
Let $b_N$ be the number of $N$-step bridges that start at the origin.  
\\
We count the 0-step walk as a half-space SAW and a bridge, so $h_0=b_0=1$.
\\
(c) If $\omega=(\omega(0),\ldots,\omega(N))$ and $\psi=(\psi(0),\ldots,\psi(M))$ are SAWs, 
then the concatenation of $\omega$ and $\psi$ is the $(N+M)$-step walk (not necessarily 
self-avoiding) denoted by $\omega\circ\psi$ obtained by combining $\omega$ with the translation of 
$\psi$ that starts at the last site of $\omega$.  
The formula for the sites of $\omega\circ\psi$ is 
\[    (\omega\circ\psi)(i) \;=\;  \begin{cases}   \omega(i)   & \hbox{if } 0\leq i \leq N,   \\
       \omega(N)+\psi(i-N)   -\psi(0) & \hbox{if }N\leq i\leq N+M.
       \end{cases}
\]
\end{defi}

We shall use the following generalized submultiplicative inequality.

\begin{lem}
    \label{lem.submult}
Let $\{a_n:  \,n\geq 1\}$ be a positive sequence and let $g: \mathbb{N}\rightarrow (0,\infty)$ be 
a function such that $\lim_{n\rightarrow\infty}g(n)^{1/n}=1$. 
Assume that 
\begin{equation}
   \label{eq.combsubmultAA}
   a_{M+N}  \;\leq \;  g(M)\,a_M  \,a_N    \hspace{5mm}\hbox{for all }M,N\geq 1 \,.
\end{equation}
Then $\lim_{n\rightarrow \infty}a_n^{1/n}$ exists and
\begin{equation}
    \label{eq.submultlim}
    \lim_{n\rightarrow \infty}a_n^{1/n}  \;=\;  \inf_{n\geq 1}[g(n)\,a_n]^{1/n}  \,.
\end{equation}
\end{lem}

As noted by others (e.g.\ Appendix II of \cite{Gr}), the proof of Lemma \ref{lem.submult} is
similar to the standard case (where $g$ is identically 1).  
For completeness, here is the argument.

\bigskip
\noindent
\textbf{Proof of Lemma \ref{lem.submult}}:
Let $V=\inf_{n\geq 1}[g(n)a_n]^{1/n}$.  
Fix $M\in\mathbb{N}$ for now.  
For each natural number $n$, let $j=j(n)$ and $r=r(n)$ be the integers such that
$1\leq  r(n)\leq M$ and $n=j(n)M+r(n)$.  Then repeated application of (\ref{eq.combsubmultAA}) gives
\[    a_n  \;\leq \;  [g(M)\,a_M]^j \,a_r \,.
\]
Let $\hat{a}_M \,=\, \max\{a_1,\ldots,a_{M}\}$.  We obtain
\[   [a_n]^{1/n}   \;\leq \;   [g(M)\,a_M]^{j/(jM+r)}\, [\hat{a}_M]^{1/(jM+r)} \,.
\]  
Let $n\rightarrow\infty$.  Then $j(n)\rightarrow\infty$ and $r(n)$ stays bounded.   We deduce
that 
\[    \limsup_{n\rightarrow\infty} [a_n]^{1/n}   \;\leq \;  [g(M)\,a_M]^{1/M} \,.
\]
Since this holds for every $M$, we see that $\limsup_{n\rightarrow\infty} [a_n]^{1/n}\,\leq \,V$.
Since $\lim_{n\rightarrow\infty}[g(n)]^{1/n}=1$, we obtain 
$\limsup_{n\rightarrow\infty} [g(n)a_n]^{1/n}\,\leq \,V$.
By definition of $V$, it follows that $\lim_{n\rightarrow\infty} [g(n)a_n]^{1/n}$ exists and equals $V$.
Therefore  $\lim_{n\rightarrow\infty} [a_n]^{1/n}\; = \;  V$.   
\hfill   $\Box$

\bigskip
Let $\rho$ be a tree and let $v$ be a vertex of $\rho$ of  degree $k$. 
Let $e_1,e_2,\ldots,e_k$ be the 
edges of $\rho$ that have $v$ as an endpoint.   For each $i$, let $T_i$ be the component of
$\rho$ containing $v$ after all of $\{e_j:j\neq i\}$ are deleted.  Then $T_1,T_2,\ldots,T_k$ are
called the subtrees obtained by breaking $\rho$ at $v$.  Observe that these subtrees partition
the set of all edges of $\rho$.
We also refer to $T_1,\ldots,T_k$ as the subtrees of $\rho$ ending at $v$.

\medskip
\noindent
\textbf{Proof of Proposition \ref{prop.combgrowth}:}  
The proof requires several inequalities, which we shall state and justify in turn.

\smallskip
(\textit{i}) 
Using the notation of Definition  \ref{def.saws}(a,b),  the first inequality is
\begin{equation}
   \label{eq.comb-bhh}
   b_{\ell}h_k \;\leq \;  h_{\ell+k}  \hspace{5mm}  \hbox{for all }\ell,k\geq 0.
\end{equation}
This holds because if $\omega$ is an ${\ell}$-step  bridge and $\psi$ is a $k$-step half-space SAW, 
then $\omega\circ\psi$ is an $(\ell+k)$-step half-space SAW, and the resulting
walk uniquely determines 
$\omega$ and $\psi$ (since there is a unique integer $D$ such that exactly $k$ sites of 
$\omega\circ\psi$ have $d^{th}$ coordinate is greater than $D$).

\smallskip
(\textit{ii})
The second inequality is that there exists a constant $\alpha$ (depending only on the dimension $d$)
such that
\begin{equation}
   \label{eq.HWineq}
      w_k   \;\leq \;  e^{\alpha \sqrt{k}}\,b_k    \hspace{5mm}\hbox{ for all }k\geq 1.
\end{equation}
This is due to  Hammersley and Welsh \cite{HW}; alternatively, see 
Theorem 3.1.1 and Corollary 3.1.6 in \cite{MS}.

\smallskip
(\textit{iii})  The third inequality is
\begin{equation}
   \label{eq.comb-chc}
   |\mathcal{C}^*_{\ell}| \,h_k \;\leq \;  c_{\ell+k}  \hspace{5mm}  \hbox{for all }\ell,k\geq 0.
\end{equation}
To verify this, let $\rho\in \mathcal{C}^*_{\ell}$. Let $v\in\hbox{Top}(\rho)$ be such that either
$v$ has degree 1 in $\rho$ or else $v$ is on the backbone of $\rho$ and has degree 2.  
Let $\psi$ be any 
$k$-step half-space SAW that starts at $v$.  Then the union of $\rho$ and $\psi$ is an $(\ell+k)$-step
comb.  (If $v$ is a labelled site of $\rho$, then the label moves to $\psi(k)$ in the union.)
As in the previous paragraph, the union uniquely determines $\psi$ and $\rho$ (for given 
$\ell$ and $k$).

\smallskip
(\textit{iv})
The fourth inequality says that
\begin{equation}
   \label{eq.comb-cstarhc}
   c_n   \;\leq \;  \sum_{j=0}^{n-1} (n-j)\, |\mathcal{C}^*_{n-j}|\,h_j  \hspace{5mm}  \hbox{for all }n\geq 1.
\end{equation}
The number of combs in $\mathcal{C}_n^*$ is bounded by the $j=0$ term in this sum.
Let $\rho\in \mathcal{C}_n\setminus \mathcal{C}^*_n$. Let $v\in\hbox{Top}(\rho)$.
Since $\rho\not\in \mathcal{C}^*_n$, one of the following two cases must hold:
\begin{verse}
(a) $v$ is on a side chain of $\rho$ and has degree 2, or
\\ 
(b) $v$ is on the backbone of $\rho$ and has degree 3.
\end{verse}
In both cases, $v$ belongs to a unique side chain of $\rho$.  
(In case (b), $v$ is the attached end of the side chain.)
Let $u$ be the free end of this side chain.
Note that $u$ is not in Top$(\rho)$ since $\rho \not\in \mathcal{C}^*_n$ and $u$ has degree 1. 
As one traverses the side chain from $v$ to $u$, let $z$ be the last site that is in Top$(\rho)$.  (It
is possible that $z$ equals $v$.)  
Let $\theta$ be the part of the side chain that starts at $z$ and ends at $u$.   
Let $j$ be the number of steps in the SAW $\theta$.
Then $0<j<n$.  Also, since $\theta_d(i)<v_d=\theta_d(0)$ for $i=1,\ldots,j$, we see that
$\theta$ is a half-space SAW that has been reflected through the hyperplane $\{x:x_d=v_d\}$
(this is the hyperplane whose intersection with $\rho$ defines Top$(\rho)$).
Let $K$ be the subtree of $\rho$ obtained by deleting all of $\theta$ except for $z$.  Observe 
that $K$ is in $\mathcal{C}^*_{n-j}$  (if we are in case (b) with $z=v$, then $z$ is on the backbone
of $K$ and has degree 2; otherwise $z$ has degree 1).   
If (up to translation) we know $K$ and $\theta$
and the site of $K$ that corresponds to $z$, then $\rho$ is uniquely determined.  
Since $K$ has no more than $n-j$ possibilities for $z$, the inequality
(\ref{eq.comb-cstarhc}) follows.

\smallskip
(\textit{v})   
Next, we claim
\begin{equation}
   \label{eq.comb-ccwc}
   c_{M+N} \;\leq \; (2M+2)\,c_N \sum_{i=0}^{M-1}  c_{M-i} \,w_i 
      \hspace{9mm}  \hbox{for all }M,N\geq 1\,.
\end{equation}
This is the crucial inequality.  It is the analogue of the relation $w_{M+N}\leq w_Mw_N$ for SAWs, 
which holds because you can always break an $(N+M)$-step SAW into an $M$-step SAW and an
$N$-step SAW.  The exact analogue is false for trees, and it is false for combs, but it does not fail as
badly for combs as it does for general trees.  
We shall show that any $(M+N)$-step comb can be broken into 
two combs and a SAW, where one comb has exactly $N$ steps.  (The SAW could have no steps.)
More precisely, we shall construct a mapping
\begin{eqnarray}
    \nonumber
    \overline{\mathcal{C}}_{M+N}  & \longrightarrow & \overline{\mathcal{C}}_N \times
         \bigcup_{i=0}^{M-1}  \left(    \overline{\mathcal{C}}_{M-i}\times  \overline{\mathcal{W}}_i\right)
         \\
         \label{eq.combmap}
        & = & \left\{  \left( \overline{\kappa}^{[1]},\overline{\kappa}^{[2]},\overline{\theta}\right)\,:  \,
        \overline{\kappa}^{[1]} \in \overline{\mathcal{C}}_N ,\,
        \overline{\kappa}^{[2]} \in \overline{\mathcal{C}}_{M-i} ,\,
        \overline{\theta} \in   \overline{\mathcal{W}}_i   \right.
        \\
        \nonumber 
        & & \hspace{44mm} \left.   \hbox{  for some }i\in [0,M-1]  \right\}  \,.   
\end{eqnarray}
Our construction will actually produce $\kappa^{[1]}$, $\kappa^{[2]}$, and $\theta$; 
each one must be translated appropriately to get $\overline{\kappa}^{[1]}$, $\overline{\kappa}^{[2]}$,
and $\overline{\theta}$.  Finally, we shall show that this mapping is at most $(2M+2)$-to-one.

Let $\rho\in \overline{\mathcal{C}}_{M+N}$.  
Let the backbone of $\rho$ be the SAW $(\tilde{\rho}(0),\tilde{\rho}(1),\ldots,\tilde{\rho}(t))$, where
$\tilde{\rho}(0)$ and $\tilde{\rho}(t)$ are the labelled sites $\rho_A$ and $\rho_B$ respectively.  
Let $b$ be the number of sites of degree 3 in $\rho$ (equivalently, the number of side chains).
Let $0<i_1<i_2<\ldots < i_b<t$ be such that $\{\tilde{\rho}(i_j): j=1,\dots,b\}$ are the sites of degree 3. 
Let $i_0=0$.
For each $j=1,\dots,b$, let $s_j$ be the number of steps in the side chain that starts at $\tilde{\rho}(i_j)$.
Observe that 
\begin{equation}
   \label{eq.MNcomb}
    M+N  \;=\;  t \,+\,\sum_{j=1}^bs_j \,.
\end{equation}
Exactly one of the following cases must occur for $\rho$ (see Figure \ref{fig.3cases}):
\begin{eqnarray*}
  \hbox{I.} & & i_b+\sum_{j=1}^b s_j \; < \; N \,,
  \\
  \hbox{II.}  & & i_J+\sum_{j=1}^{J-1}s_j  \;<\; N \;\leq i_J+\sum_{j=1}^Js_j   \hspace{4mm}
     \hbox{for some }J\in [1,b]\,,  \hbox{ or}
   \\
  \hbox{III.}  & & i_{J-1}+\sum_{j=1}^{J-1}s_j  \;<\; N \;\leq i_J+\sum_{j=1}^{J-1}s_j   \hspace{4mm}
     \hbox{for some }J\in [1,b] \,.   
\end{eqnarray*}

\setlength{\unitlength}{1mm}
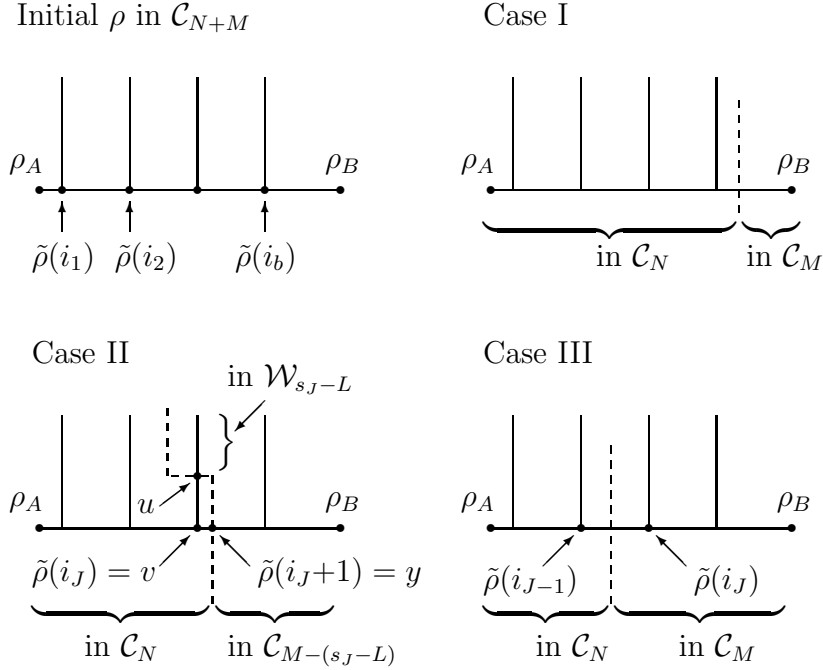
\begin{figure}[h]
\begin{center}
\begin{picture}(100,90)
\put(60,50){
\begin{picture}(40,40)
\put(0,32){Case I}
\put(1,10){\line(1,0){40}}
\put(1,10){\circle*{1.0}}
\put(41,10){\circle*{1.0}}
\put(-3,13){$\rho_A$}
\put(39,13){$\rho_B$}
\multiput(4,10)(9,0){4}{\line(0,1){15}}
\multiput(34,7)(0,2){8}{\line(0,1){1}}
\put(0,6){$\underbrace{\hbox{\hspace{33.5mm}}}$}
\put(15,0){in $\mathcal{C}_N$}
\put(34.5,6){$\underbrace{\hbox{\hspace{5.5mm}}}$}
\put(35,0){in $\mathcal{C}_M$}
\end{picture}
}
\put(0,5){
\begin{picture}(40,40)
\put(0,32){Case II}
\put(1,10){\line(1,0){40}}
\put(1,10){\circle*{1.0}}
\put(41,10){\circle*{1.0}}
\put(-3,13){$\rho_A$}
\put(39,13){$\rho_B$}
\multiput(4,10)(9,0){4}{\line(0,1){15}}
\multiput(24,0)(0,2){9}{\line(0,1){1}}
\multiput(18,17)(0,2){5}{\line(0,1){1}}
\multiput(18.5,17)(2,0){3}{\line(1,0){1}}
\multiput(22,10)(2,0){2}{\circle*{1}}
\put(0,3){$\tilde{\rho}(i_J)=v$}   
\put(17,5){\vector(1,1){4}}
\put(30,3){$\tilde{\rho}(i_J{+}1)=y$}
\put(29,5){\vector(-1,1){4}}
\put(26,29){in $\mathcal{W}_{s_J-L}$}  
\put(31,27){\vector(-1,-1){4}}
\put(14,12){$u$}
\put(17,13){\vector(4,3){4}}
\put(22,17){\circle*{1}}
\put(24,20.5){$\left.  \rule{0mm}{4mm}  \right\}$}
\put(0,0){$\underbrace{\hbox{\hspace{23.5mm}}}$}
\put(7,-7){in $\mathcal{C}_N$}
\put(24.5,0){$\underbrace{\hbox{\hspace{15.5mm}}}$}
\put(26,-7){in $\mathcal{C}_{M-(s_J-L)}$}
\end{picture}
}
\put(60,5){
\begin{picture}(40,40)
\put(0,32){Case III}
\put(1,10){\line(1,0){40}}
\put(1,10){\circle*{1.0}}
\put(41,10){\circle*{1.0}}
\put(-3,13){$\rho_A$}
\put(39,13){$\rho_B$}
\multiput(4,10)(9,0){4}{\line(0,1){15}}
\multiput(17,0)(0,2){11}{\line(0,1){1}}
\put(0,0){$\underbrace{\hbox{\hspace{16.5mm}}}$}
\put(7,-7){in $\mathcal{C}_N$}
\put(17.5,0){$\underbrace{\hbox{\hspace{22.5mm}}}$}
\put(26,-7){in $\mathcal{C}_M$}
\put(13,10){\circle*{1}}
\put(22,10){\circle*{1}}
\put(0,2){$\tilde{\rho}(i_{J-1})$}   
\put(8,5){\vector(1,1){4}}
\put(28,2){$\tilde{\rho}(i_J)$}
\put(27,5){\vector(-1,1){4}}
\end{picture}
}
\put(0,50){
\begin{picture}(40,40)
\put(-2,32){Initial $\rho$ in $\mathcal{C}_{N+M}$}
\put(1,10){\line(1,0){40}}
\put(1,10){\circle*{1.0}}
\put(41,10){\circle*{1.0}}
\put(-3,13){$\rho_A$}
\put(39,13){$\rho_B$}
\multiput(4,10)(9,0){4}{\line(0,1){15}}
\multiput(4,10)(9,0){4}{\circle*{1.0}}
\put(0,0){$\tilde{\rho}(i_1)$}
\put(4,4.5){\vector(0,1){4}}
\put(11,0){$\tilde{\rho}(i_2)$}
\put(13,4.5){\vector(0,1){4}}
\put(27,0){$\tilde{\rho}(i_b)$}
\put(31,4.5){\vector(0,1){4}}
\end{picture}
}
\end{picture}
\end{center}
\caption{\label{fig.3cases}   Cases I, II, and III in the proof of Theorem \ref{prop.combgrowth}.
The initial comb $\rho$ with $b$ side chains is shown in the upper left diagram. (Here, $b=4$.)}
\end{figure}

In case I, we know that $t-M>i_b$  (by Equation (\ref{eq.MNcomb})), so $\tilde{\rho}(t-M)$ has 
degree 2 in $\rho$.
Let $\kappa^{[1]}$ (respectively, $\kappa^{[2]}$) be the subtree of $\rho$ ending at $\tilde{\rho}(t-M)$
that contains $\rho_A$ (respectively, $\rho_B$).  
Observe that  $\kappa^{[2]}$ contains no site of degree 3; it is simply
the part of the backbone from $\tilde{\rho}(t-M)$ to $\tilde{\rho}(t)$.  
By labelling $\tilde{\rho}(t-M)$ as $\kappa^{[2]}_A$ and $\tilde{\rho}(t)$ as $\kappa^{[2]}_B$,  
we view $\kappa^{[2]}$ as an $M$-step comb.
And $\kappa^{[1]}$ is an $N$-step comb (labelled $\kappa^{[1]}_A$ at the original $\rho_A$ and 
$\kappa_B^{[1]}$ at $\tilde{\rho}(t-M)$).   Lastly, let $\theta$ be a 0-step SAW.

In case II, let $L=N-i_J-\sum_{j=1}^{J-1}s_j-1$.   Observe that $0\leq L< s_J$.
Let $v=\tilde{\rho}(i_J)$, let $y=\tilde{\rho}(i_J+1)$, and let $u$ be the site
of the side chain at $v$ that is $L$ steps from $v$.  
Let $\theta$ be the $(s_J-L)$-step SAW that forms the part of the side chain starting at $u$ and
ending at the free end of the side chain.  
Let $\kappa^*$ be the subtree of $\rho$ ending at $y$ that contains $v$.  Let $\kappa^{[1]}$ be
the comb formed by removing all of $\theta$ except $u$ from $\kappa^*$, with labelled sites
at $\kappa^{[1]}_A=\rho_A$ and $\kappa^{[1]}_B=y$.  
Let  $\kappa^{[2]}$ be the comb formed by removing all of $\kappa^*$ except $y$ from $\rho$.  
In $\kappa^{[2]}$, the site $\rho_B$
gets labelled $\kappa^{[2]}_B$.  If $y$ had degree 2 in $\rho$, then $y$ becomes the site labelled 
$\kappa^{[2]}_A$, but if $y$ had degree 3 in $\rho$ then the free end of the side chain of $\rho$ at
$y$ gets labelled $\kappa^{[2]}_A$.  Observe that $\kappa^{[1]}$ consists of the first $i_J+1$ steps of
the backbone of $\rho$, the first $J-1$ side chains of $\rho$, and the first $L$ steps of the $J^{th}$
side chain.  Therefore the number of edges in  $\kappa^{[1]}$
is $i_J+1+\sum_{j=1}^{J-1}s_j+L$, which equals $N$.  Hence $\kappa^{[2]}$ has $M-(s_J-L)$ edges.

In case III, let $I=N-\sum_{j=1}^{J-1}s_j$.  Then $i_{J-1}<I\leq i_J$.  
If $I=i_J$, then $\tilde{\rho}(I)$ has degree 3, and the three subtrees of $\rho$ ending at 
$\tilde{\rho}(I)$ are an $N$-step comb ($\kappa^{[1]}$),
an $s_J$-step SAW $\theta$ (the side chain, starting at the attached end), 
and an $(M-s_J)$-step comb $\kappa^{[2]}$.  
Otherwise, $i_{J-1}<I<i_J$
and $\tilde{\rho}(I)$ has degree 2, so we get two subtrees ending at $\tilde{\rho}(I)$:  
an $N$-step comb $\kappa^{[1]}$ and an $M$-step comb $\kappa^{[2]}$; in this second situation 
we also let $\theta$ be a 0-step SAW.
In both situations, the break site $\tilde{\rho}(I)$ becomes the new labelled site in each smaller comb
($\kappa^{[1]}_B$ and $\kappa^{[2]}_A$).
And $\rho$ is  uniquely determined by the resulting pieces (up to translation).

In each of Cases I, II, and III, we end up with three subgraphs of $\rho$:  
an $N$-step comb $\kappa^{[1]}$, an $(M-i)$-step comb $\kappa^{[2]}$, and 
an $i$-step SAW $\theta$, for some $i\in [0,M-1]$.  These determine a unique 
$\left( \overline{\kappa}^{[1]},\overline{\kappa}^{[2]},\overline{\theta}\right)$
in $\overline{\mathcal{C}}_N \times
\bigcup_{i=0}^{M-1}  \left(    \overline{\mathcal{C}}_{M-i}\times  \overline{\mathcal{W}}_i\right)$
(recall Equation (\ref{eq.combmap})) such that 
$\overline{\kappa}^{[1]}$ is a translate of $\kappa^{[1]}$, $\overline{\kappa}^{[2]}$ is a translate
of $\kappa^{[2]}$, and $\overline{\theta}$ is a translate of $\theta$.  
We must now ask to what extent does the triple
$\left( \overline{\kappa}^{[1]},\overline{\kappa}^{[2]},\overline{\theta}\right)$ determine the
original $\rho$ from $\overline{\mathcal{C}}_{M+N}$?
To begin with, we cannot generally know which of Cases I, II, or III applied to $\rho$.
If it was Case I, then $\rho$ is determined uniquely by joining $\kappa_B^{[1]}$ to
$\kappa_B^{[2]}$.  In Case II, if we also know
the site on $\kappa^{[1]}$ that played the role of $u$ and 
the site on $\kappa^{[2]}$ that played the role of $y$, then we can 
reconstruct $\rho$.  Note that there are no more than $M$ possibilities for the site on $\kappa^{[2]}$,
and at most 2 possibilities for the site on $\kappa^{[1]}$ 
(either the end of the side chain of $\kappa^{[1]}$ closest to $\kappa^{[1]}_B$, or the 
site of $\kappa^{[1]}$ that is adjacent to $\kappa^{[1]}_B$ (the latter case occurs when $L=0$)).
Therefore in Case II there are at most $2M$ possibilities for $\rho$.
In Case III, we see that $\rho$ is uniquely determined.  
Combining the three cases, we see that there are at most $2M+2$ possible combs $\rho$
for a given $\left( \overline{\kappa}^{[1]},\overline{\kappa}^{[2]},\overline{\theta}\right)$.
The claimed inequality (\ref{eq.comb-ccwc}) follows.

\smallskip
(\textit{vi})
Now we can put the ingredients together.  For $M,N\geq 1$, we have
\begin{eqnarray}
   \nonumber
   c_{M+N}  & \leq & (2M+2) c_N \sum_{i=0}^{M-1}  c_{M-i} \,w_i  \hspace{35mm} 
         \hbox{(by (\ref{eq.comb-ccwc}))}
   \\
   \nonumber
      & \leq & (2M+2)c_N \sum_{i=0}^{M-1}  c_{M-i} \,e^{\alpha\sqrt{i}} \,b_i
        \hspace{30mm}  \hbox{(by (\ref{eq.HWineq}))}
        \\
   \nonumber
      & \leq & (2M+2)c_N \sum_{i=0}^{M-1}  \sum_{j=0}^{M-i-1} (M-i-j)\,|\mathcal{C}^*_{M-i-j}|\, h_j
         \,e^{\alpha\sqrt{M}} \,b_i
        \hspace{5mm}  \hbox{(by (\ref{eq.comb-cstarhc}))}
        \\
        \nonumber
      & \leq & (2M+2)c_N \,M\,e^{\alpha\sqrt{M}}\sum_{i=0}^{M-1} 
       \sum_{j=0}^{M-i-1} |\mathcal{C}^*_{M-i-j}|\, h_{i+j}
        \hspace{11mm}  \hbox{(by (\ref{eq.comb-bhh}))}    
        \\
        \nonumber
        & \leq & (2M+2)c_N \,M\,e^{\alpha\sqrt{M}}\sum_{i=0}^{M-1} 
       \sum_{j=0}^{M-i-1} c_M
        \hspace{30mm}  \hbox{(by (\ref{eq.comb-chc}))}    
        \\
        \label{eq.combsubmult}
      & \leq & (2M+2)c_N \,M^3\,e^{\alpha\sqrt{M}} \,c_M \,.
\end{eqnarray}
Inequality (\ref{eq.combsubmult})  gives us a generalized submultiplicative inequality for combs:
\begin{equation}
   \label{eq.combsubmultA}
   c_{M+N}  \;\leq \;  g(M)\,c_M  \,c_N    \hspace{5mm}\hbox{for all }M,N\geq 1
\end{equation}
where
\[       g(M) \;=\;(2M+2)M^3 e^{\alpha\sqrt{M}}    \,.
\]
By Lemma \ref{lem.submult}, we obtain
\[    \lim_{n\rightarrow\infty} [c_n]^{1/n}\; = \;      
    \inf_{k\geq 1} [(2k+2)k^3 e^{\alpha\sqrt{k}} c_k]^{1/k} \,.
\]
\hfill   $\Box$

\subsection{The limiting comb free energy for negative $\beta$}
    \label{sec.combbetaneg}

Proposition \ref{prop.combgrowth} proves that the limit $\mathcal{F}(0\,|\,{\cal C}^X)$ exists.
We shall use this fact in this subsection to prove that the limiting annealed
free energy for combs exists whenever $\beta$ is negative.

\begin{prop}
    \label{prop.Fcombneglim}    
Consider the lattice $\mathbb{L}^d$ with $d\geq 2$, and let $X$ be either $+$ or $\pen$.
Then  the limit $\mathcal{F}(\beta\,|\,{\cal C}^X)$ exists and equals $\log \lambda_{d,C}$
for every $\beta\leq 0$.
\end{prop}

\noindent
\textbf{Proof:}
Since $F_N(\beta\,|\,\mathcal{C}^+)   \,\leq \,  F_N(\beta\,|\,\mathcal{C}^{\pen})  \,\leq \,
\frac{1}{N} \log |\mathcal{C}^{\pen}_N| \,\leq \,\frac{1}{N}\log (N c_N)$ for all $\beta \leq 0$, it suffices
to prove that the limit  $\mathcal{F}(\beta\,|\,{\cal C}^+)$ exists and equals $\log \lambda_{d,C}$. 

We shall construct a one-to-one map $F$ from  $\mathcal{C}_N^+$ into  $\mathcal{C}_{N+2}^+$.  
Let $\rho\in \mathcal{C}_N^+$.  Then the origin is a site of $\rho$.  We consider two cases.

First, if the origin has degree 1 in $\rho$, then let $\hat{\rho}$ be the union of $\rho$ and the 2-step
SAW from the origin to $-2e^{(1)}$.  If the origin was a labelled site in $\rho$, then move that label 
to $-2e^{(1)}$ in $\hat{\rho}$.  

Second, if the origin does not have degree 1 in $\rho$, then there is a site $v$ of $\rho$ that is
connected to the origin by an edge of $\rho$ and such that $v_1=0$.  Let $\hat{\rho}$ be formed
from $\rho$ by deleting the edge from $v$ to the origin and inserting the 3-step SAW from $v$
to $v-e^{(1)}$ to $-e^{(1)}$ to the origin.  

In both cases, $\hat{\rho}$ is an $(N+2)$-step comb.  Let $z$ be the lexicographically smallest
site of $\hat{\rho}$, and let $F(\rho)$ be the translation of $\hat{\rho}$ by $-z$.  Then $F(\rho)$ is in 
$\mathcal{C}_{N+2}^+$.  By looking at whether 
$F(\rho)$ has one or two sites in the hyperplane $x_1=0$, one can tell which of the above two cases 
applied to $\rho$, and hence one can determine $\rho$ from $F(\rho)$.  

Let $\beta \leq 0$.  Since $F$ is one-to-one, and since $F(\rho)$ has at most 2 sites in the
surface $x_1=0$ for each $\rho$, we have
\begin{eqnarray*}
   (N+2) \, c_{N+2}   & \geq &  Z_{N+2}(\beta\,|\,\mathcal{C}^+)  
       \\
          & \geq & \sum_{\rho\in \mathcal{C}_N^+}  \exp(\beta \sigma(F(\rho)))
      \\
      & \geq & |\mathcal{C}_N^+|\,e^{2\beta}    \hspace{5mm}  \geq \;  c_N\,e^{2\beta}.
\end{eqnarray*}
It follows that $\lim_{N\rightarrow\infty}F_N(\beta\,|\,\mathcal{C}^+)$ exists and equals
$\log \lambda_{d,C}$.  This completes the proof.
\hfill $\Box$

\subsection{The limiting comb free energy for positive $\beta$}
    \label{sec.combbetapos}

The following result, together with Proposition \ref{prop.Fcombneglim}, completes the proof
of Theorem \ref{thm.Fannexists}.

\begin{prop}
    \label{prop.Fcombann}
Consider the lattice $\mathbb{L}^d$ with $d\geq 2$, and let $X$ be either $+$ or $\pen$.
Then  the limit $\mathcal{F}(\beta\,|\,{\cal C}^X)$ exists 
for every $\beta \geq  0$. 
\end{prop}

\medskip
\noindent
\textbf{Proof:}   Fix $X$.  Consequently, we shall suppress $X$ in our new notation below (e.g.\ 
$\mathcal{B}_N$).
Fix $\beta\geq 0$.

Recall that $ \mathcal{W}^{X}_N$ consists of $N$-step SAWs $\omega$ (in $\mathbb{L}^d$ 
or $\mathbb{L}^d_+$) such  that  $\omega(i)=0$ for some $i\in [0,N]$.   
We shall also use the following sets of SAWs, for each $N\geq 0$ and $A\geq 1$:
\begin{eqnarray}
   \label{eq.defcalW}
    \mathcal{W}^{0X}_N  & = & \{ \omega\in\mathcal{W}^{X}_N\,:\,   \omega(0)=0   \} \,,
 \\
      \label{eq.defcalB}
    \mathcal{B}_N  & = & \{ \omega\in\mathcal{W}^{0X}_N\,:\, \omega \text{ is a bridge } \} \,,
   \\
   \label{eq.defcalBA}
    \mathcal{B}^{[A]}_N  & = & \{ \omega\in\mathcal{B}_N\,: \,|\omega_i(N)|\leq A \text{ for all }
    i=1,\ldots,d-1 \, \} \,.
\end{eqnarray}
There is an obvious bijection via translations 
between $\mathcal{W}_N^{0\pen}$ and $\overline{\mathcal{W}}_N$.  The set $\mathcal{B}^{[A]}_N$
consists of bridges that start at the origin and stay in the ``square tube of radius $A$'' that is centred along the $x_d$-axis.

Define the projection $\pi: \mathbb{Z}^d\rightarrow \mathbb{Z}^{d-1}$ by $\pi(z_1,\ldots,z_{d-1},z_d)
\,=\, (z_1,\ldots,z_{d-1})$.  For $y\in \mathbb{Z}^{d-1}$, let
\begin{equation}  
   \label{eq.defcalBAy}
    \mathcal{B}^{[A;y]}_N  \; = \; \{ \omega\in\mathcal{B}^{[A]}_N\,: \,\pi(\omega(N))=y \}\,.
\end{equation}
The case $y=0$ is of particular importance:  
\begin{equation}  
   \label{eq.defcalBA0}
    \mathcal{B}^{[A;0]}_N  \; = \; \{ \omega\in\mathcal{B}^{[A]}_N\,: \,\omega_i(N)=0 \text{ for all }
    i=1,\ldots,d-1 \, \}\,.
\end{equation}
Thus $\mathcal{B}^{[A;0]}_N$ is the set of $N$-step bridges that start at the origin, end 
on the $x_d$-axis, and stay in the ``tube.''

When 
$\mathcal{P}$ is any of $\mathcal{W}^{X}$, $\mathcal{W}^{0X}$, $\mathcal{B}$, $\mathcal{B}^{[A]}$,
$\mathcal{B}^{[A;y]}$,    
we also define the following:
\begin{eqnarray*}
    Z_N(\beta\,|\,\mathcal{P})  & = & \sum_{\rho\in \mathcal{P}_N} e^{\beta \sigma(\rho)}
        \hspace{15mm}  \hbox{(as in Equation (\ref{eq.defZPgen}))},
   \\
   F_N(\beta\,|\,\mathcal{P}) & = &\frac{1}{N}\log Z_N(\beta\,|\,\mathcal{P}) \,, \hspace{9mm}\hbox{and}
     \\
   \mathcal{F}(\beta \,|\,\mathcal{P}) & = &  
\lim_{N\rightarrow\infty}F_N(\beta \,|\,\mathcal{P})  \hspace{10mm}\hbox{\em{if this limit exists}.}
\end{eqnarray*}

It is proved in \cite{HTW} that $\mathcal{F}(\beta\,|\,\,\mathcal{P})$ exists when $\mathcal{P}$ is
$\mathcal{W}^{0X}$ or $\mathcal{B}$ (whether $X$ is $+$ or $\pen$),
and that
\begin{equation}
    \label{eq.HTW-WeqB}
     \mathcal{F}(\beta \,|\,\mathcal{W}^{0X}) \;=\;  \mathcal{F}(\beta \,|\,\mathcal{B}) 
     \hspace{4mm}\hbox{for every real $\beta$.}
\end{equation}

For $\omega\in \mathcal{W}^{X}_N$ with $\omega(i)=0$, consider the subwalk from 
$\omega(i)$ to $\omega(N)$ and the subwalk from $\omega(i)$ back to $\omega(0)$.  We then see
that
\[     \sum_{i=0}^N Z_{N-i}(\beta\,|\mathcal{W}^{0X})\,Z_i(\beta\,|\,\mathcal{W}^{0X})  \;\geq \;
    e^{\beta} Z_N(\beta\,|\,\mathcal{W}^{X})   \;\geq \;  e^{\beta} Z_N(\beta \,|\,\mathcal{W}^{0X}) \,,
\]
from which it follows that the limiting free energy for $\mathcal{W}^{X}$ exists and
\begin{equation}
    \label{eq.WeqWvee}
     \mathcal{F}(\beta \,|\,\mathcal{W}^{X}) \;=\;  \mathcal{F}(\beta \,|\,\mathcal{W}^{0X}) 
     \hspace{4mm}\hbox{for every real $\beta$.}
\end{equation}

Fix $A\geq 1$.  For two bridges $\omega\in \mathcal{B}^{[A;0]}_N$ and
$\psi\in\mathcal{B}^{[A;0]}_M$, their concatenation $\omega\circ\psi$ is in 
$\mathcal{B}^{[A;0]}_{N+M}$ and $\sigma(\omega\circ\rho)=\sigma(\omega)+\sigma(\rho)-1$
(since $\omega(N)$ is in the surface $x_1=0$).   It follows that
\begin{equation}
   \label{eq.ZB0concat}
Z_{N+M}(\beta\,|\,\mathcal{B}^{[A;0]})  \;\geq \;e^{-\beta} Z_{N}(\beta\,|\,\mathcal{B}^{[A;0]})\,
      Z_{M}(\beta\,|\,\mathcal{B}^{[A;0]}), 
\end{equation}
and hence that 
\begin{equation}
   \label{eq.FBA0}
   \hbox{the limit $\mathcal{F}(\beta\,|\,\mathcal{B}^{[A;0]})$ exists for every real $\beta$.}
\end{equation}
Now we turn to $\mathcal{F}(\beta\,|\,\mathcal{B}^{[A]})$.
Define the norm $\|\cdot \|_{\infty}$ on $\mathbb{Z}^{d-1}$ by 
$\|y\|_{\infty}:=\max\{|y_1|,\ldots,|y_{d-1}|\}$.
Observe that 
\begin{equation}
    \label{eq.BAsumBAy}
    Z_N(\beta \,|\, \mathcal{B}^{[A]})  \;=\; \sum_{y\in \mathbb{Z}^{d-1}:  \|y\|_{\infty}\leq A}
    Z_{N}(\beta\,|\,\mathcal{B}^{[A;y]}) \,.
\end{equation}
Consider two bridges $\omega$ and $\psi$ in $Z_{N}(\beta\,|\,\mathcal{B}^{[A;y]})$.  Let 
$\omega\oslash\psi$ be the $(2N+1)$-step walk defined by $\omega\circ e^{(d)}\circ R(\psi)$,
where $e^{(d)}$ represents the one-step SAW from 0 to $e^{(d)}$ and $R(\psi)$ is the 
reversal of the reflection
of $\psi$ through the hyperplane $x_d=0$, 
i.e.
\[    (R(\psi))(i)   \;=\; (\psi_1(N-i),\ldots,\psi_{d-1}(N-i), -\psi_d(N-i)) \,,    \hspace{5mm}i=0,\ldots,N.  
\]
Then we see that  $\omega\oslash\psi\in  \mathcal{B}_{2N+1}^{[A;0]}$, 
$\sigma(\sigma\oslash\psi)=\sigma(\omega)+\sigma(\psi)$, and 
\begin{equation}
   \label{eq.Z0geqZBy2}
   Z_{2N+1}(\beta \,|\, \mathcal{B}^{[A;0]})  \;\geq \; \sum_{y\in \mathbb{Z}^{d-1}:  \|y\|_{\infty}\leq A}
    \left(Z_{N}(\beta\,|\,\mathcal{B}^{[A;y]})\right)^2 \,.
\end{equation}
Applying the Cauchy-Schwarz inequality to the right-hand side of Equation (\ref{eq.Z0geqZBy2})
gives
\begin{eqnarray}
    \nonumber
    Z_{2N+1}(\beta \,|\, \mathcal{B}^{[A;0]}) & \geq & 
       \frac{1}{(2A+1)^{d-1}}   \left(  \sum_{y\in \mathbb{Z}^{d-1}:  \|y\|_{\infty}\leq A}
    Z_{N}(\beta\,|\,\mathcal{B}^{[A;y]})  \right)^2
      \\
      \label{eq.ZB0geqZB2}
      & = &  \frac{  \left(  Z_N(\beta\,|\,\mathcal{B}^{[A]})\right)^2  }{(2A+1)^{d-1}}   
       \hspace{6mm}\hbox{(by Equation (\ref{eq.BAsumBAy}))}
       \\
       \nonumber
       & \geq & \frac{  \left(  Z_N(\beta\,|\,\mathcal{B}^{[A;0]})\right)^2  }{(2A+1)^{d-1}}  \,.       
\end{eqnarray}
Hence, by Equation (\ref{eq.FBA0}), we see that
\begin{equation}
   \label{eq.FBAeqA0}
   \hbox{the limit $\mathcal{F}(\beta\,|\,\mathcal{B}^{[A]})$ exists 
   and equals  $\mathcal{F}(\beta\,|\,\mathcal{B}^{[A;0]})$   for every $\beta\geq 0$.}
\end{equation}
We also claim that 
\begin{equation}
    \label{eq.limBAeqW}
    \lim_{A\rightarrow\infty}  \mathcal{F}(\beta\,|\,\mathcal{B}^{[A]})  \;=\; 
      \mathcal{F}(\beta\,|\,\mathcal{W}^{0X}) \,.
\end{equation}    
Since $\mathcal{F}(\beta\,|\,\mathcal{B}^{[A]})$ is 
non-decreasing in $A$, the limit on the left-hand side exists.
By Equation   (\ref{eq.HTW-WeqB}),
it suffices to prove that this limit is greater than or equal to
$\mathcal{F}(\beta\,|\,\mathcal{B})$. 
Temporarily fix $n\geq 1$.
By the definition (\ref{eq.defcalBA}), we see that 
\[   Z_n(\beta\,|\,\mathcal{B})  \,=\,  Z_n(\beta\,|\,\mathcal{B}^{[n]}) \,.
\] 
Using this relation and Equations (\ref{eq.ZB0concat}) and (\ref{eq.ZB0geqZB2}), we see that 
for every $k\geq 1$ we have
\begin{eqnarray*}
   Z_{k(2n+1)}(\beta\,|\,\mathcal{B}^{[n;0]})   & \geq & 
        e^{-\beta k}  \left(Z_{2n+1}(\beta\,|\,\mathcal{B}^{[n;0]})   \right)^k
        \\
        & \geq &   
      \frac{  \left(  Z_n(\beta\,|\,\mathcal{B}^{[n]})\right)^{2k}  }{e^{\beta k}(2n+1)^{k(d-1)} } \,.     
         \\
        & = &   
      \frac{  \left(  Z_n(\beta\,|\,\mathcal{B})\right)^{2k}  }{e^{\beta k}(2n+1)^{k(d-1)} } \,.               
\end{eqnarray*}
Take logarithms of the above inequality, divide by $k(2n+1)$, and let $k\rightarrow \infty$.  We get
\[    
    \mathcal{F}(\beta\,|\,\mathcal{B}^{[n;0]})   \;  \geq \; 
     \frac{2}{2n+1}\log  Z_n(\beta\,|\,\mathcal{B})  \,-\, \frac{\beta+(d-1)\log(2n+1)}{2n+1}
\]
Therefore $\lim_{n\rightarrow\infty} \mathcal{F}(\beta\,|\,\mathcal{B}^{[n;0]}) \,\geq \,  
\mathcal{F}(\beta\,|\,\mathcal{B})$.   By our previous comments and Equation (\ref{eq.FBAeqA0}),
the claim (\ref{eq.limBAeqW}) follows.

Now that we have dealt with these classes of SAWs, we return to discussing combs.

The following inequality is the direct analogue of the crucial inequality (\ref{eq.comb-ccwc}) in 
the proof of    the existence of $\lim_{N\rightarrow\infty}c_N^{1/N}$ in Proposition 
\ref{prop.combgrowth}.
\begin{eqnarray}
   \nonumber
    Z_{N+M}(\beta\,|\,\mathcal{C}^X)   & \leq &  3\,(2M+2) \, Z_N(\beta\,|\,\mathcal{C}^X)
        \sum_{i=0}^{M-1}  Z_{M-i}(\beta\,|\,\mathcal{C}^X) \,Z_i (\beta\,|\,\mathcal{W}^{X}) 
         \\
         \label{eq.ZCCCW} 
        & &  \hspace{25mm}  \hbox{for all }M,N\geq 1, \,\beta\geq 0\,.  
\end{eqnarray} 
The proof of this inequality is based on essentially the same mapping of combs into 
triples of two combs and a SAW that we used to prove Equation (\ref{eq.comb-ccwc}), 
with one important difference:  in Proposition \ref{prop.combgrowth},
we only needed to consider configurations up to translation, but now translation may not
preserve the function $\sigma$.  Instead of the mapping of Equation (\ref{eq.combmap}),
we shall use a mapping
\begin{eqnarray}
    \nonumber
   \mathcal{C}^X_{M+N}  & \longrightarrow & \mathcal{C}^X_N \times
         \bigcup_{i=0}^{M-1}  \left(   \mathcal{C}^X_{M-i}\times \mathcal{W}^{X}_i\right)
         \\
         \label{eq.combmapZ}
        & = & \left\{  \left( \underline{\kappa}^{[1]},\underline{\kappa}^{[2]},\underline{\theta} \right)\,:  \,
        \underline{\kappa}^{[1]} \in \mathcal{C}^X_N ,\,
        \underline{\kappa}^{[2]} \in \mathcal{C}^X_{M-i} ,\,
        \underline{\theta} \in   \mathcal{W}^{X}_i   \right.
        \\
        \nonumber 
        & & \hspace{44mm} \left.   \hbox{  for some }i\in [0,M-1]  \right\}  \,.   
\end{eqnarray}
For a given $\rho\in \mathcal{C}^X_{M+N}$,
the method in the proof of Proposition \ref{prop.combgrowth}   
creates three subgraphs of ${\kappa}^{[1]}$, $\kappa^{[2]}$, and $\theta$ of $\rho$ 
whose union is $\rho$, such that no two subgraphs share an edge.   Note that one or two of 
these subgraphs may not intersect the surface $x_1=0$.
To define the mapping (\ref{eq.combmapZ}), we apply the following procedure, with $\phi$ being
$\kappa^{[1]}$, $\kappa^{[2]}$, or $\theta$ in turn.
\begin{verse}
\underline{Shift Procedure}
\\
(P1) If $\phi$ contains the origin, then let $\underline{\phi}$ equal $\phi$.
\\
(P2) If $\phi$ does not contain any site of the surface $x_1=0$, then let $\underline{\phi}$ 
be the translation of $\phi$ whose lexicographically smallest site is the origin.  In this case,
$\sigma(\phi)=0 <\sigma(\underline{\phi})$.
\\
(P3) If neither (1) nor (2) holds, then let $v$ be the lexicographically smallest site of $\phi$ that
is contained in the surface $x_1=0$, and let $\underline{\phi}$ be the translation $\phi-v$.
Since this translation is parallel to the surface, we have $\sigma(\underline{\phi})=\sigma(\phi)$.
\end{verse}
Observe that we always have
\begin{equation}
    \label{eq.sigma3leq}
    \sigma(\rho)  \;\leq \;  \sigma(\kappa^{[1]}) +\sigma(\kappa^{[2]})+\sigma(\theta)
     \;\leq \;  \sigma(\underline{\kappa}^{[1]}) +\sigma(\underline{\kappa}^{[2]})+
     \sigma(\underline{\theta}) \,.
\end{equation}

Given $\left( \underline{\kappa}^{[1]},\underline{\kappa}^{[2]},\underline{\theta} \right)$,   
how many different combs $\rho$ in $\mathcal{C}^X_{M+N}$ could produce this triple?
First, we need to know which of the three original subgraphs satisfied (P1).  (If more than one 
subgraph satisfied (P1), then we only need to know one of them to proceed.)  
Given this information, the upper bound $2M+2$ from Proposition \ref{prop.combgrowth} still
applies for us.  
Finally, since Equation (\ref{eq.sigma3leq}) always holds and since $\beta\geq 0$, the 
inequality (\ref{eq.ZCCCW}) follows.

In each of parts (\textit{iii}) and (\textit{iv}) of the proof of Proposition \ref{prop.combgrowth},  
we had to attach a SAW to a comb.  We did this
at a site of the comb with largest $x_d$ coordinate.  That method assumed 
that we could translate the SAW to any site of the lattice, but that does not work here because
the surface interaction only permits translations parallel to the surface.  We now describe
our modified approach.

For $z\in \mathbb{Z}$ and $A\geq 1$, let $T[A,z]$ be the ``semi-infinite square tube''
\begin{eqnarray*} 
 T[A,z] & := &  \{x\in \mathbb{Z}^d:  \, |x_i|\leq A \hbox{ for $i=1,\ldots,d-1$, and $x_d\geq z$}\}
     \\
     & = & [-A,A]^{d-1}\times [z,\infty)\,.
\end{eqnarray*}

Let $\mathcal{C}_m^{*A}$ be the set of combs $\rho\in \mathcal{C}^X_m$ such that there there is 
a nonnegative integer $u\equiv u(\rho)$ with the properties that 
$(0,0,\ldots,0,u)$ ($=ue^{(d)}$) is a site of 
degree 1 in $\rho$ and  $\rho\cap T[A,u+1] \,=\,\emptyset$.  

\setlength{\unitlength}{0.93mm}
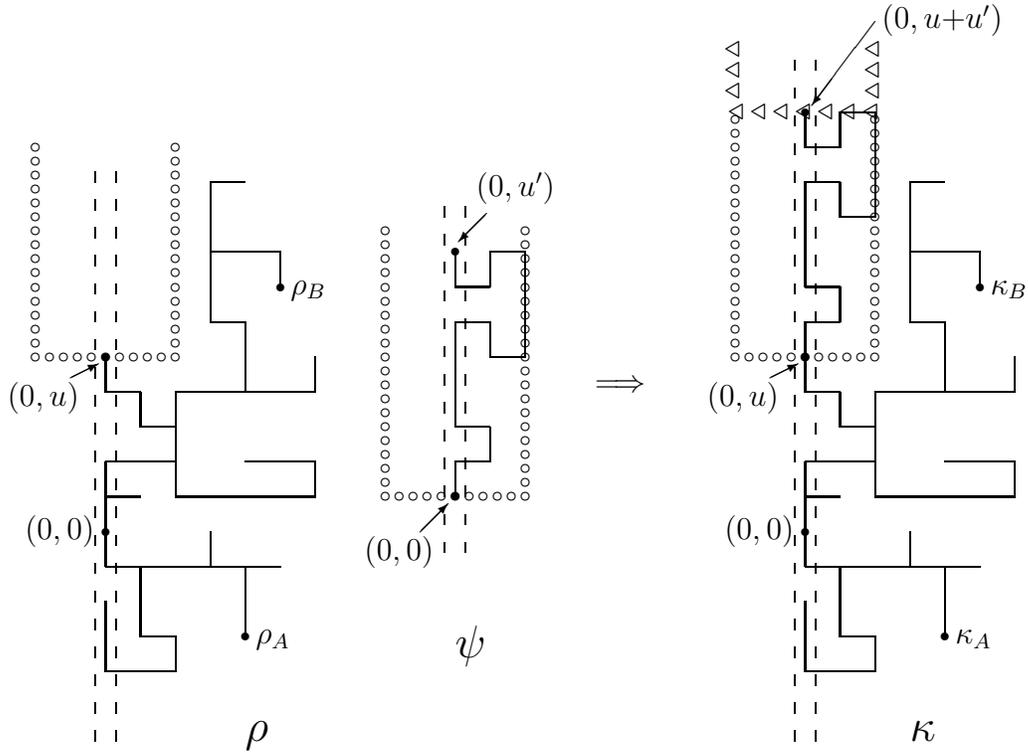
\begin{figure}
\begin{center}
\begin{picture}(145,105)
\put(10,0){\begin{picture}(35,80)
\put(0,10){\line(1,0){10}}
\put(0,10){\line(0,1){10}}
\put(10,10){\line(0,1){5}}
\put(5,15){\line(1,0){5}}
\put(5,15){\line(0,1){10}}
\put(0,25){\line(1,0){25}}
\put(0,25){\line(0,1){15}}
\put(15,25){\line(0,1){5}}
\put(20,25){\line(0,-1){10}}
\put(0,35){\line(1,0){5}}
\put(0,40){\line(1,0){10}}
\put(10,35){\line(1,0){20}}
\put(10,35){\line(0,1){15}}
\put(30,35){\line(0,1){5}}
\put(30,40){\line(-1,0){10}}
\put(5,45){\line(1,0){5}}
\put(5,45){\line(0,1){5}}
\put(5,50){\line(-1,0){5}}
\put(0,50){\line(0,1){5}}
\put(10,50){\line(1,0){20}}
\put(30,50){\line(0,1){5}}   
\put(20,50){\line(0,1){10}}
\put(15,60){\line(1,0){5}}
\put(15,60){\line(0,1){20}}   
\put(15,70){\line(1,0){10}}
\put(25,70){\line(0,-1){5}}
\put(15,80){\line(1,0){5}}
\put(20,15){\circle*{1.2}}
\put(25,65){\circle*{1.2}}
\put(21.5,14){$\rho_A$}
\put(26.5,64){$\rho_B$}
\put(0,30){\circle*{1.2}}
\put(-11.5,29){$(0,0)$}
\put(0,55){\circle*{1.2}}
\put(-14,48){$(0,u)$}
\put(-5,52){\vector(2,1){4}}
\multiput(-10,55)(2,0){11}{\circle{1}}
\multiput(-10,57)(0,2){15}{\circle{1}}
\multiput(10,57)(0,2){15}{\circle{1}}
\end{picture}
}
\put(50,35){\begin{picture}(20,50)
\put(10,0){\line(0,1){5}}
\put(10,5){\line(1,0){5}}
\put(15,5){\line(0,1){5}}
\put(10,10){\line(1,0){5}}
\put(10,10){\line(0,1){15}}
\put(10,25){\line(1,0){5}}
\put(15,20){\line(1,0){5}}
\put(15,20){\line(0,1){5}}
\put(20,20){\line(0,1){15}}
\put(20,35){\line(-1,0){5}}
\put(15,30){\line(0,1){5}}
\put(10,30){\line(1,0){5}}
\put(10,30){\line(0,1){5}}
\multiput(0,0)(2,0){11}{\circle{1}}
\multiput(0,2)(0,2){19}{\circle{1}}
\multiput(20,2)(0,2){19}{\circle{1}}
\put(10,0){\circle*{1.2}}
\put(10,35){\circle*{1}}
\put(-3,-9){$(0,0)$}
\put(5,-5){\vector(1,1){4}}
\put(13,43){$(0,u')$}
\put(14.5,40.5){\vector(-1,-1){4}}
\end{picture}
}
\put(30,0){\Large{$\rho$}}
\put(60,12){\Large${\psi}$}
\put(125,0){\Large{$\kappa$}}
\put(110,0){\begin{picture}(35,80)
\put(0,10){\line(1,0){10}}
\put(0,10){\line(0,1){10}}
\put(10,10){\line(0,1){5}}
\put(5,15){\line(1,0){5}}
\put(5,15){\line(0,1){10}}
\put(0,25){\line(1,0){25}}
\put(0,25){\line(0,1){15}}
\put(15,25){\line(0,1){5}}
\put(20,25){\line(0,-1){10}}
\put(0,35){\line(1,0){5}}
\put(0,40){\line(1,0){10}}
\put(10,35){\line(1,0){20}}
\put(10,35){\line(0,1){15}}
\put(30,35){\line(0,1){5}}
\put(30,40){\line(-1,0){10}}
\put(5,45){\line(1,0){5}}
\put(5,45){\line(0,1){5}}
\put(5,50){\line(-1,0){5}}
\put(0,50){\line(0,1){5}}
\put(10,50){\line(1,0){20}}
\put(30,50){\line(0,1){5}}   
\put(20,50){\line(0,1){10}}
\put(15,60){\line(1,0){5}}
\put(15,60){\line(0,1){20}}   
\put(15,70){\line(1,0){10}}
\put(25,70){\line(0,-1){5}}
\put(15,80){\line(1,0){5}}
\put(20,15){\circle*{1.2}}
\put(25,65){\circle*{1.2}}
\put(21.5,14){$\kappa_A$}
\put(26.5,64){$\kappa_B$}
\put(0,30){\circle*{1.2}}
\put(-11.5,29){$(0,0)$}
\put(0,55){\circle*{1.2}}
\put(-14,48){$(0,u)$}
\put(-5,52){\vector(2,1){4}}
\put(10,103){\vector(-3,-4){9}}
\put(11,102){$(0,u{+}u')$}
\multiput(-10,55)(2,0){11}{\circle{1}}
\multiput(-10,57)(0,2){17}{\circle{1}}
\multiput(10,57)(0,2){17}{\circle{1}}
\put(-10,55){\begin{picture}(20,35)
\put(10,0){\line(0,1){5}}
\put(10,5){\line(1,0){5}}
\put(15,5){\line(0,1){5}}
\put(10,10){\line(1,0){5}}
\put(10,10){\line(0,1){15}}
\put(10,25){\line(1,0){5}}
\put(15,20){\line(1,0){5}}
\put(15,20){\line(0,1){5}}
\put(20,20){\line(0,1){15}}
\put(20,35){\line(-1,0){5}}
\put(15,30){\line(0,1){5}}
\put(10,30){\line(1,0){5}}
\put(10,30){\line(0,1){5}}
\put(10,35){\circle*{1.2}}
\multiput(-1,34)(3.2,0){7}{$\triangleleft$}
\multiput(-1.5,37)(0,3){3}{$\triangleleft$}
\multiput(18.5,37)(0,3){3}{$\triangleleft$}
\end{picture}
}
\end{picture}
}
\put(80,50){$\Longrightarrow$}
\multiput(8.5,0)(0,4){21}{\line(0,1){1.5}}
\multiput(11.5,0)(0,4){21}{\line(0,1){1.5}}
\multiput(58.5,27)(0,4){13}{\line(0,1){1.5}}
\multiput(61.5,27)(0,4){13}{\line(0,1){1.5}}
\multiput(108.5,0)(0,4){25}{\line(0,1){1.5}}
\multiput(111.5,0)(0,4){25}{\line(0,1){1.5}}
\end{picture}
\end{center}
\caption{\label{fig.Zcomb}  The derivation of Equation (\ref{eq.ZCZbleqZCstar}) for an impenetrable
surface when $d=2$.
In each sketch, the parallel dashed lines indicate the surface $x_1=0$.
\textit{Left}:  A comb $\rho\in \mathcal{C}_m^{*A}$; the open circles $\circ\circ\circ$ indicate the 
boundary of $T[A,u]$.
\textit{Middle}:  A bridge $\psi \in \mathcal{B}_t^{[A;0]}$, with endpoints $\psi(0)=(0,0)$ and 
$\psi(t)=(0,u')$;
the open circles $\circ\circ\circ$ indicate the 
boundary of $T[A,0]$.
\textit{Right}:  The comb $\kappa$ produced by translating $\psi$ by $(0,u)$
($=ue^{(2)}$) and joining it to $\rho$; observe that 
$\kappa$ is in  $\mathcal{C}_{m+t}^{*A}$.  
The triangles $\triangleleft\triangleleft\triangleleft$ indicate the boundary of $T[A,u+u']$.
}
\end{figure}

Suppose $\rho\in \mathcal{C}_m^{*A}$, and let $u=u(\rho)$. 
Let $\psi\in \mathcal{B}_t^{[A;0]}$.  Then the translated bridge
$\psi + ue^{(d)}$ does not intersect $\rho$ except at the site $ue^{(d)}$, so the union of
$\rho$ and  $\psi + ue^{(d)}$ is a comb in $\mathcal{C}_{m+t}^{*A}$ (see Figure \ref{fig.Zcomb}).  
Since this translation does not
change $\sigma(\psi)$, we deduce that
\begin{equation}
    \label{eq.ZCZbleqZCstar}
   Z_{m}(\beta\,|\,\mathcal{C}^{*A}) \,Z_t(\beta\,|\,\mathcal{B}^{[A;0]})   \;\leq \;
     Z_{m+t}(\beta\,|\,\mathcal{C}^{*A})\,.
\end{equation}
This is our analogue of the inequality (\ref{eq.comb-chc}) of part (\textit{iii}).

Now we proceed to our analogue of part (\textit{iv}).
Let $\rho\in \mathcal{C}_m$.  Let $r$ be the smallest value of $z$ such that 
$\rho \cap T[A,z]$ is empty.  Then $0< r<\infty$ since $\rho$ contains the origin and
is finite.
Let $v$ be a site of $\rho\cap T[A,r-1]$.  In particular, we must have $v_d=r-1$.
One of the following cases must hold:
\begin{verse}
(a)  $v$ is on a side chain of $\rho$ and has degree 2, or
\\
(b) $v$ has degree 3, or
\\
(c)  $v$ is on the backbone of $\rho$ and has degree 2, or 
\\
(d)  $v$ has degree 1.
\end{verse}
In cases (a) or (b), let $u$ be the free end of the side chain of $\rho$ that contains $v$, and
let $\theta$ be the SAW that consists of the part of the side chain from $v$ to $u$.
In cases (c) and (d), let $\theta$ be the 0-step SAW consisting of the site $v$.
Let $j$ be the number of steps in $\theta$.
Next, let $\chi$ be a $(d A)$-step  SAW whose first step is from $v$ to $v+e^{(d)}$, then 
takes $|v_1|+|v_2|+\cdots +|v_{d-1}|$ steps to the point $(0,0,\ldots,0,z)$, and then takes the rest of its
steps along the $x_d$ axis in the $+e^{(d)}$ direction.  
Let $K$ be the graph obtained from $\rho$ by deleting $\theta$ and adding $\chi$:
that is, $K\,=\, (\rho\setminus \theta)\cup \chi$.
Apply the Shift Procedure described above to obtain
a translation $\underline{\theta}\in \mathcal{W}^X_j$.  Then apply the Shift Procedure to 
$K$, except in (P2) and (P3) we replace ``the surface $x_1=0$'' by ``the $x_d$-axis,''
obtaining $\underline{K}\in \mathcal{C}^{*A}_{m+dA-j}$.  
This replacement ensures that we know which site of $\underline{K}$ is the translation
of the site $\chi(dA)$ of $K$:  it is
the lexicographically largest site of $\underline{K}$ on the $x_d$-axis.
Similarly to Equation (\ref{eq.sigma3leq}), we see that 
$\sigma(\rho)\leq \sigma(\underline{K})+\sigma(\underline{\theta})$.
Given $(\underline{K},\underline{\theta})$, there are at most $2$ possibilities for $\rho$
(depending on which of $K$ or $\theta$ contained the origin; note that 
since we know which site of $\underline{K}$ is the 
translation of $\chi(dA)$, we know which site of 
$\underline{K}$ is the translation of the site $v$, where $\theta$ is attached to $K$).
It follows that
\begin{equation}
    \label{eq.ZCleqZCZW}
   Z_{m}(\beta\,|\,\mathcal{C}^X)   \;\leq \;
    2\,   \sum_{j=0}^{m-1}  Z_{m+dA-j}(\beta\,|\,\mathcal{C}^{*A})\, Z_j(\beta\,|\,\mathcal{W}^{X})
\end{equation}
for $m\geq 1$, $A\geq 1$, and $\beta\geq 0$.

\smallskip
Next we prove the following bound.  
\begin{equation}
    \label{eq.combmonotone}
    Z_m(\beta \,|\,\mathcal{C}^X) \;\leq \;   2dm\,e^{\beta} \,Z_{m+t}(\beta \,|\,\mathcal{C}^X)
       \hspace{5mm}\hbox{for all $m,t\geq 1$ and $\beta\geq 0$}.
\end{equation}    
Write $t=2k-\ell$, where $k\geq 1$ and $\ell\in \{0,1\}$.
Let $\rho\in \mathcal{C}^X_m$.  Let $v$ be the lexicographically largest site of $\rho$.
We consider two cases.
In the first case, if
$v$ has degree 1 in $\rho$ or if $v$ is a site of degree 2 on the backbone of $\rho$, then 
let $\hat{\rho}$ be the union of $\rho$ with 
the straight $t$-step SAW from $v$ to $v+te^{(1)}$.  Evidently, $\rho\in \mathcal{C}^X_{m+t}$
and $\sigma(\hat{\rho})=\sigma(\rho)$.
If the first case does not hold, then $v$ must have a neighbour $u$ in $\rho$ such that 
$v_1=u_1$.  Let $\rho_0$ be obtained by the union of $\rho$ with the $(2k+1)$-step 
SAW from $v$ to $v+ke^{(1)}$ to $u+ke^{(1)}$ to $u$, followed by the deletion of
the edge from $u$ to $v$.  On the one hand, if $t$ is even, then let $\hat{\rho}=\rho_0$.   
On the other hand, if $t$ is odd, then let $\hat{\rho}$ be obtained from $\rho_0$ by deleting 
a site of degree 1 in $\rho_0$ (that is not the origin) as well as its incident edge
(and relabelling if $\rho_A$ or $\rho_B$ was deleted). 
Then $\hat{\rho}\in \mathcal{C}^X_{m+t}$ and $\sigma(\hat{\rho}) \geq \sigma(\rho)-1$.
The map $\rho \mapsto \hat{\rho}$ is at most $(2d-1)m$-to-one. 
The result (\ref{eq.combmonotone}) follows.

\smallskip

Now we can put the pieces together.  We write the sum on the right side of Equation 
(\ref{eq.ZCCCW}) as
\begin{eqnarray}
       \label{eq.sumZZbreak}
   \sum_{i=0}^{M-1}
         Z_{M-i}(\beta\,|\,\mathcal{C}^X) \,Z_i (\beta\,|\,\mathcal{W}^{X}) 
        & = &  \Sigma_M^< \,+\,\Sigma_M^>  
        \hspace{6mm}\hbox{where}
\end{eqnarray}
\begin{eqnarray}
   \label{eq.sumless}
   \Sigma_M^<   & = &    \sum_{i=0}^{\min\{dA,M-1\}}
         Z_{M-i}(\beta\,|\,\mathcal{C}^X) \,Z_i (\beta\,|\,\mathcal{W}^{X}) 
        \hspace{6mm}\hbox{and}
     \\
   \label{eq.sumgreater}
   \Sigma_M^>   & = &    \sum_{i=\min\{dA+1,M\}}^{M-1}
         Z_{M-i}(\beta\,|\,\mathcal{C}^X) \,Z_i (\beta\,|\,\mathcal{W}^{X}) \,.
\end{eqnarray}
By Equation (\ref{eq.combmonotone}), we have
\begin{equation}
    \label{eq.sumlessbound}
       \Sigma_M^<   \; \leq \;  2dMe^{\beta} \, Z_{M}(\beta\,|\,\mathcal{C}^X) \,   \sum_{i=0}^{dA}
        Z_i (\beta\,|\,\mathcal{W}^{X}) \,.
\end{equation}
To bound $\Sigma_M^>$, consider an $i$ such that $dA<i<M$.  
By Equation (\ref{eq.ZCleqZCZW}), we have
\begin{eqnarray}
   \nonumber
     \lefteqn{Z_{M-i}(\beta\,|\,\mathcal{C}^X) \, Z_i (\beta\,|\,\mathcal{W}^{X}) }
     \\
     \nonumber
     & \leq & 
    2\,   \sum_{j=0}^{M-i-1}  Z_{M-i+dA-j}(\beta\,|\,\mathcal{C}^{*A})\, Z_j(\beta\,|\,\mathcal{W}^{X})
        \,Z_i (\beta\,|\,\mathcal{W}^{X})
        \\
        \label{eq.sumZZZmiddle}
        & \leq &  
         2\,   \sum_{j=0}^{M-i-1}  Z_{M-i+dA-j}(\beta\,|\,\mathcal{C}^{*A})\, \times 
         \\
          \nonumber
      & &   \hspace{15mm}   
         Z_j(\beta\,|\,\mathcal{B}^{[A;0]}) \,
        Z_{i-dA} (\beta\,|\,\mathcal{B}^{[A;0]}) \,  h_1(M;A)^2 h_2(M;A)
\end{eqnarray}
where
\begin{eqnarray*}
   h_1(M;A)  & = & \max\left\{   \frac{ Z_k(\beta\,|\,\mathcal{W}^{X}) }{ 
   Z_k(\beta \,|\,\mathcal{B}^{[A;0]}) } \;:\,1\leq k\leq M \right\}
   \hspace{5mm}\hbox{and}
   \\
     h_2(M;A)  & = & \max\left\{   \frac{ Z_{k+dA}(\beta\,|\,\mathcal{W}^{X}) }{ 
   Z_k(\beta \,|\,\mathcal{W}^{X}) } \;:\,1\leq k\leq M \right\}   \,.
\end{eqnarray*}
Using Equation (\ref{eq.ZCZbleqZCstar}) twice on the inequality (\ref{eq.sumZZZmiddle}), 
we obtain
\begin{eqnarray}
   \nonumber
   Z_{M-i}(\beta\,|\,\mathcal{C}^X) \, Z_i (\beta\,|\,\mathcal{W}^{X})
      & \leq &   2 \sum_{j=0}^{M-i-1}  Z_M(\beta \,|\,  \mathcal{C}^{*A})  \,  h_1(M;A)^2 \,h_2(M;A) 
      \\
      \nonumber
      & \leq &   2 \,M\, Z_M(\beta \,|\,  \mathcal{C}^X)  \, h_1(M;A)^2 \,h_2(M;A)       
\end{eqnarray}   
when $dA<i<M$.  Therefore
\begin{equation}
    \label{eq.sumgreatbound}
       \Sigma_M^>   \; \leq \;  2M^2 \, Z_{M}(\beta\,|\,\mathcal{C}^X) \,  h_1(M;A)^2 \,h_2(M;A)  
\end{equation}
for every $M,A\geq 1$.  

Now choose $A=\lfloor \sqrt{M}\rfloor$.
Combining Equations (\ref{eq.ZCCCW}), (\ref{eq.sumZZbreak}),   (\ref{eq.sumlessbound}), and
(\ref{eq.sumgreatbound}), we obtain
\begin{equation}
    \label{eq.Zsubmult}
    Z_{M+N}(\beta\,|\,\mathcal{C}^X)   \;\leq \; g(M)  \, Z_{M}(\beta\,|\,\mathcal{C}^X) \,
    Z_{N}(\beta\,|\,\mathcal{C}^X) 
    \hspace{5mm}\hbox{ for all }M,N\geq 1 \,,
\end{equation}
where the function $g$ is defined by
\begin{eqnarray}
   \nonumber
    g(M)  & = &   (6M+6)\left( 2dMe^{\beta}  \sum_{i=0}^{d\lfloor \sqrt{M}\rfloor}
        Z_i (\beta\,|\,\mathcal{W}^{X})  \;+\,  \right.
        \\
         \label{eq.gdefsqrtM}
       & &   \hspace{15mm}\left.  2M^2 \,h_1\!\left(M;\lfloor \sqrt{M}\rfloor \right)^2\,
        h_2\!\left(M;\lfloor \sqrt{M}\rfloor\right)  \right) \,.
\end{eqnarray}

We claim that $\lim_{M\rightarrow\infty}g(M)^{1/M}=1$.  To prove this, observe first that
$Z_i(\beta\,|\,\mathcal{W}^{X}) \,\leq \,(i+1)(2d)^i e^{\beta(i+1)}$ for every $i$, which implies that
the $M^{th}$ root of 
$\sum_{i=0}^{d\lfloor \sqrt{M}\rfloor} Z_i (\beta\,|\,\mathcal{W}^{X})$ converges to 1.
To deal with $h_1$ and $h_2$, we shall use the following elementary
fact about nonnegative sequences $\{u_n\}$:
\begin{equation}
    \label{eq.lemlimmax}
   \hbox{If }\lim_{n\rightarrow\infty}u_n^{1/n}=L\geq 1, \hbox{ then} 
   \lim_{n\rightarrow\infty} \left[  \max\{ u_i \,:\, 1\leq i\leq n\} \right]^{1/n}   \;=\; L.
\end{equation}
By this fact and the existence of the limit in Equation (\ref{eq.WeqWvee}), we see that 
$\lim_{M\rightarrow\infty}h_2(M;\lfloor \sqrt{M}\rfloor)^{1/M}=1$.
By the fact (\ref{eq.lemlimmax})  and Equation (\ref{eq.FBAeqA0}), we see that
$\lim_{M\rightarrow\infty}h_1(M;A)^{1/M}\,=\,\exp[\mathcal{F}(\beta\,|\,\mathcal{W}^X)- 
\mathcal{F}(\beta \,|\,\mathcal{B}^{[A;0]})]$ for every fixed $A$.  Using Equation (\ref{eq.limBAeqW})
and the fact that $h_1(M;A) \,\geq \,  h_1\!\left(M;\lfloor \sqrt{M}\rfloor \right) \,\geq\, 1$ for all
$M\geq A^2$, we can deduce that  $\lim_{M\rightarrow\infty}h_1(M;\lfloor \sqrt{M}\rfloor)^{1/M}=1$.
It now follows that $\lim_{M\rightarrow\infty}g(M)^{1/M}=1$.

The preceding paragraph shows  that Lemma \ref{lem.submult} applies to the inequality of Equation
(\ref{eq.Zsubmult}).  It follows that $\lim_{N\rightarrow\infty}Z_N(\beta\,|\,\mathcal{C}^X)^{1/N}$ exists.
This proves the Proposition.
\hfill   $\Box$

\section{Properties of Combs}
     \label{sec.combprops}

\medskip
\begin{prop}
  \label{prop.combneqsaw}
For dimension $d\geq 2$, the growth constant for combs is strictly greater than 
the growth constant for self-avoiding walks.  That is,
\[     \lambda_{d,C}   \; > \;   \mu_d \,. 
\]
\end{prop}

\medskip
\noindent
\textbf{Proof:}  We shall apply Kesten's Pattern Theorem 
(\cite{Ke}; see also Theorem 7.2.3 of \cite{MS}).
Let $\tilde{P}$ be the 4-step walk from  $-2e^{(1)}$ to $+2e^{(1)}$.
Let $\tilde{Q}$ be the hypercube $[-2,2]^d$ of side 4 centred at the origin.
We say that $(\tilde{P},\tilde{Q})$ occurs at the $j^{th}$ step of an $N$-step SAW $\omega$ 
if the 4-step subwalk $(\omega(j-2),\omega(j-1),\omega(j),\omega(j+1),\omega(j+2))$ 
is a translate of $\tilde{P}$ and 
if the translated set $\omega(j)+\tilde{Q}$ contains no other site of $\omega$.
(Here it is implicit that $2\leq j\leq N-2$.)
For each $N,m\geq 1$, let $\mathcal{W}_N[>m, (\tilde{P},\tilde{Q})]$ be the set of 
SAWs $\omega$ in $\mathcal{W}_N$ such that  $(\tilde{P},\tilde{Q})$ occurs at more than 
$m$ sites of $\omega$.  
Then Kesten's Pattern Theorem implies that there exists a $\delta>0$ such
\[     \limsup_{N\rightarrow\infty}  \left( \left|\mathcal{W}_N\right| \,-\,
   \left|\mathcal{W}_N[>\delta N, (\tilde{P},\tilde{Q})] \right|  \right)^{1/N}     \;<\;  \mu_d  \,;
\]
that is,  $(\tilde{P},\tilde{Q})$ occurs at more than $\delta N$ sites of $\omega$ 
for all but exponentially few $\omega$ in $\mathcal{W}_N$.  

Suppose that $(\tilde{P},\tilde{Q})$ occurs at the $j$-th step of an $N$-step SAW $\omega$.
Label the ends of $\omega$ to make it a comb with no side chains.
Then we can create an $(N+1)$-step comb by appending the edge from $\omega(j)$
to any site of the form $\omega(j)\pm e^{(i)}$ for $2\leq i\leq d$.

For $\omega\in \mathcal{W}_N$, let $\mathcal{B}(\omega)$ be the set of all $j$ in $[2,N-2]$ such that 
$(\tilde{P},\tilde{Q})$ occurs at the $j$-th step of $\omega$.
For any nonempty $D\subset \mathcal{B}(\omega)$, let $h$ be any function from $D$ into 
$\{ \pm e^{(i)} : 2 \leq i\leq d\}$, and let $\omega^{D,h}$ be the graph obtained by adding
the edge from $\omega(j)$ to $\omega(j)+h(j)$ for every $j\in D$.  By the choice of $\tilde{Q}$,
we must have $\omega(j)+h(j)\neq \omega(k)+h(k)$ whenever $j$ and $k$ are distinct elements of
$D$.  Therefore, with the sites $\omega(0)$ and $\omega(N)$ labelled, we conclude that 
$\omega^{D,h}$ is a comb, and it has $N+|D|$ edges.  Furthermore, given a comb $\kappa$
that was produced in this way, we can uniquely determine $\omega$, $D$, and $h$.
Restricting our attention to SAWs in $\mathcal{W}_N[>\delta N, (\tilde{P},\tilde{Q})]$ and 
sets $D$ of size $\lfloor \delta N/2\rfloor$, we obtain
\[
    c_{N+\lfloor \delta N/2\rfloor}  \;\geq \;   \left| \mathcal{W}_N[>\delta N, (\tilde{P},\tilde{Q})] \right|
    \,  \binom{\lfloor \delta N\rfloor}{  \lfloor \delta N/2\rfloor}  (2d-2)^{\lfloor \delta N/2\rfloor} \,.
\]
Therefore
\begin{eqnarray*}
   \liminf_{N\rightarrow\infty} (c_{N+\lfloor \delta N/2\rfloor})^{1/N}
     & \geq & \mu_d \, 2^{\delta}\,(2d-2)^{\delta/2}    
\;  > \;  \mu_d \, (2d)^{\delta/2}. 
\end{eqnarray*} 
The rightmost inequality holds because $4(2d-2) > 2d$ for $d\geq 2$.   Finally, since $2d>\mu_d$,
\[     \lambda_{d,C}  \;=\;  \lim_{N\rightarrow\infty}   
     (c_{N+\lfloor \delta N/2\rfloor})^{1/(N+\lfloor \delta N/2\rfloor)}  
     \;> \;  \left(\mu_d \,(2d)^{\delta/2}\right)^{1/(1+(\delta/2))}
           \;  >  \: \mu_d \,.
\] 
\hfill   $\Box$  

\bigskip
We now prove that adsorption of combs at an impenetrable boundary has a strictly positive 
critical value.  This will complete the proof of Theorem \ref{thm.betacanngt0}.

\begin{prop}
    \label{prop.Fcombcrit}    
For combs in $\mathbb{L}_+^d$ with $d\geq 2$, we have
$\beta_c(\mathcal{C}^+)\geq (4\lambda_{d,C}^2)^{-1}>0$.
\end{prop}

\medskip
\noindent
\textbf{Proof:}
We use the method of \cite{Ma}, as follows.  

Consider a comb $\rho\in \mathcal{C}_N^+$.
Let $G_0(\rho)$ be the graph that is the intersection of $\rho$ with the hyperplane $x_1=0$.
Then $\mathcal{H}(\rho)$ is the set of sites of $G_0(\rho)$, and we 
let $E \equiv E(G_0(\rho))$ be the set of edges of $G_0(\rho)$.  
Each connected component of $G_0(\rho)$ is a tree (perhaps an isolated site or a SAW).     
Let $D_s\equiv D_s(\rho)$ be the set of    
isolated sites in $G_0(\rho)$.

It is not hard to show that in any tree with $n_e$ 
edges in which no vertex has degree above 3, there is a subset $D$ of edges such that  
$|D|\geq \lceil n_e/3\rceil$ and no two edges in $D$ have an endpoint in common.
Moreover, the number $n_s$ of sites in such a tree equals $n_e+1$, and if $n_e\geq 1$, then 
$|D|\geq n_s/4$.
It follows that there is a $D_e\equiv D_e(\rho)\subset E$ such that 
$|D_e|\geq (|\mathcal{H}(\rho)| - |D_s|)/4$ and no two edges in $D_e$ have an endpoint in common.
In particular, we have 
\begin{equation}
   \label{eq.DDsigma4}
    |D_e(\rho)|  \,+\, |D_s(\rho)|  \;\geq \; \frac{\sigma(\rho)}{4} \,.
\end{equation}

We now follow the calculations of \cite{Ma}.  For $N,k\geq 1$, let left$_N(k)$ be the number of 
combs $\rho\in \mathcal{C}^+_N$ such that $|\mathcal{H}(\rho)|\,=k$.  Observe that 
\begin{equation}
   \label{eq.Zdef2}
     |{\cal C}^+_N| \;=\; \sum_{k=1}^{N+1} \textrm{left}_N(k)   
          \hspace{5mm}  \hbox{and} \hspace{5mm}   
      Z_N(\beta\,|\,\mathcal{C}^+)  
                  \;=\; \sum_{k=1}^{N+1} \textrm{left}_N(k)\, e^{\beta k}\,. 
\end{equation}

Fix $\beta$ such that $0<\beta<1/(4\lambda_{d,C}^2)$.  Choose $\epsilon>0$ sufficiently small
so that $4\beta(\lambda_{d,C}+\epsilon)^2<1$, and choose $A>0$ so that 
\begin{equation}
   \label{eq.Clambdabd}
       |\mathcal{C}^{\pen}_m|  \;\leq \;  A\, (\lambda_{d,C}+\epsilon)^m
        \hspace{5mm}\hbox{for all }m\geq 1.
\end{equation}
From Equation (\ref{eq.Zdef2}) we have
\begin{equation}
   \label{eq.ZkNexp}
     Z_N(\beta\,|\,\mathcal{C}^+)  
       \;=\;  \sum_{k=1}^{N+1} \, \sum_{j=0}^{\infty} \, \frac{\beta^jk^j}{j!} \,\textrm{left}_N(k)  \,.
\end{equation}
For any $j\geq 0$ and $\ell\geq 1$, we have
\begin{equation}
   \label{eq.combin}
         \binom{\ell+j-1}{j} \; \geq \;  \frac{\ell^j}{j!}  \,.
\end{equation}
The left hand side of inequality (\ref{eq.combin}) is the number of $\ell$-tuples $(m_1,\ldots,m_{\ell})$ 
of nonnegative integers such that $m_1+\cdots+m_{\ell}=j$.  

We shall define a \textit{marked comb} (with $N$ edges) to be a comb $\rho$ in ${\cal{C}}_N^+$ 
that has a nonnegative integer $m(\rho;x)$ assigned to each $x$ in $D_s(\rho)\cup D_e(\rho)$.  
(We think of $m(\rho;x)$ as the number of ``marks'' on the the site or edge $x$.)
Let ${\cal C}_N^{(j)}$ be the set of all marked combs $\rho$ with $N$ edges (in $\mathbb{L}^d_+$) 
such that the total
number of marks on $D_s(\rho)\cup D_e(\rho)$ is $j$ (that is, $\sum_{x}m(\rho;x)=j$).
Using Equations (\ref{eq.DDsigma4}) and (\ref{eq.combin}), we obtain
\begin{equation}
   \label{eq.rhomark}
      \left| {\cal C}_N^{(j)} \right|  \;\geq \;  \sum_{k=1}^{N+1} 
       \binom{\lceil k/4\rceil +j-1}{j}  \,  \textrm{left}_N(k)  \; \geq \; 
        \sum_{k=1}^{N+1} \frac{(k/4)^j}{j!}   \,  \textrm{left}_N(k)   \,.      
\end{equation}      

Now consider an arbitrary marked comb $\rho\in \mathcal{C}^{(j)}_N$.
We modify $\rho$ to get a new comb as follows.  For each site $x\in D_s(\rho)$, append
the straight SAW with $2m(\rho;x)$ steps from $x$ to $x-2m(\rho;x)e^{(1)}$.  
For each edge $x\in D_e(\rho)$, call its endpoints $u_x$ and $v_x$.
Delete the edge $x$ and replace it by the $(2m(\rho;x)+1)$-step
SAW from $u_x$ to $u_x-m(\rho;x)e^{(1)}$ to $v_x-m(\rho;x)e^{(1)}$ to $v_x$.
The result is a comb $\rho[m]$ with $N+2j$ edges that contains the origin (indeed, it contains every 
site of $\rho$).  This mapping from $\mathcal{C}^{(j)}_N$
to $\mathcal{C}^{\pen}_{N+2j}$ is evidently one-to-one.  Thus
\begin{equation}
    \label{eq.CjnCN2j}
      \left|\mathcal{C}^{(j)}_N \right|   \;\leq \;  \left|\mathcal{C}^{\pen}_{N+2j} \right| 
        \;\leq \;   A\, (\lambda_{d,C}+\epsilon)^{N+2j}.
\end{equation}
(recall Equation (\ref{eq.Clambdabd})).
Now we have
\begin{eqnarray*}
     Z_N(\beta\,|\,\mathcal{C}^+)  
          & = &   \sum_{j=0}^{\infty} \, \sum_{k=1}^{N+1} \,\frac{(4 \beta)^j(k/4)^j}{j!} \,\textrm{left}_N(k)  
              \hspace{5mm}\hbox{(by Eq.\ (\ref{eq.ZkNexp}))}
              \\
      & \leq &    \sum_{j=0}^{\infty}  (4\beta)^j  \left| {\cal C}_N^{(j)} \right| 
       \hspace{25mm}\hbox{(by Eq.\ (\ref{eq.rhomark}))}
              \\
              & \leq &  \sum_{j=0}^{\infty}  (4\beta)^j A (\lambda_{d,C}+\epsilon)^{N+2j}
       \hspace{25mm}\hbox{(by Eq.\ (\ref{eq.CjnCN2j}))}
              \\
              & = & \frac{A\, (\lambda_{d,C}+\epsilon)^N}{1-4\beta ( \lambda_{d,C}+\epsilon)^2} 
\end{eqnarray*}
(recall that $4\beta ( \lambda_{d,C}+\epsilon)^2<1$, so the above series converges).
Therefore $\mathcal{F}(\beta\,|\,\mathcal{C}^+)\,\leq \, \log(\lambda_{d,C}+\epsilon)$.
Since $\epsilon$ can be made arbitrarily small, and since 
$\mathcal{F}(\beta\,|\,\mathcal{C}^+) \,\geq \, \mathcal{F}(0\,|\,\mathcal{C}^+) \,=\,\log \lambda_{d,C}$,
we conclude that $\mathcal{F}(\beta\,|\,\mathcal{C}^+)= \log \lambda_{d,C}$.  
As we have proved this for every positive $\beta$ less than $1/(4\lambda_{d,C}^2)$, it follows
that $\beta_c(\mathcal{C}^+)\geq 1/(4\lambda_{d,C}^2)$.
\hfill   $\Box$  

\medskip

\medskip
The next result says that in most $N$-edge combs, there are at least $\delta N$ side chains for 
some positive $\delta$ that is independent of $N$.  This is in the spirit of the pattern theorem 
for lattice trees in \cite{Ma99}, but the set of combs does not readily fit the assumptions needed 
for that theorem.  
Instead, our proof is similar to that of a result in \cite{LipW} which says that the class of lattice trees 
in which the number of sites of degree three or more is  $o(N/\log N)$ has growth constant 
equal to $\mu_d$.

\begin{lem}
   \label{lem.manychains}
Fix $d\geq 2$.  Let $\overline{\mathcal{C}}_N\langle [<m;\cdot;\cdot ]$ denote the set of combs in 
$\overline{\mathcal{C}}_N$ having fewer than $m$ side chains.  Then there exists $\delta>0$
such that 
\[    \limsup_{N\rightarrow\infty}  \left|  \overline{\mathcal{C}}_N[ < \delta N;\cdot;\cdot]  \right|^{1/N} 
   \,<\, \lambda_{d,C} \,.
\] 
\end{lem}
\textbf{Proof:}  
Choose $\epsilon>0$ small enough so that $\mu_d+2 \epsilon<\lambda_{d,C}$; this is possible
by Proposition \ref{prop.combneqsaw}.  Fix an ${\epsilon}$-dependent constant $A$ such that
$w_k \leq A (\mu_d+\epsilon)^k$ for all $k\geq 1$.

For any graph $\langle b;\vec{n};\vec{s}\rangle$ in $\mathcal{G}_N^C$, we have
\[
    \left|  \overline{\mathcal{C}}_N[ b;\vec{n};\vec{s} ]  \right|   \;\leq \;  
        w_{(\sum_{i=0}^bn_i)}  \prod_{j=1}^b w_{s_j} 
         \;  \leq \;  A^{b+1}  (\mu_d+\epsilon)^N\,.
\]
Therefore, for any possible $b$,
\[
   \sum_{\vec{n},\vec{s}}   \left|  \overline{\mathcal{C}}_N[ b;\vec{n};\vec{s} ]  \right|   \;\leq \;  
    \binom{N}{2b}  \,  A^{b+1}  (\mu_d+\epsilon)^N \,.
\]
It follows that for any $\delta \in (0,1/8)$ we have
\begin{eqnarray}
   \nonumber
     \left|  \overline{\mathcal{C}}_N[ < \delta N;\cdot;\cdot]  \right|   & = & 
          \sum_{b=0}^{\lceil \delta N \rceil -1}  
            \sum_{\vec{n},\vec{s}}   \left|  \overline{\mathcal{C}}_N[ b;\vec{n};\vec{s} ]  \right|            
      \\
      \nonumber  
   & \leq &  \sum_{b=0}^{\lceil \delta N \rceil -1}     \binom{N}{2b}  \,  A^{b+1}  (\mu_d+\epsilon)^N
      \\
      \nonumber 
        & \leq & \lceil \delta N \rceil \binom{N}{2\lceil \delta N \rceil}  
       A^{\delta N}  (\mu_d+\epsilon)^N.
\end{eqnarray}
From the above bound and Stirling's formula, we obtain
\begin{equation}
    \label{eq.StirlingCbd} 
 \limsup_{N\rightarrow\infty}  \left|  \overline{\mathcal{C}}_N[ < \delta N;\cdot;\cdot]  \right|^{1/N} 
 \;\leq \;  \left(  \frac{A^{\delta}}{(2\delta)^{2\delta} (1-2\delta)^{1-2\delta} }  \right)  \,(\mu_d+\epsilon).
\end{equation}
Since $\lim_{\delta\rightarrow 0+}A^{\delta}/[(2\delta)^{2\delta} (1-2\delta)^{1-2\delta} ] \,=1$, 
we can choose $\delta>0$ so that the right side of Equation (\ref{eq.StirlingCbd}) is less than 
$\mu_d+2\epsilon$, which in turn is less than $\lambda_{d,C}$.  
The proposition follows.
\hfill $\Box$

\medskip
In the notation of Equation (\ref{eq.betacQdef}), Corollary \ref{cor.negbetaquench}
tells us that $\beta_c^Q(\mathcal{C}^{\pen})$ and $\beta_c^Q(\mathcal{C}^+)$ are both nonnegative.
The next result tells us that these quenched critical points are both finite.
This is something that we cannot yet prove rigorously for our other branched models, and indeed which 
we believe to be \textit{false} for trees and animals in two dimensions  (Conjecture 
\ref{conj.2dnoads}). 
Proposition \ref{prop.negbetaQcomb} completes the proof
of Theorem \ref{thm-slopelim}.

\begin{prop}
   \label{prop.negbetaQcomb}
Let $X$ be either $+$ or $\pen$.
\\
(a) 
Assume $d\geq 3$.  Then $\beta^Q_c(\mathcal{C}^X)\,\leq \, 
\log \lambda_{d,C}$  and
\begin{equation}
    \label{eq.propQlimE3}   
     \lim_{\beta\rightarrow\infty}   \liminf_{N\rightarrow\infty}
       \frac{E_{\beta}^Q\left(\sigma(\rho)\,|\,\mathcal{C}^X_N \right)}{N}  \;=\;     
     \lim_{\beta\rightarrow\infty}   \limsup_{N\rightarrow\infty}
       \frac{E_{\beta}^Q\left(\sigma(\rho)\,|\,\mathcal{C}^X_N \right)}{N}  \;=\; 1.
\end{equation}
(b) Assume $d=2$.  Let $\delta$ be as defined in the statement of Lemma \ref{lem.manychains}.
Then $\beta^Q_c(\mathcal{C}^X)\,\leq \, \delta^{-1}\log \lambda_{2,C}$.  Also,
\begin{eqnarray}
   \label{eq.propQlimE2a}
     \lim_{\beta\rightarrow\infty}   \limsup_{N\rightarrow\infty}
       \frac{E_{\beta}^Q \left(\sigma(\rho)\,|\,\mathcal{C}^X_N \right)}{N}  & \leq & 1-\frac{\delta}{2}
         \hspace{5mm}\hbox{and}
   \\
     \label{eq.propQlimE2b}
     \lim_{\beta\rightarrow\infty}   \liminf_{N\rightarrow\infty}
       \frac{E_{\beta}^Q\left(\sigma(\rho)\,|\,\mathcal{C}^X_N \right)}{N}  & \geq & \delta \,.
\end{eqnarray}
\end{prop}

\bigskip
\noindent
\textbf{Proof of Proposition \ref{prop.negbetaQcomb}:}  
(a) 
Fix $d\geq 3$. Let $N\geq 1$.   
For each topology $g  =\langle b;\vec{n};\vec{s}\rangle   \in \mathcal{G}^C_N$, 
there is at least one $\rho\in \mathcal{C}^{X}_N[g]$ that lies entirely in the hyperplane $x_1=0$ 
(e.g.\ let the backbone be a straight SAW along the $x_2$-axis, and let each side chain be a
straight SAW parallel to the $x_3$-axis).  
Therefore, for $\beta\geq 0$, we have
\begin{equation}
    \label{eq.ZQcombbound1}
      e^{\beta (N+1)} \;\leq \;  Z_N(\beta \,|\,  \mathcal{C}^{X}[g])
        \;\leq \;   \left|\mathcal{C}^{X}[g] \right|  \,e^{\beta(N+1)} \,.
\end{equation}
Taking logs and then averaging over $g$ as in Equation (\ref{eq.FQbranchdef}),  we obtain
\begin{equation}
    \label{eq.ZQcombbound2}
    \frac{1}{N}\,\beta (N+1)  \;\leq \;  F^Q_N(\beta \,|\,  \mathcal{C}^{X})
        \;\leq \;   F^Q_N(0 \,|\,\mathcal{C}^{X})  +  \frac{1}{N}  \,\beta(N+1) \,.
\end{equation}
Since 
\begin{equation}
   \label{eq.FQ0bound}
       F^Q_N(0 \,|\,\mathcal{C}^{X})   \;\leq \;   \frac{1}{N}\log   \left| \mathcal{C}^{X}_N\right| \,,
\end{equation}
it follows from the left inequality of (\ref{eq.ZQcombbound2}) that
\[
      F^Q_N(\beta \,|\,  \mathcal{C}^{X})  \,-\, F^Q_N( 0\,|\,  \mathcal{C}^{X})   \;\geq \;  \beta \,-\,
         \frac{1}{N}\log   \left| \mathcal{C}^{X}_N\right|\,.
\]
Thus if $\beta > \log \lambda_{d,C}$, we see that 
$\liminf_{N\rightarrow\infty} F^Q_N(\beta \,|\, \mathcal{C}^{X})  \,-\, F^Q_N( 0\,|\,  \mathcal{C}^{X}) 
 \,>\,0$.     This proves that $\beta_c^Q(\mathcal{C}^X) \,\leq \, \log \lambda_{d,C}$.   
      
Now apply Lemma \ref{lem.convderiv}  to Equation (\ref{eq.ZQcombbound2}) with 
$f(\beta)=F_N^Q( \beta \,|\,\mathcal{C}^X)$.  
Using Equation (\ref{eq.FQ0bound}), we obtain 
\[   \frac{N+1}{N} \,-\,   \frac{ \log   \left| \mathcal{C}^{X}_N\right| }{N\beta}    
    \;\leq \;   \frac{d}{d\beta}\,F_N^Q( \beta \,|\,\mathcal{C}^X) \;\leq \;  \frac{N+1}{N} \,.
\]
The result (\ref{eq.propQlimE3}) follows from these bounds and Equation  (\ref{eq.expvisitsquench}).

\smallskip
\noindent
(b)     Assume $d=2$.
Consider  a topology $g=\langle b;\vec{n};\vec{s}\rangle \in \mathcal{G}^C_N$.
Observe that every comb in $\mathcal{C}^X[g]$ has at least $b+2$
sites in its backbone.  Therefore there is a comb $\rho^*\in \mathcal{C}^X[g]$ such that 
$\sigma(\rho) \; \geq \; b+2$ (the backbone lies along the $x_2$-axis, and each side chain 
is perpendicular to it).  
Also observe that in any comb $\rho$ of  $\mathcal{C}^X[g]$,  for each site $v$ of degree 3 in $\rho$,
at least one of the three adjacent sites of $v$ in $\rho$ does not lie in the line $x_1=0$; 
call this site $g(v)$.  
 There can be at most one other site $w$ of degree 3 such that $g(v)=g(w)$.  
This shows that at least $b/2$ sites of $\rho$ are not in the line $x_1=0$.  
That is, $\sigma(\rho)\leq N+1-\frac{b}{2}$ for every $\rho \in \mathcal{C}^X[g]$.  These 
considerations show that
\begin{equation}
    \label{eq.ZQcombbound3}
       e^{\beta (b+2)} \;\leq \;  Z_N(\beta \,|\,  \mathcal{C}^{X}[g])
        \;\leq \;   \left|\mathcal{C}^{X}[g] \right|  \,e^{\beta(N+1-b/2)}
         \hspace{5mm} \hbox{ for }\beta\geq 0\,.
\end{equation}
      
Let $\delta$  be as defined in the statement of Lemma \ref{lem.manychains}.      
Let $\mathcal{C}^X_N(<m)$  be the set of combs in $\mathcal{C}^X_N$
that have fewer than $m$ side chains.  
Let $\epsilon_N=| \mathcal{C}^X_N(<\delta N)|/| \mathcal{C}^X_N|$, so that  
$\lim_{N\rightarrow\infty}\epsilon_N=0$ by Lemma \ref{lem.manychains}.
Observe that for each $g\in \mathcal{G}_N^C$, the set $\mathcal{C}^X[g]$ is either contained in
or disjoint from $\mathcal{C}^X_N(<\delta N)$.  
Thus from Equation (\ref{eq.ZQcombbound3}) we obtain for all $\beta\geq 0$ that
\begin{equation}
   \label{eq.FQcomblowbd}
   F_N^Q(\beta \,|\,\mathcal{C}^X) \; \geq \; \frac{1}{N}\,
   \frac{\left |\mathcal{C}^X_N \setminus \mathcal{C}^X_N(<\delta N) \right|}{
      |\mathcal{C}^X_N|} 
   \,\beta \delta N  \;\;  =  \;  (1-\epsilon_N)\delta \beta
\end{equation}
and   
\begin{eqnarray}
   \nonumber
   F_N^Q(\beta \,|\,\mathcal{C}^X)     & \leq & 
    \frac{1}{N}\,\frac{\left|\mathcal{C}^X_N \setminus \mathcal{C}^X_N(<\delta N) \right|
        }{ |\mathcal{C}^X_N|} 
   \,\left(\log |\mathcal{C}_N^X| \,+\, \beta N\left(1-\frac{\delta}{2}\right)\,+\,\beta  \right)
   \\
   \nonumber
   & & \hspace{15mm}
    \,+\,  \frac{1}{N}\, \frac{| \mathcal{C}^X_N(<\delta N) |}{|\mathcal{C}^X_N|} 
   \,\left(  \log |\mathcal{C}_N^X| + \beta (N+1)\right)   
   \\
   \label{eq.FQcombupbd}
   & \leq & \frac{\log |\mathcal{C}_N^X| }{N}  \,+\beta \left(1-\frac{\delta}{2}\right) \,+\,\frac{\beta}{N}
   \,+\,  \epsilon_N \,\frac{N{+}1}{N}\,\beta\,.
\end{eqnarray}    
By the above inequalities, Equation (\ref{eq.expvisitsquench}), 
and Lemma \ref{lem.convderiv}, we obtain
\[
            (1-\epsilon_N)\delta \,-\,\frac{\log |\mathcal{C}_N^X| }{N\,\beta} 
            \;\leq \;  \frac{E_{\beta}^Q(\sigma(\rho)\,|\,\mathcal{C}^X_N)}{N} 
              \;\leq \;  1\,-\,\frac{\delta}{2}\,+\,\epsilon_N\,+ \,\frac{1+\epsilon_N}{N}   
\] 
for every $\beta>0$.       
The results of Equations (\ref{eq.propQlimE2a})  and (\ref{eq.propQlimE2b}) follow.  To prove
the bound on $\beta_c^Q(\mathcal{C}^X)$,
observe that Equations (\ref{eq.FQ0bound}) and (\ref{eq.FQcomblowbd}) imply that
\[      
       F_N^Q(\beta \,|\,\mathcal{C}^X) \,-\,  F_N^Q(0 \,|\,\mathcal{C}^X)   \; \geq \; 
  (1-\epsilon_N)\delta \beta \,- \,\frac{1}{N} \,\log \left|\mathcal{C}^X_N  \right| 
  \hspace{5mm}\hbox{for }\beta\geq 0.
\]         
Since the limit as $N\rightarrow\infty$ of the right-hand side is strictly positive 
whenever $\delta\beta>  \log\lambda_{2,C}$,
we conclude that $\beta_c^Q(\mathcal{C}^X)\,\leq \,\delta^{-1}\log \lambda_{2,C}$.       
\hfill $\Box$

\section{Properties of rings and trees}
   \label{sec-ringtree}
   
\subsection{Adsorption of quenched knotted rings}    
    \label{sec-ringQfin}
    
In this subsection we prove that  adsorption for rings with quenched knot type occurs at 
a finite value of $\beta$.  That is, we prove $\beta_c^Q(\mathcal{R}^X)<\infty$.

\begin{prop}
   \label{prop.ringQfin}
Let $X$ be either $+$ or $\pen$.  There exists a positive number $\overline{\beta}$ such that 
$\liminf_{N\rightarrow\infty}\left(F^Q_N(\beta\,|\,\mathcal{R}^X) \,-\,F^Q_N(0\,|\,\mathcal{R}^X)\right)
 \,>\,0$ for all $\beta>\overline{\beta}$.
\end{prop}   
\textbf{Proof:}
Consider the 30-step self-avoiding polygon $\phi$ in $\mathbb{Z}^3$
shown in Figure \ref{fig.knotQ} where we have translated the site A to be at the origin.
Then H is $(1,0,0)$ and C is (0,0,4).
Let $\psi$ be the 29-step SAW that starts at A, ends at H, and uses every edge of $\phi$ except
the edge joining A to H.  Let $Q$ be the box $[0,3]\times[-2,1]\times [0,4]$, which contains every 
site of $\psi$.  All of the labelled sites in Figure \ref{fig.knotQ} except E  are  on the boundary of $Q$.

\setlength{\unitlength}{1mm}
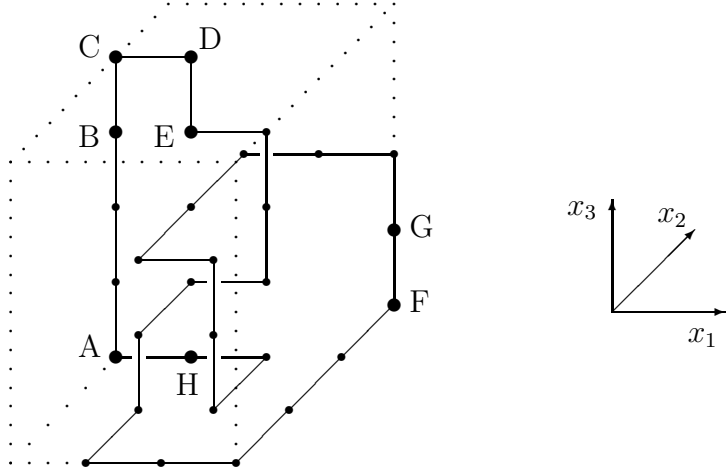
\begin{figure}[h]
\begin{center}
\begin{picture}(100,70)(0,0)
\put(10,0){\line(1,0){20}}
\put(10,0){\line(1,1){7}}
\put(30,0){\line(1,1){21}}
\put(17,7){\line(0,1){10}}
\put(51,21){\line(0,1){20}}
\put(31,41){\line(1,0){2}}
\put(35,41){\line(1,0){16}}
\put(17,27){\line(1,1){14}}
\put(17,27){\line(1,0){10}}
\put(27,7){\line(0,1){20}}
\put(27,7){\line(1,1){7}}
\put(14,14){\line(1,0){2}}
\put(18,14){\line(1,0){8}}
\put(28,14){\line(1,0){6}}
\put(14,14){\line(0,1){40}}
\put(14,54){\line(1,0){10}}
\put(24,44){\line(0,1){10}}
\put(24,44){\line(1,0){10}}
\put(34,24){\line(0,1){20}}
\put(24,24){\line(1,0){2}}   
\put(28,24){\line(1,0){6}}
\put(17,17){\line(1,1){7}}
\multiput(10,0)(10,0){3}{\circle*{1.2}}
\multiput(37,7)(7,7){2}{\circle*{1.2}}
\multiput(31,41)(10,0){3}{\circle*{1.2}}
\multiput(51,21)(0,10){2}{\circle*{1.6}}
\put(24,34){\circle*{1.2}}
\multiput(17,27)(10,0){2}{\circle*{1.2}}
\multiput(27,7)(0,10){2}{\circle*{1.2}}
\multiput(14,14)(10,0){2}{\circle*{1.6}}
\put(34,14){\circle*{1.2}}
\multiput(14,24)(0,10){2}{\circle*{1.2}}
\multiput(14,44)(0,10){2}{\circle*{1.6}}
\multiput(24,44)(0,10){2}{\circle*{1.6}}
\multiput(34,24)(0,10){3}{\circle*{1.2}}
\multiput(17,17)(7,7){2}{\circle*{1.2}}
\put(17,7){\circle*{1.2}}
\multiput(0,0)(2.5,0){4}{\circle*{0.5}}
\multiput(0,0)(0,2.5){16}{\circle*{0.5}}
\multiput(0,40)(2.5,0){12}{\circle*{0.5}}
\multiput(30,2.5)(0,2.5){16}{\circle*{0.5}}
\multiput(0,40)(1.75,1.75){12}{\circle*{0.5}}
\multiput(34,44)(1.75,1.75){10}{\circle*{0.5}}
\put(32.5,42.5){\circle*{0.5}}
\multiput(21,61)(2.5,0){13}{\circle*{0.5}}
\multiput(51,43.5)(0,2.5){7}{\circle*{0.5}}
\multiput(2.5,2.5)(2.5,2.5){5}{\circle*{0.5}}
\put(9,14){A}
\put(22,9){H}
\put(19,42){E}
\put(9,42){B}
\put(9,54){C}
\put(25,55){D}
\put(53,20){F}
\put(53,30){G}
\put(80,20){\vector(1,0){15}}
\put(80,20){\vector(0,1){15}}
\put(80,20){\vector(1,1){11}}
\put(90,16){$x_1$}
\put(86,32){$x_2$}
\put(74,33){$x_3$}
\end{picture}
\end{center}
\caption{\label{fig.knotQ}   A   30-step self-polygon $\phi$ whose knot type is the trefoil knot $3_1$.
The coordinate directions are indicated on the right.
The dotted lines describe a box ${Q}$ that contains the polygon.
Setting A=(0,0,0), we have  C=(0,0,4), D=(1,0,4), F=(3,1,0), G=(3,1,1), and 
 ${Q}=[0,3]\times [-2,1]\times [0,4]$.
}
\end{figure}

For a site $x$ on a self-avoiding polygon $\rho$, we say that $(\psi,Q)$ occurs on $\rho$ at $x$ 
if the intersection of $\rho$ with the translated cube $Q+x$ is equal to the 
the translated SAW $\psi+x$. 
For each $N\geq k\geq  1$, let $\mathcal{R}^X_{N,>k}$ be the set of $\rho\in \mathcal{R}_N^X$ 
such that $(\psi,Q)$ occurs at more than $k$ distinct sites of $\rho$.
It follows from Kesten's Pattern Theorem (see \cite{Ke} or Theorem 7.2.3 of \cite{MS})
that there exists an $\epsilon>0$ such that $\limsup_{N\rightarrow\infty}|\mathcal{R}_N^X\setminus
\mathcal{R}^X_{N,>\epsilon N}|^{1/N}<\mu_3$, and hence that
\begin{equation}
    \label{eq.limRoverR}
       \lim_{N\rightarrow\infty}  \frac{\left|\mathcal{R}_{N,>\epsilon N}^X\right|}{
         \left|\mathcal{R}_{N}^X\right|}   \;=\; 1 \,.
\end{equation}

Let $\frak{K}_N$ be the collection of knot types $K$ such that 
$\mathcal{R}^X_N[K] \cap \mathcal{R}^X_{N,>\epsilon N} \neq \emptyset$.
The following assertion is key to the proof.  
\begin{verse}
    \underline{Claim}:  If the knot type $K$ is in $\frak{K}_N$, then there is a $\rho$
    in $\mathcal{R}^X_N[K]$ such that $\sigma(\rho) \geq \epsilon N$.
\end{verse}
Suppose that the claim is true.  It follows that for every $K\in \frak{K}_N$ we have 
$Z_N(\beta \,|\,\mathcal{R}^X[K]) \,\geq \,\exp(\beta \epsilon N)$ for every positive $\beta$.
Therefore 
\begin{eqnarray*}
   F_N^Q( \beta \,|\,\mathcal{R}^X)  & \geq & \frac{1}{N}  \sum_{K\in \frak{K}_N}
     \frac{| \mathcal{R}^X_N[K]|}{| \mathcal{R}^X_N |}   
      \log Z_N(\beta \,|\,\mathcal{R}^X[K]|)
      \\
     &  \geq &  \frac{1}{N}  \sum_{K\in \frak{K}_N}
     \frac{| \mathcal{R}^X_N[K] \cap \mathcal{R}_{N,>\epsilon N}^X  |}{| \mathcal{R}^X_N| }   
    \, \beta \epsilon N
      \\
    & = &    \frac{| \mathcal{R}_{N,>\epsilon N}^X  |}{| \mathcal{R}^X_N| }   \, \beta \epsilon  
    \\
    & \rightarrow &  \beta \epsilon \hspace{12mm}\hbox{as $N\rightarrow \infty$, by Equation 
     (\ref{eq.limRoverR}).}
\end{eqnarray*}
Since $\log \mu_3\geq \limsup_{N\rightarrow\infty} F_N^Q(0\,|\,\mathcal{R}^X)$, the above
inequalities imply that   $\liminf_{N\rightarrow\infty}\left(F^Q_N(\beta\,|\,\mathcal{R}^X) \,-\,
F^Q_N(0\,|\, \mathcal{R}^X)\right) \,>\,0$ whenever  $\beta>(\log \mu_3)/\epsilon$.  
This would prove the proposition.

It remains only to prove the Claim.  First, for each integer $t\geq 1$, 
we shall construct a special $(28t)$-step self-avoiding polygon $\Phi^{\langle t\rangle}$.  
The idea is that we will join $t$ copies of $\phi$ together, one on top of the other, and then 
slightly truncate the very top. 
For each integer $j$, let $\phi^{[j]}=\phi+4je^{(3)}$.   Thus $\phi^{[j]}$ is the result of translating of the
polygon $\phi$ of Figure \ref{fig.knotQ} so that site A has been moved to $(0,0,4j)$.
Let $u_j$ be the edge joining $(0,0,4j)$ to $(1,0,4j)$.  Then $u_j$ is the unique edge 
in $\phi^{[j-1]}\cap \phi^{[j]}$.
Let $\Theta^{\langle t\rangle}$ be the subgraph of $\mathbb{Z}^3$ obtained by 
the union of $\{\phi^{[j]}:0\leq j<t\}$
followed by the deletion of the edges $\{u_j:  1\leq j<t\}$.   
Then we see that $\Theta^{\langle t\rangle}$ is a self-avoiding polygon with $28t+2$ steps.  
Finally, to get $\Phi^{\langle t\rangle}$ from $\Theta^{\langle t\rangle}$, 
replace the 3-step SAW from $(0,0,4t-1)$ to $(0,0,4t)$ to 
$(1,0,4t)$ to $(1,0,4t-1)$ by the 1-step SAW from $(0,0,4t-1)$ to $(1,0,4t-1)$.  
(In the suitable translation of Figure \ref{fig.knotQ}, we are replacing the SAW BCDE by the SAW BE.) 
The result $\Phi^{\langle t\rangle}$ is a polygon in $\mathcal{R}^+_{28t}$. 
Since $\phi$ is a trefoil knot ($3_1$), the knot type of $\Phi^{\langle t\rangle}$ is the product of $t$
trefoils, written $(3_1)^t$.
Moreover, $\sigma(\Phi^{\langle t\rangle})=4t$.

Now assume that $K\in \frak{K}_N$ and let 
$\xi\in \mathcal{R}^X_N[K] \cap \mathcal{R}^X_{N,>\epsilon N}$.  Let $t=\lceil \epsilon N\rceil$, and 
let $x_1,x_2,\ldots,x_t$ be distinct sites of $\xi$ at which $(\psi,Q)$ occurs.   
Let $\Xi$ be obtained from $\xi$ by replacing each 29-step SAW $\psi+x_i$ by the one-step SAW from 
$x_i$ to $x_i+e^{(1)}$.  Then $\Xi$ is a self-avoiding polygon having $N-28t$ steps. 
Let $\hat{K}$ be the knot type of $\Xi$.   Then we must have $K \,=\, (3_1)^t\sharp \hat{K}$.

\setlength{\unitlength}{1mm}
\begin{figure}[h]
\begin{center}
\begin{picture}(110,70)
\put(0,0){\line(0,1){40}}
\put(10,0){\line(0,1){40}}
\put(19,12){\line(0,1){37}}
\multiput(0,0)(0,10){5}{\line(1,0){10}}
\multiput(10,0)(0,10){5}{\line(1,1){9}}
\put(0,40){\line(1,1){9}}
\put(9,49){\line(1,0){10}}
\put(0,59){$\overbrace{\hbox{\hspace{20mm}}}$}
\put(8,63){\Large{$\Phi^{(t)}$}}
\put(19,9){\line(1,0){3}}
\put(19,12){\line(1,0){3}}
\put(19,9){\circle*{1.3}}
\put(19,12){\circle*{1.3}}
\put(22,9){\circle*{1.3}}
\put(22,12){\circle*{1.3}}
\multiput(22,14)(0,3){9}{\circle*{.7}}
\multiput(26,9)(3,0){11}{\circle*{.7}}
\multiput(23,40)(3,3){3}{\circle*{.7}}
\multiput(38,55)(4,0){8}{\circle*{.7}}
\multiput(56,9)(3,3){5}{\circle*{.7}}
\multiput(71,24)(0,4){8}{\circle*{.7}}
\multiput(22,24)(5,0){6}{\circle*{.7}}
\multiput(33,50)(5,0){6}{\circle*{.7}}
\multiput(53,14)(0,5){6}{\circle*{.7}}
\multiput(33,25)(0,5){5}{\circle*{.7}}
\multiput(60,26)(0,5){6}{\circle*{.7}}
\multiput(25,19)(5,5){5}{\circle*{.7}}
\multiput(40,13)(5,5){6}{\circle*{.7}}
\put(23,59){$\overbrace{\hbox{\hspace{48mm}}}$}
\put(15,7){F}
\put(15,12){G}
\put(21,4){F*}
\put(23,12){G*}
\put(45,63){\Large{$\Xi^*$}}
\put(90,10){\vector(1,0){15}}
\put(90,10){\vector(0,1){15}}
\put(90,10){\vector(1,1){11}}
\put(100,6){$x_1$}
\put(96,22){$x_2$}
\put(84,23){$x_3$}
\end{picture}
\end{center}
\caption{\label{fig.knotQPsi}   Conclusion of the proof of Proposition \ref{prop.ringQfin}:
construction of the self-avoiding polygon $\rho$ by joining $\Phi^{(t)}$ to $\Xi^*$.
The coordinate directions are indicated on the right.
In this sketch, $t=4$.  The stack of boxes at the left represent the boxes $Q+4je^{(3)}$, 
which contain the polygons $\phi^{[j]}$,  for $j=0,\ldots,t-1$.  The box $Q$ is the same as
in Figure \ref{fig.knotQ}, as are the points F=(3,1,0) and G=(3,1,1).
The other labelled points are  F*=(4,1,0) and G*=(4,1,1).
The left face of each box $Q+4je^{(3)}$ lies in the surface $x_1=0$.
The polygon $\Xi^*$ is indicated very imprecisely by a cloud of dots.
}
\end{figure}
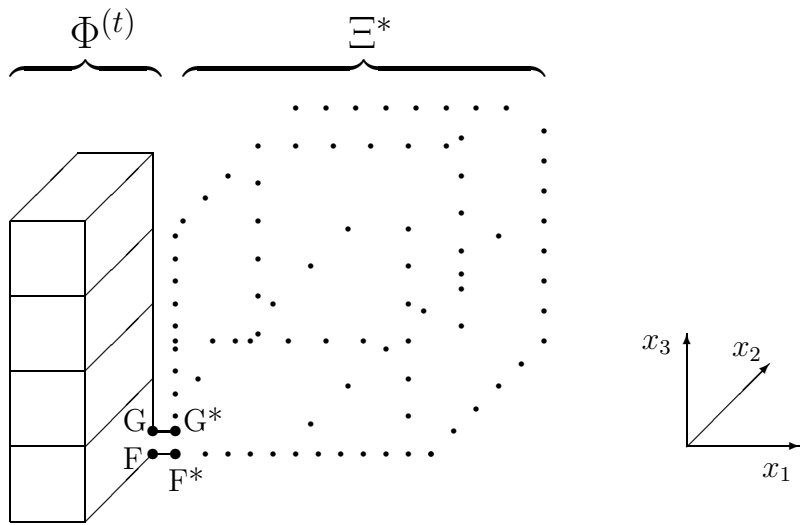

Let $\Xi^*$ be obtained by translating and rotating $\Xi$ so that it is contained in the half-space
$\{x:x_1\geq 4\}$ and has an edge from (4,1,0) to (4,1,1).  Then $\Xi^*$ has knot type $\hat{K}$.
Note that (3,1,0) and (3,1,1) are joined by an edge in $\Phi^{\langle t\rangle}$ 
(sites F and G in Figure \ref{fig.knotQ}).
Finally, let $\rho$ be obtained from $\Phi^{\langle t\rangle}\cup\Xi^*$ by adding 
the edges from (3,1,0) to (4,1,0) 
and from (3,1,1) to (4,1,1), and deleting the edges from (3,1,0) to (3,1,1) and from (4,1,0) to (4,1,0).
See Figure \ref{fig.knotQPsi}.
Then $\rho$ is a self-avoiding polygon with $28t+(N-28t)$ edges and has knot type 
$(3_1)^t\sharp \hat{K}$.  Thus $\rho\in \mathcal{R}_N^+[K]$.  Finally, 
$\sigma(\rho)=\sigma(\Phi^{\langle t\rangle}) =4t \geq 4\epsilon N$, and the Claim follows.

The proof of the proposition is now complete.
\hfill $\Box$

\subsection{The number of polymers with a given topology}
   \label{sec-sizes}

In this section we investigate the commonest topologies in specific models.  Since our focus will
be on exponential growth rates of cardinalities, we shall only deal with the penetrable models
(which have the same growth rates as the impenetrable models).

For a polymer model $\mathcal{P}$, 
we write $\mathcal{P}_N[\tau]$ for the 
set of configurations in $\mathcal{P}^{\pen}_N$ with topology $\tau$ (whether $\tau$ is 
an abstract graph or a knot type).   
Let $M_N(\mathcal{P})$ be the number of configurations in the 
commonest topology in $\mathcal{P}^{\pen}_N$:
\begin{equation}
   \label{eq.defMaxN}
      M_N(\mathcal{P})  \;=\; \max_{\tau}\, \left |\mathcal{P}_N[\tau] \right|   
\end{equation}
where $\tau$ ranges over $\mathcal{G}_N$ or $\mathcal{G}_N^C$ or the set of knot types, as 
appropriate.

Now we consider  whether the commonest topology is exponentially rare.  
This is a situation where ring polymers  differ significantly from branched polymers.
First we show that in $\mathcal{R}_N$, the commonest knot type is \underline{not} exponentially rare.
\begin{prop}[Rings]
    \label{prop.MNeqmu}
The limit  $\lim_{N\rightarrow\infty}|M_N(\mathcal{R})|^{1/N}$ 
exists and equals $\mu_3$.
\end{prop}
\textbf{Proof:}   Recall $r_N=|\overline{\mathcal{R}}_N |$.
Let $\epsilon >0$, and fix an even $M$ such that   $r_M> 2\,(\mu_3-\epsilon)^M$. 
 (Such an $M$ exists
because $\lim_{m\rightarrow \infty}r_{2m}^{1/2m}=\mu_3$.)

Let $\mathcal{K}$ be the set of all prime knots that can occur as the knot type of a
self-avoiding polygon of at most $M$ steps.  We shall write this set of knots as $\mathcal{K}\,=\,
\{\pi_1,\ldots,\pi_p\}$.
For each $N$, let $\mathcal{R}_N((\mathcal{K}))$ be the set of all  polygons in $\mathcal{R}^{\pen}_N$
whose knot type is a product of members in $\mathcal{K}$ (with repetitions allowed).
That is, each polygon in $\mathcal{R}_N((\mathcal{K}))$ has a knot type of the form
\begin{equation}
   \label{eq.knotcomp}
      \pi_1^{m_1}\,\sharp\, \pi_2^{m_2} \, \sharp \,\cdots\, \sharp \,  \pi_p^{m_p}   
\end{equation}
for some choice of nonnegative integers $m_1,\ldots,m_p$; in particular, these $p$ integers determine
the knot type.

We now recall the standard concatenation of self-avoiding polygons, specialized to 
$\mathbb{L}^3$ (\cite{Ham}; also described in Theorem 3.2.3 of \cite{MS}).  
For  $i=2,3$ and even $M$, 
let $\mathcal{Q}_i[M]$ be the set of polygons in $\overline{\mathcal{R}}_M$ that contain the 
bond from the origin to $e^{(i)}$.  By symmetry, $|\mathcal{Q}_2[M]|=|\mathcal{Q}_3[M]|\geq r_M/2$.
For even $N$, let $\alpha$ be a polygon in $\overline{\mathcal{R}}_N$. 
Let $v$ be the lexicographically largest site of $\alpha$.  Let $I$ be the larger of the two indices
$i\in \{1,2,3\}$  such that $\alpha$ contains the bond from $v$ to $v-e^{(i)}$.  We can now 
concatenate $\alpha$ to any $\beta\in \mathcal{Q}_I[M]$ by taking 
the union of $\alpha$ and $\beta+v+e^{(1)}-e^{(I)}$, and then performing an ``exclusive or''
with the 
edges in the square whose corners are $v$, $v-e^{(I)}$, $v+e^{(1)}-e^{(I)}$, and $v+e^{(1)}$.
The result is in $\overline{\mathcal{R}}_{N+M}$.
This operation leads to a proof that $r_{N+M}\geq r_N \,r_M/2$.  
Notice also that the knot type of the concatenation is the 
product of the knot types of $\alpha$ and $\beta$.
By repeated concatenation, this reasoning shows that for every positive integer $t$
we have 
\begin{equation}
    \label{eq.Rtmineq}
      | \mathcal{R}_{tM}((\mathcal{K}))| \;\geq \; \frac{1}{2^{t-1}}\,(r_M)^t
       \;>\;  (\mu_3-\epsilon)^{tM}\,.
\end{equation}
More generally, since $|  \mathcal{R}_{N+2}((\mathcal{K})) |\,\geq \, | \mathcal{R}_N((\mathcal{K}))|$
for every $N$ (the argument is the same as for the proof given for 
$r_{N+2}\geq r_N$, which 
is Equation (3.2.3) of \cite{MS}), we find from Equation (\ref{eq.Rtmineq}) that there is a constant 
$A$ such that
\begin{equation}
    \label{eq.RNineq}
       | \mathcal{R}_{N}((\mathcal{K}))| \;\geq \; A\, (\mu_3-\epsilon)^{N}   \hspace{6mm}
         \hbox{for every even $N$}
\end{equation}
(for example, we can take $A=(\mu_3-\epsilon)^{-M}$).

Next, we claim that for any $N$-step polygon whose knot type is Equation (\ref{eq.knotcomp}), 
we must 
have $m_1+m_2+\cdots +m_p\leq N/6$.  We shall prove this using properties of \textit{bridge numbers}
from knot theory.  In a given regular projection of a knot,  an overpass is a subarc of the knot that 
only contains overcrossings but no undercrossings.  The bridge number of a knot type $K$ is the 
smallest possible number of disjoint maximal overpasses among all projections of the knot, and is 
denoted $b(K)$.  (The bridge number of the unknot is defined to be 1.)
Schubert \cite{Sch} proved that $b(K_1\,\sharp\,K_2) \,=\, b(K_1)+b(K_2)-1$ for any knots $K_1$ and 
$K_2$, and that $b(K)\geq 2$ for every nontrivial knot.  Therefore the knot given by
Equation (\ref{eq.knotcomp}) has bridge number at least $m_1+m_2+\cdots +m_p$.  
Now use the fact that the bridge number of an $N$-step self-avoiding polygon is 
at most $N/6$, which is Theorem 1 of \cite{DET}.  The claim follows.

Since the number of $p$-tuples $(m_1,\ldots,m_p)$ of nonnegative integer solutions of 
the inequality
$m_1+m_2+\cdots+m_p<N$ is at most $N^p$, we see that there are at most $N^p$ knot types 
in $\mathcal{R}_N((\mathcal{K}))$.  Therefore the number of polygons having 
the commonest knot type in $\mathcal{R}_N((\mathcal{K}))$ satisfies
\begin{equation}
   \label{eq.knotmaxbound}
     M_N(\mathcal{R})   \;\geq \;  N^{-p} \,|\mathcal{R}_N((\mathcal{K}))|  \;\geq \;
      A N^{-p}\, (\mu_3-\epsilon)^N
\end{equation}
for every even $N$.   Therefore
\[  \mu_3 \;\geq \;    \limsup_{N\rightarrow\infty} M_N(\mathcal{R})^{1/N} \;\geq \; 
    \liminf_{N\rightarrow\infty} M_N(\mathcal{R})^{1/N} \;\geq \;\mu_3-\epsilon.
\]
Since $\epsilon$ is arbitrary, it follows that the limit exists and equals $\mu_3$.
This proves the proposition.
\hfill $\Box$

\bigskip

We note that Proposition \ref{prop.MNeqmu}
disproves a conjecture of Janse van Rensburg and Rechnitzer \cite{RR}
which said that for all sufficiently large $N$, the commonest knot type is a product of some 
number of left-handed  and right-handed trefoils.  
To see why the conjecture cannot be true, observe that 
all such knots lack a figure-8 knot (called $4_1$ in the standard knot table), 
for example, and hence  polygons with only trefoils
are contained in an exponentially small subset of all self-avoiding polygons, by 
Corollary 2.5 of \cite{SSW}.  This contradicts Proposition \ref{prop.MNeqmu}.
This argument also shows that the knot type $K$ that maximizes $|\mathcal{R}_N[K]|$ 
must become more complex as $N$ increases.

For the same reason, for any {\em fixed} knot type $K$, the size of $\mathcal{R}_N[K]$ is
exponentially smaller than $|\overline{\mathcal{R}}_N|$.  This leads to an open question of 
long standing.
Define $\mu[K]$ to be $\lim_{N\rightarrow\infty}  |\mathcal{R}_N[K]|^{1/N}$ if this limit exists.
This limit exists when $K$ is the unknot $0_1$ \cite{Pip}, essentially because the product of two 
unknots is an unknot.  It is not known whether this limit exists for any other knot type $K$, but
it is conjectured to exist and equal $\mu[0_1]$ for every fixed $K$.  
(It is not hard to show that the lim inf is greater than or equal to $\mu[0_1]$ \cite{Whi}.)
See \cite{OTJW} for intuition and numerical support for this conjecture.

In contrast to Proposition \ref{prop.MNeqmu}, it turns out that the largest tree topology is
an exponentially small fraction of all trees.  To see this, we shall use the following result.

\begin{prop}[Trees]
    \label{prop.MNtree}
$M_N(\mathcal{T})  \,\leq \, N (2d)(2d-1)^{N-2}$ for every $N\geq 2$.
\end{prop}
\textbf{Proof:}   Let $g$ be an abstract tree in $\mathcal{G}_N^T$.  We need to bound the 
number of ways to embed $g$ into $\mathbb{L}^d$ so that the origin is one of its sites.

Suppose $\rho$ is a lattice tree in $\mathcal{T}_N^{\pen}$ that is isomorphic to $g$.
Let $V(\rho)$ (respectively, $V(g)$) be the set of sites of $\rho$ (respectively, $g$). 
In particular, $V(\rho)$ is a subset of $\mathbb{Z}^d$ that contains the origin.
Then there is a bijection $\Psi$ from $V(g)$ to $V(\rho)$ 
such that $\Psi(v)$ and $\Psi(w)$ are joined by an edge of $\rho$ if and only if $v$ and $w$ are 
joined by an edge of $g$.

Let $v_0\in V(g)$ such that $\Psi(v_0)$ is the origin.  For each $v\in V(g)\setminus\{v_0\}$, 
the \textit{parent} of $v$ is the unique neighbour of $v$ that is on the path in 
$g$ from $v$ to $v_0$, and is denoted pa$(v)$. 
Also for each $v\in V(g)\setminus\{v_0\}$, let $\Theta(v)$ be the unit vector 
$\Psi(v)-\Psi(\text{par}(v))$.
Then the function $\Theta$ has the property that 
for any two neighbours $v_1$ and $v_2$ of $v_0$ in $g$,
we must have $\Theta(v_1)\neq \Theta(v_2)$.  
Similarly,  for every $v\in V(g)\setminus \{v_0\}$ such that par$(v)\neq v_0$, we must have 
$\Theta(v)\neq -\Theta(\text{par}(v))$  (since $\Psi(v)\neq \Psi(\text{par}(\text{par}(v)))$).

Given $g$, the choice of $v_0$ and the function $\Theta$ completely 
determine the lattice tree $\rho$.
There are $N$ possible choices for $v_0$, and then the properties described in the preceding
paragraph imply that  there are at most $(2d)(2d-1)^{N-2}$ 
possibilities for $\Theta$.  This proves that $\left|\mathcal{T}_N[g]\right|\leq (2d)(2d-1)^{N-2}$,
and the proposition follows.
\hfill  $\Box$

\begin{cor}[Trees]
    \label{cor.MNtree}
For dimensions $d=2,\ldots,12$, and for sufficiently large $d$, we have
\[    \limsup_{N\rightarrow\infty}M_N(\mathcal{T})^{1/N}  \;\leq \; 2d-1  \;<\,  \lambda_{d,T} \,.
\]      
\end{cor}    

We expect that Corollary \ref{cor.MNtree} holds for every dimension $d$.

\medskip
\noindent
\textbf{Proof:}
The first inequality follows immediately from Proposition \ref{prop.MNtree}.    
The best lower bounds on growth constants in specific dimensions are typically based on
exact enumeration.
For 2 and 3 dimensions, we shall use the bound
\begin{equation}
   \label{eq.Ma95}
      \lambda_{d,T}  \;\geq \;   \left(  d(2N)^{(d-1)/d} t_N\right)^{1/N}
\end{equation}
from Theorem 1.1 of \cite{Ma95}.
Using the known values $t_{15}=$ 338,158,676 for the square lattice and 
$t_{11}=$ 248,160,162 for the simple cubic lattice \cite{WTG}, 
we obtain the bounds 
\[     \lambda_{2,T}  \;\geq \;  4.3442  \;>\;3
  \hspace{5mm}\text{and}\hspace{5mm}  \lambda_{3,T} \;\geq \; 7.7248  \;>\; 5\,.
\]
For $d\geq 4$, enumerations are scarcer.  As an alternative, we use the inequality $t_N\geq s_N$ 
where $s_N$ is the number of site animals (in $\mathbb{L}^d$).  A site animal is a lattice animal such
that every edge of $\mathbb{L}^d$ having both endpoints in the animal must be in the animal.
We see that $t_N\geq s_N$ because every site animal contains a spanning tree, and a spanning 
tree uniquely determines the site animal.  Let $\lambda_{d,S} := \lim_{N\rightarrow\infty}s_N^{1/N}$
 (the limit exists by \cite{Kla}).
Hence we have $\lambda_{d,S} \,\leq \,\lambda_{d,T}$ for each $d$.   
Table 2 of \cite{BBO} provide suitable lower bounds for $\lambda_{d,S}$:
$\lambda_{4,S}\geq 9.7714$, 
$\lambda_{5,S}\geq 12.9569$,
$\lambda_{6,S}\geq 16.3087$,
$\lambda_{7,S}\geq 19.2927$,
$\lambda_{8,S}\geq 21.5298$, and
$\lambda_{9,S}\geq 24.6416$.  In each case, we easily have $\lambda_{d,T}\geq \lambda_{d,S}>2d-1$.
Moreover, since $\lambda_{12,S}\geq \lambda_{11,S}\geq\lambda_{10,S}\geq \lambda_{9,S}
>23$, we also see that $\lambda_{d,T}\geq \lambda_{d,S}>2d-1$ for $d=10,11,12$.
(As noted in \cite{BBO}, site animals in $\mathbb{L}^d$ are identified with $d$-dimensional polycubes,
which are connected unions of hypercubes of side 1 centred at each point of $\mathbb{Z}^d$).
Thus the corollary is proved for $d\leq 12$.

Finally, for large $d$, it is known that $\lambda_{d,T}\sim 2de$ as $d\rightarrow\infty$
\cite{MirS}.
\hfill  $\Box$

\bigskip
The preceding results have some implications for the possible limiting values of 
the quenched free energy.  In general, we have
\begin{eqnarray}
  \nonumber
    F_N^Q(0\,|\,\mathcal{P}^X)  & = &  \frac{1}{N}  
    \sum_{\tau} \frac{|\mathcal{P}_N^X[\tau]|}{ |\mathcal{P}^X_N|}  \,\log \left|\mathcal{P}^X_N[\tau]\right|
  \\
   \label{eq.FN0bound}
    & \leq &     \frac{1}{N}  
    \sum_{\tau} \frac{|\mathcal{P}_N^X[\tau]|}{ |\mathcal{P}^X_N|}  \,\log |M_N(\mathcal{P})|    
    \;= \; \frac{1}{N}\, \log |M_N(\mathcal{P})| \,.  \hspace{4mm}\hbox{ }
\end{eqnarray}
For trees, Equation (\ref{eq.FN0bound})
and Corollary \ref{cor.MNtree} imply 
that $\limsup_{N\rightarrow\infty} F_N^Q(0\,|\,\mathcal{T}^X)$ $\leq \,\log(2d-1)$ which is 
strictly less that $\mathcal{F}(0\,|\,\mathcal{T}^X) \,=\, \log \lambda_{d,T}$. 
In contrast, for rings in 3 dimensions, Proposition \ref{prop.MNeqmu} does not rule out the possibility
that the limiting quenched free energy equals the limiting annealed free energy.  
Indeed, if most $N$-step self-avoiding polygons  have knot types close to the commonest knot type,
and if these similar knot types $K$ all have $|\mathcal{R}_N^X[K]|\approx M_N(\mathcal{R})$,
then, since $M_N(\mathcal{R})\approx \mu_3^N$, 
it is certainly possible that $\lim_{N\rightarrow\infty} F_N^Q(0\,|\,\mathcal{R}^X)$ could equal
$\log \mu_3$ (which is $\mathcal{F}(0\,|\,\mathcal{R})$).  Note that it is highly plausible that 
the number of knot types $K$ such that $\mathcal{R}_N^X[K]\neq \emptyset$ grows exponentially in 
$N$, since it is known that the number of prime knots with crossing number of order $N$ grows
exponentially in $N$ \cite{ES}; however, the vast majority of these knot types might have relatively few
embeddings of size $N$.  Thus, even assuming its existence, the value of 
$\lim_{N\rightarrow\infty}F_N^Q(0\,|\,\mathcal{R}^X)$ is far from clear.

\section{Discussion}
    \label{sec.discussion}

We have introduced a mathematical framework for quenched polymer topology, with a focus on 
adsorption to surfaces, both penetrable and impenetrable.  
Our motivation is the physical situation that a large polymer molecule first forms
in a good solvent in the bulk, where all allowable configurations of a given size are 
approximately equally likely.
This fixes the polymer's topology but otherwise leaves the configuration flexible.  The polymer is
then subject to an interaction with a surface, whereupon it can move flexibly but cannot change its
(quenched) topology.  
This contrasts with the more commonly studied case of adsorption with annealed topology, where 
a polymer's topology is a product of its interaction with the surface, and is not fixed during the
adsorption process.

Our framework applies to the topology of branched polymer architecture in general dimension
as well as to the knot type of ring polymers in three dimensions.
The classes we consider are lattice animals, lattice trees, lattice combs, and self-avoiding polygons.
Since lattice combs have mainly been studied with restrictions on their branching, we prove 
some fundamental properties of the class of all lattice combs,
including the existence of its growth constant and of the annealed free energy.

When the surface attraction is strong, quenched topology 
maintains the natural complexity of a polymer with respect to 
local topological features that prevent it from adsorbing completely to a surface,  
in marked contrast to annealed topology.
This leads to provably different behaviour
of the expected energy per monomer in the limit as the attraction increases without bound. 

Although we are unable to prove that the quenched free energy per site 
$F^Q_N(\beta \,|\,\mathcal{P}^X)$ 
converges as $N\rightarrow\infty$ in our models, we can prove that 
$F^Q_N(\beta\,|\,\mathcal{P}^X)-F^Q_N(0\,|\,\mathcal{P}^X)$ converges whenever $\beta$ is negative, 
and in fact that the limit is 0 in this situation.
This permits a natural definition of the critical point for each model's quenched adsorption transition,
enabling comparison with the analogous annealed point.  
One major difference is that we cannot prove rigorously that there
is always a finite quenched adsorption transition; indeed, we conjecture that it does not exist 
for animals and trees in two dimensions.   We can prove that it exists 
for lattice combs in every dimension and for rings in three dimensions, while for animals and trees
above two dimensions we merely conjecture that it exists.

We consider the subclass of all polymers with a given topology for a given size $N$, and 
we prove bounds on the size of the largest such subclass.  In particular, we prove that the 
commonest tree structure is exponentially rare among all $N$-site lattice trees, but that 
the analogous statement for the commonest knot type is false.  The latter assertion disproves a 
conjecture of Janse van Rensburg and Rechnitzer \cite{RR}.

Many open questions remain.  The existence of the limiting quenched free energy appears to be
challenging to prove, requiring stronger techniques than concatenation and pattern theorems.
Proving that the quenched critical points for trees and animals 
are finite in three dimensions, and whether or not they
equal the annealed critical points, will likely need
an improved understanding of probabilities of nonlocal topological properties.

\subsection{Other classes of lattice polymers}
   \label{sec-discother}

To conclude, we mention some examples of  other kinds of polymers whose quenched behaviour
could also be studied by  the methods of this paper.
We expect that analogous results would hold for the following classes, although specific  
details may need further thought for some of the variants described.

For a first  example, a  \textit{bottlebrush polymer}  consists of side chains attached to a single 
backbone chain, like a comb polymer, except that two (or perhaps more) side chains attach to a single 
backbone site \cite{Den,PSPR}.  
One could also consider classes with different constraints
on the comb or bottlebrush side chains (e.g.\ lengths, spacings between chains, possibilities of 
different numbers attached to different points).

As mentioned in Section \ref{sec-quench}, 
we chose to ignore knotting properties of quenched animals in three dimensions, for simplicity.
Alternatively, one could define each class $\mathcal{A}_N^X[\cdot]$
of $N$-site animals to be such that two animals $\alpha_1$ and $\alpha_2$ 
are in the same class if and only if there exists a continuous deformation of $\mathbb{R}^3$ (more 
precisely, an ambient isotopy) that carries each edge of $\alpha_1$ to a distinct edge of 
$\alpha_2$.    In particular, in this situation we would 
necessarily have $G(\alpha_1)=G(\alpha_2)$, and hence
the modified class containing $\alpha_1$ would be a subset of the class 
$\mathcal{A}_N^X[G(\alpha_1)]$  that we use in the present paper.  This definition would make the 
treatment of three-dimensional animals consistent with our treatment of knots.

Another example is the set of entangled clusters in three dimensions,
which extends the idea of linking from knots to animals.  
Briefly, an entangled cluster  is a finite (not necessarily connected)
subgraph of  $\mathbb{L}^3$ 
that cannot be separated by a deformed sphere.  
They arise in entanglement percolation \cite{GH,KH}.
Since the number of $N$-site entangled clusters have a finite exponential growth rate \cite{AM}, 
one could 
use the quenched topology framework by putting two entangled clusters in the same class
if some ambient isotopy transforms one into the other.

Our final example is the set of (non-splittable) links in three dimensions.  A non-splittable 
link  can be described as an 
entangled cluster in which each connected component is a ring \cite{Ad,OrWh}. 
A link with one component is  a knot.  
Polymeric links with unrestricted numbers of components have been referred to as 
Olympic ring networks or Olympic gels \cite{DeG,LFWS}.  
Alternatively, one could consider the set of non-splittable links with a fixed number of connected 
components, which is closer to the traditional domain of knot theory \cite{Ad,Fla,OrWh}.   

These and other models are worthy of exploration.  It would be particularly interesting to find
differences in qualitative behaviours between similar-looking models.

\section*{Acknowledgments}
I am pleased to thank Stu Whittington for several discussions, including those that started off this
entire project.
I also thank Mahshid Atapour and Gill Barequet for helpful pointers to the literature
that contributed to the proofs of Proposition \ref{prop.MNeqmu} and Corollary \ref{cor.MNtree} respectively.  I also thank two referees for useful comments and suggestions.
 
%  For Springer: 
\section*{Statements and Declarations}
The author has no relevant financial or non-financial interests to disclose.  
 There is no conflict of interest.

\smallskip
\noindent
\underline{Funding:}  This research was supported in part by a Discovery Grant to the author from 
the Natural Sciences and Engineering Research Council of Canada.

\end{document}